\documentclass{article}

\usepackage{arxiv}

\usepackage{amsmath, amssymb}
\usepackage{amsfonts}       
\usepackage[utf8]{inputenc}
\usepackage[T1]{fontenc}   
\usepackage{hyperref}      
\usepackage{url}            
\usepackage{booktabs}    
\usepackage{nicefrac}      
\usepackage{microtype}     
\usepackage{lipsum}         
\usepackage{graphicx}
\usepackage{natbib}
\usepackage{doi}
\usepackage[title]{appendix}
\usepackage{booktabs, tabularx}
\usepackage{enumitem}
\usepackage{verbatim}
\usepackage{longtable}
\usepackage{amsthm}
\usepackage{bbm}
\usepackage{acro}
\usepackage{svg}
\usepackage{subfig}
\usepackage{circuitikz}
\usepackage{multirow}
\usepackage{needspace}
\usepackage{xargs}
\usepackage{algorithm}
\usepackage{algorithmicx}
\usepackage{cleveref}       
\usepackage{algpseudocode}

\theoremstyle{plain}
\newtheorem{theorem}{Theorem}
\newtheorem{proposition}[theorem]{Proposition}
\newtheorem{lemma}{Lemma}

\newtheorem*{remark}{Remark}

\theoremstyle{definition}
\newtheorem{definition}{Definition}
\newtheorem{assumption}{Assumption}
\crefname{assumption}{Assumption}{Assumptions}

\crefname{assumptionA}{Assumption}{Assumptions}
\newcommand{\E}{\mathbb{E}}

\newcommand{\M}{50}
\newcommand{\B}{200}

\newcommand{\beginsupplement}{%
        \setcounter{table}{0}
        \renewcommand{\thetable}{\arabic{table}}%
        \renewcommand{\theHtable}{S\arabic{table}}%
        \renewcommand{\tablename}{Web Table}%
        \setcounter{figure}{0}
        \renewcommand{\thefigure}{\arabic{figure}}%
        \renewcommand{\theHfigure}{S\arabic{figure}}%
        \renewcommand{\figurename}{Web Figure}%
     }

\makeatletter
\newenvironment{breakablealgorithm}
  {
   \begin{center}
     \refstepcounter{algorithm}
     \hrule height.8pt depth0pt \kern2pt
     \renewcommand{\caption}[2][\relax]{
       {\raggedright\textbf{\ALG@name~\thealgorithm} ##2\par}
       \ifx\relax##1\relax
         \addcontentsline{loa}{algorithm}{\protect\numberline{\thealgorithm}##2}
       \else
         \addcontentsline{loa}{algorithm}{\protect\numberline{\thealgorithm}##1}
       \fi
       \kern2pt\hrule\kern2pt
     }
  }{
     \kern2pt\hrule\kern2pt
   \end{center}
  }
\makeatother

\renewcommand{\shorttitle}{DR-Learners for Generalizing CATEs}

\hypersetup{
pdftitle={DR-Learners for Generalizing CATEs from a Randomized Controlled Trial},
pdfsubject={stat.ME, stat.ML},
pdfauthor={Vanessa Rodriguez, Brieuc Lehmann, Karla Diaz-Ordaz},
pdfkeywords={Heterogeneous treatment effects, Generalizability, Causal machine learning, Covariate shift},
}

\acsetup{
  uppercase/list,
  make-links = false,
  pages/display = none,
  pages/fill = {, },
  list/sort = true
}

\DeclareAcronym{ml}{
  short = ML,
  long = machine learning
}

\DeclareAcronym{cate}{
  short = CATE,
  long = conditional average treatment effect
}

\DeclareAcronym{scate}{
  short = sCATE,
  long = sample conditional average treatment effect
}

\DeclareAcronym{ate}{
  short = ATE,
  long = average treatment effect
}

\DeclareAcronym{ef}{
  short = EF,
  long = ejection fraction
}

\DeclareAcronym{ols}{
  short = OLS,
  long = ordinary least squares
}

\DeclareAcronym{pcate}{
  short = pCATE,
  long = target-population conditional average treatment effect
}

\DeclareAcronym{dgp}{
  short = DGP,
  long = data-generating process
}

\DeclareAcronym{rct}{
  short = RCT,
  long = randomized controlled trial
}

\DeclareAcronym{mi}{
  short = MI,
  long = myocardial infarction
}

\DeclareAcronym{dag}{
  short = DAG,
  long = directed acyclic graph
}

\DeclareAcronym{ipw}{
  short = IPW,
  long = inverse probability weighting
}

\DeclareAcronym{iptw}{
  short = IPTW,
  long = inverse probability of treatment weighting
}

\DeclareAcronym{ipsw}{
  short = IPSW,
  long = inverse probability of sampling weighting
}

\DeclareAcronym{hte}{
  short = HTE,
  long = heterogeneous treatment effect
}

\DeclareAcronym{eif}{
  short = EIF,
  long = efficient influence function
}

\DeclareAcronym{aipw}{
  short = AIPW,
  long = augmented inverse probability weighting
}

\DeclareAcronym{ite}{
  short = ITE,
  long = individual treatment effect
}

\DeclareAcronym{mse}{
  short = MSE,
  long = mean squared error
}

\title{DR-Learners for Generalizing CATEs from a Randomized Controlled Trial}

\newif\ifuniqueAffiliation
\uniqueAffiliationfalse

\usepackage{authblk}

\newbox{\orcid}\sbox{\orcid}{\includegraphics[scale=0.06]{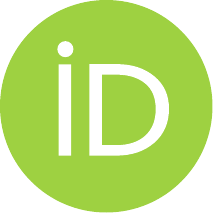}}

\author[1]{%
  \href{https://orcid.org/0000-0002-4643-2920}{\usebox{\orcid}\hspace{1mm}Vanessa Rodriguez\thanks{Corresponding author. \texttt{vanessa.rodriguez.22@ucl.ac.uk}}}%
}
\author[1]{%
  \href{https://orcid.org/0000-0002-7302-4391}{\usebox{\orcid}\hspace{1mm}Brieuc Lehmann}%
}
\author[1]{%
  \href{https://orcid.org/0000-0003-3155-1561}{\usebox{\orcid}\hspace{1mm}Karla Diaz-Ordaz}%
}
\affil[1]{Department of Statistical Science, University College London, Gower Street, London WC1E 6BT, United Kingdom}

\begin{document}
\maketitle

\begin{abstract}
Randomized clinical trials are essential for evaluating interventions, yet average treatment effects can obscure treatment effect heterogeneity, making conditional average treatment effects (CATEs) a natural target. However, restrictive eligibility criteria and differential trial participation can limit the generalizability of these effects to broader target populations. Building on existing DR-Learners, we develop the gDR-Learner for generalizing RCT CATEs to a target population, using cross-fitted machine-learning nuisance estimation. We derive excess-risk bounds in which nuisance errors enter only as products, showing that the estimator achieves oracle-rate convergence despite slower individual nuisance estimation.
The variables needed to generalize are rarely measured across the entire target population. We therefore extend the methods to two settings with incomplete covariate information: when certain covariates are measured only in the trial (e.g., biomarkers), and covariates driving trial participation measured only in a subsample of the target population, such as a representative cohort or registry. 
We construct weighted conformal prediction intervals for individual treatment effects, accounting for covariate shift between the trial and target population. In simulations, the gDR-Learner achieved the lowest estimation error when all covariates were observed relative to standard T- and DR-Learners and an inverse-probability-weighted T-Learner, and also outperformed the latter under incomplete covariate observation.
We illustrate the methods by reanalyzing the Coronary Artery Surgery Study, examining treatment effect heterogeneity by prior myocardial infarction and baseline ejection fraction.
\end{abstract}


\section{Introduction}
\Acp{rct} remain the gold standard for estimating causal treatment effects. However, their findings do not always generalize to broader populations relevant to clinicians and policymakers. For instance, trial participants are frequently unrepresentative of the target population due to restrictive eligibility criteria, self-selection, and logistical constraints \citep{van_spall_eligibility_2007, giuliano_participation_2000}. This gives rise to covariate shift: systematic differences in the distribution of baseline patient characteristics between the trial sample and the target population. Such shift implies that even a well-conducted trial may yield misleading treatment effect estimates for under-represented patients, particularly when effects are heterogeneous across individuals. 

While methods for generalizing average treatment effects to target populations have been extensively studied \citep{cole_generalizing_2010, dahabreh_generalizing_2019-1}, here we focus on data-adaptive methods to estimate complex functions of possibly high dimensional data for generalizing heterogeneous treatment effects which may not generalize when the distribution of the effect modifiers differ between populations. 
In particular, we consider a nested trial design \citep{dahabreh_generalizing_2019-1}, in which an \ac{rct} is nested within a cohort from the target population, such as a registry or observational cohort of trial-eligible individuals in which baseline covariates are observed for the entire cohort, whereas treatment and outcome are observed only for trial participants.
The estimand of interest is the \ac{pcate}, defined as $\tau(X_\tau) = \mathbb{E}[Y^1 - Y^0 | X_\tau]$, where $Y^a$ denotes potential outcome under a binary treatment level $a\in\{0,1\}$, $X_\tau$ are the baseline pre-specified effect modifiers, and where, crucially, the expectation is taken over the target population
thus distinguishing it from the \ac{scate}, defined with respect to the sample, $\tau^{\text{sCATE}}= \mathbb E[Y^1 - Y^0\mid X_\tau=x_\tau,S=1]$, where $S=1$ for trial participants. Because the \ac{ite}, $Y^1-Y^0$, is fundamentally unobservable for any individual, the pCATE provides an estimable, covariate-specific summary of treatment-effect heterogeneity that can be used to inform personalized treatment decisions.

The \ac{scate} and the \ac{pcate} will differ whenever the covariates driving sample selection (i.e. trial inclusion) are also treatment effect modifiers. Earlier literature focused on generalizing CATEs defined by discrete variables \citep{mehrotra_transporting_2021,robertson_estimating_2024}, by either IPW or g-computation and doubly robust (DR) methods. These DR methods could be readily implemented using data-adaptive estimation of the nuisance models, by applying the established literature of debiased machine learning, based on efficient influence functions \citep{chernozhukov_doubledebiased_2018, van_der_laan_foundations_2011}. 
Several data-adaptive estimators for a \ac{cate} indexed by continuous covariates are available in the case where generalizability is not of interest, such as the T-Learner, DR-Learner and other meta-learners \citep{kunzel_metalearners_2019, kennedy_towards_2023}.  
\citet{robertson_regression-based_2023} extended the DR-Learner to generalizability settings. Their implementation using parametric nuisance models is consistent provided that either the outcome model or both the treatment and trial-sample selection models are correctly specified, an increasingly difficult task when the number of baseline covariates grows -- for example, in trials collecting omics data. While they noted that the nuisance components can be estimated using \ac{ml}, and that cross-fitting should be used to control overfitting bias, they neither implemented such a scheme nor derived the asymptotic properties of such an estimator.

Here, we implement this generalized DR-Learner (denoted as gDR-Learner) with non-parametric nuisance estimation and cross-fitting and prove that under reasonable conditions on the convergence of the nuisance estimators, the non-parametric gDR-Learner performs asymptotically as well as if the true nuisance functions were known. We also propose a generalized T-Learner (gT-Learner), a weighted-regression plug-in estimator without this orthogonal correction, as a simpler comparator.
Furthermore, covariate availability often differs across data sources, for example (i) trials may collect detailed biomarkers and laboratory measurements that are absent from the target population data, or (ii) covariates driving trial participation may be available only in a subsample of the target population. We therefore present extensions of the gDR-Learner that handle these settings in Section \ref{sec:extensions}.

We then consider uncertainty quantification in Section~\ref{sec:conformal}, constructing weighted conformal prediction intervals for ITEs using the estimated pCATE as the predictor \citep{tibshirani_conformal_2019, lei_conformal_2021}. This approach accounts for covariate shift between the trial and target population and provides approximate marginal coverage guarantees without relying on smoothness or distributional assumptions. 

In Section~\ref{sec:simulation}, we evaluate these methods' performance via simulation. Section~\ref{sec:cass} illustrates the methods by re-analysing the Coronary Artery Surgery Study \citep{killip_coronary_1985}, to estimate pCATEs conditional on history of myocardial infarction and baseline ejection fraction value for the target population of trial-eligible patients with stable ischemic heart disease, as previously done by \citet{robertson_regression-based_2023}, before concluding in Section~\ref{sec:discussion}.

\section{Estimands and identification}
\label{sec:notation}

\subsection{Set-up and notation}

Consider an \ac{rct} nested within a target population. 
Let $n_{1}$ and $n_{0}$ denote the numbers of trial participants ($S=1$) and non-participants ($S=0$). Let $A\in\{0,1\}$ be the binary treatment and $Y$ the (binary or continuous) outcome. We observe $n=n_{1}+n_{0}$ i.i.d.\ units with data $O=(X,S,S\cdot A,S\cdot Y)\sim P$, where $P$ denotes the true observed data generating distribution. Until Section~\ref{sec:extensions}, we assume that the covariates $X$ are observed for all units, whereas $A$ and $Y$ are observed only when $S=1$. 
Let $X_\tau \subseteq X$ denote the effect modifiers of interest, i.e. the observed covariates over which the CATE is defined. Analogously, let $X_S \subseteq X$ be the set of measured covariates that drive trial participation. The set of the remaining variables in $X$, which are neither driving selection into the study nor defining the estimand, is denoted by  $X_P$. Note that $X_S$ and $X_\tau$ are not mutually exclusive, but $X_P$ is disjoint from both $X_S$ and $X_\tau$. For example, if age predicts $Y$ and also influences trial participation, then age should be included in $X_S$ but not $X_P$. If age is additionally a treatment effect modifier of interest, it should also be included in $X_\tau$.

Formally, our estimand of interest, the \ac{pcate} conditional on $X_\tau$, can be written as
\begin{equation}
\label{eq:estimand}
\tau(x_\tau) = \mathbb E[Y^1 - Y^0 \mid X_\tau = x_\tau].
\end{equation} 

\subsection{Identification}
We now outline the key assumptions required to identify $\tau(x_\tau)$ using observed data.
Throughout, we assume \textit{no interference}, and \textit{consistency}: for $a \in \{0,1\}$, if $A_i=a$, then the corresponding observed outcome is $Y_i=Y_i^a$. 
In addition, \vspace{-\topsep}
\begin{assumption}[Identification]
\label{assump:identification}
For $a \in \{0,1\}$:
\begin{enumerate}[label=(\alph*), ref=\theassumption(\alph*), leftmargin=*, itemsep=2pt, topsep=2pt]
\item \label{assump:mean_exchangeability_trial} \textbf{Mean conditional exchangeability in the trial.} $\mathbb{E}[Y^a \mid X_S \cup X_\tau, A=a, S=1] = \mathbb{E}[Y^a \mid X_S \cup X_\tau, S=1]$. That is, conditional on $X_S \cup X_\tau$, treatment assignment $A$ is unconfounded in the randomized trial. This condition holds by design in a randomized trial.
\item \label{assump:treatment_positivity} \textbf{Positivity of treatment assignment in the trial.} For all $x \in \text{supp}(X_S \cup X_\tau)$ with positive density in the trial, $0 < \mathbb{P}(A=1 \mid X_S \cup X_\tau = x, S=1) < 1$. This holds automatically in fully randomized trials by design.
\item \label{assump:participation_positivity} \textbf{Positivity of trial participation.} For all $x \in \text{supp}(X_S \cup X_\tau)$, $\mathbb{P}(S=1 \mid X_S \cup X_\tau = x) > 0$. This guarantees overlap in the distribution of $X_S \cup X_\tau$ across trial and target populations.
\item \label{assump:mean_exchangeability_populations} \textbf{Mean conditional exchangeability across populations.} $\mathbb{E}[Y^a \mid X_S \cup X_\tau, S=1] = \mathbb{E}[Y^a \mid X_S \cup X_\tau]$. Conditional on $X_S \cup X_\tau$, the mean potential outcomes are the same in the randomized trial and in the target population.
\end{enumerate}
\end{assumption}
\begin{remark}
    Note that when treatment is completely randomized, $\mathbb{P}(A=a \mid S=1)= \mathbb{P}(A=a \mid  X_S, S=1) \triangleq \pi_a(X_S)$. Moreover, all variables that drive selection are included in $X_S$, implying that $\mathbb P(S=1|X_S \cup X_\tau) = \allowbreak \mathbb P(S=1|X_S) \triangleq \omega(X_S)$.
\end{remark}
Importantly, there could be other variables not in $X_\tau$ which are true treatment effect modifiers. They do not imply lack of generalizability of sCATE, provided they do not also influence trial participation. 
Under Assumptions~\ref{assump:mean_exchangeability_trial}-\ref{assump:mean_exchangeability_populations}, the \ac{pcate} eq. \eqref{eq:estimand} can be identified as,
\begin{equation}
      \tau(X_\tau)=  \allowbreak \mathbb E[\mathbb E[Y|X_S\cup X_\tau, S=1, A=1] \allowbreak - \mathbb E[Y|X_S\cup X_\tau, S=1, A=0]\vert X_\tau].
     \label{eq:g-comp-identification}
\end{equation}
Equivalently, the \ac{pcate} can be identified using \ac{ipw} representation as
\begin{align}
    \tau(X_\tau)= \mathbb E\left[\frac{\mathbb{I}(S=1)\mathbb{I}(A=1)}{\omega(X_S)\pi_1(X_S)}Y - \frac{\mathbb{I} (S=1)\mathbb{I} (A=0)}{\omega(X_S)\pi_0(X_S)}Y \middle \vert X_\tau\right]. \label{eq:ipw-identification}
\end{align}
See Web Appendix~\ref{app:identification} for the derivation of both identification results. 

\section{Estimating the Population CATE}
\label{sec:estimation}

\subsection{Generalized T-Learner}\label{sec:tlearner}
A first natural strategy to estimate pCATE is a plug-in estimator, simply substituting fitted nuisance functions into eq. \eqref{eq:g-comp-identification}. 
This two-step strategy proceeds by first estimating the conditional outcome functions $\mu_a(X_S \cup X_\tau) = \mathbb{E}[Y|A=a, X_S \cup X_\tau, S=1]$ for $a \in \{0,1\}$ using the trial data ($S=1$).
Then, use the trained model $\widehat{\mu}_a(X_S \cup X_\tau)$ to predict the potential outcomes for every unit in the target population, both trial participants ($S=1$) and non-participants ($S=0$), and marginalize over the distribution of the selection covariates $X_S$ in the population obtaining $\tau(X_\tau)$.
Marginalizing over the distribution of these covariates can be a complex operation if it involves continuous variables, often requiring numerical methods. 

Instead, here we extend the T-Learner \citep{kunzel_metalearners_2019}, based on the \ac{ipw} identification eq.~\eqref{eq:ipw-identification}. This generalized T-Learner (gT-Learner) is defined as the difference between two weighted outcome regressions,
$\widehat \tau_{\text{gTL}}(X_\tau) = \widehat{\mu}_1(X_\tau;\widehat{\omega},\widehat{\pi}) - \widehat{\mu}_0(X_\tau;\widehat{\omega},\widehat{\pi})$, each estimating the arm-specific regression 
\begin{equation} 
\mu_a(X_\tau;\omega,\pi) = {\mathbb{E}}\left[\frac{\mathbb{I}(S=1)\mathbb{I}(A=a)}{\omega(X_S)\pi_a(X_S)}Y \,\middle|\, X_\tau\right].
\end{equation} 
We fit $\widehat\omega$ on the combined trial and target sample and $\widehat\pi_a$ on the trial data, then regress $Y$ on $X_\tau$ among the arm-$a$ participants, weighting each unit by $1/\{\widehat\omega(X_S)\,\widehat\pi_a(X_S)\}$. 
We prefer the gT-Learner based on the \ac{ipw} representation, rather than the g-computation representation in eq.~\eqref{eq:g-comp-identification}, as it allows the final regression to be performed directly on $X_\tau$ while accounting for differences in the distribution of the selection covariates $X_S$. 

An outcome-regression approach could instead estimate
$
\mathbb{E}[Y \mid A=a, X_S \cup X_\tau, S=1],
$
but would then require a second regression of the fitted values on $X_\tau$ over the target population to marginalize over $X_S$, with $X_S$ observed for every unit. Alternatively, regressing the outcome directly on $X_\tau$ alone would require the additional assumption,
$
\mathbb{E}[Y \mid A=a, X_S \cup X_\tau, S=1]
=
\mathbb{E}[Y \mid A=a, X_\tau, S=1].
$
The assumption holds automatically if $X_S \subseteq X_\tau$. The \ac{ipw} formulation avoids the need for this assumption.

Under either representation, the gT-Learner bias term depends on first-order nuisance estimation errors in $\widehat{\mu}_a(X_S\cup X_\tau)$, $\widehat{\omega}(X_S)$ and $\widehat{\pi}(X_S)$ (Lemma~\ref{lem:gtl-error} in Web Appendix~\ref{app:error-bounds}). Consequently, using data adaptive nuisance estimators that converge more slowly than the oracle rate, the rate achieved by an estimator with access to the true nuisance functions, induces plug-in bias that propagates directly to the pCATE estimate. Under correct specification, parametric nuisance models can avoid the resulting plug-in bias due to achieving parametric convergence rates, although they can still be subject to misspecification bias.  

\subsection{Generalized DR-Learner}
A general strategy to construct estimators that are insulated from slow nuisance model convergence is to use the estimand's \ac{eif}, which measures its sensitivity to perturbations of the data-generating distribution. This strategy does not work in the general \ac{cate} setting, when $X_\tau$ contains continuous covariates, as the \ac{cate} is not pathwise differentiable and therefore a finite-variance \ac{eif} does not exist \citep{kennedy_towards_2023}. Estimators can instead be constructed from the \ac{eif} of a well-chosen risk function, such as a measure of counterfactual prediction error over the target population \citep{foster_orthogonal_2023}. The resulting loss is said to be Neyman orthogonal, meaning that its first-order derivative with respect to the nuisance functions vanishes at the true nuisance values, making the resulting estimator nearly insensitive to errors in nuisance estimation.

The squared-error loss is a natural criterion for estimating a conditional mean contrast. Substituting estimated nuisance functions directly into the squared-error risk introduces first-order bias, since the risk is not orthogonal to errors in nuisance function estimation. For squared-error loss, the resulting orthogonal criterion is, up to terms that do not depend on $\tau$, equivalent to 
\begin{equation} \label{eq:orthogonal-risk} 
L(\tau,\eta) = \mathbb{E}\left[ \left\{ \varphi_{\mathrm{gDR}}(O) - \tau(X_\tau) \right\}^2 \right], 
\end{equation} where 
\begin{equation} 
\label{eq:pseudo-outcome} 
\varphi_{\mathrm{gDR}}(O) = \frac{\mathbb{I}(S=1)\{A-\pi_1(X_S)\}} {\omega(X_S)\pi_1(X_S)\pi_0(X_S)} \left\{Y-\mu_A(X_S\cup X_\tau)\right\} 
+ \mu_1(X_S\cup X_\tau) - \mu_0(X_S\cup X_\tau), 
\end{equation} and $\eta=(\pi,\omega,\mu_1,\mu_0)$ denotes the nuisance functions. See Web Appendix~\ref{app:eif} for the derivation.

In practice, estimating the minimizer of this loss proceeds in two stages. First, the nuisance functions $\eta=(\pi,\omega,\mu_1,\mu_0)$ are estimated and used to construct the pseudo-outcome $\varphi_{\mathrm{gDR}}(O)$. Second, this pseudo-outcome is regressed on $X_\tau$ using a chosen regression learner.
Thus, the second-stage regression targets the conditional expectation of the pseudo-outcome given $X_\tau$. This corresponds to the gDR-Learner of \citet{robertson_regression-based_2023}.
This gDR-Learner not only minimizes the mean squared error loss for the pCATE, but, as it uses the EIF of the loss to construct its pseudo-outcomes, it is also less sensitive to errors in the nuisance function estimation, justifying the use of data-adaptive nuisance estimators. This is formalized as follows:
\begin{theorem}[Excess risk bound for the gDR-Learner]
\label{thm:gdr-excess-risk-main}
Let $\eta=(\pi,\omega,\mu_1,\mu_0)$ denote the nuisance vector, with true value $\eta_0$ and
cross-fitted estimate $\widehat\eta$, and write $L(\tau,\eta)$ for the risk (eq.~\eqref{eq:orthogonal-risk}) evaluated
under $\eta$. Let $\tau^*$ minimize $L(\cdot,\eta_0)$.
Under Assumptions~\ref{assump:mean_exchangeability_trial}--\ref{assump:mean_exchangeability_populations}, bounded outcomes, bounded selection and propensity scores,
and regularity conditions (Assumptions~A--D in Web Appendix~\ref{app:error-bounds}),
with nuisance estimates obtained via cross-fitting, the excess risk of the gDR-Learner satisfies
\begin{align}
L(\widehat{\tau}, \eta_0) - L(\tau^*, \eta_0)
\;\lesssim\; &\;\mathrm{Rate}_\tau(\widehat{\eta}) \notag + \mathrm{Rate}_\pi^2\mathrm{Rate}_{\mu_1}^2
 + \mathrm{Rate}_\pi^2\mathrm{Rate}_{\mu_0}^2 + \mathrm{Rate}_\omega^2\mathrm{Rate}_{\mu_1}^2\notag\\
&
 + \mathrm{Rate}_\omega^2\mathrm{Rate}_{\mu_0}^2 + \mathrm{Rate}_\pi^2\mathrm{Rate}_\omega^2
 + \mathrm{Rate}_\pi^4 + \mathrm{Rate}_\omega^4,
 \label{eq:gdr-excess-risk-main}
\end{align}
where $\mathrm{Rate}_\tau(\widehat{\eta})$ bounds the excess risk of the second-stage regression under estimated nuisances, and $\mathrm{Rate}_\pi = \|\widehat{\pi} - \pi_0\|_2$,
$\mathrm{Rate}_\omega = \|\widehat{\omega} - \omega_0\|_2$,
$\mathrm{Rate}_{\mu_a} = \|\widehat{\mu}_a - \mu_{a,0}\|_2$ are the nuisance
estimation errors. The full statement with explicit constants, and the proof are given in
Theorem~\ref{thm:gdr-excess-risk} in Web Appendix~\ref{app:error-bounds}.
\end{theorem}

Unlike the gT-Learner (Web Appendix Lemma~\ref{lem:gtl-error}), where nuisance errors have a first order impact, the gDR-Learner remainder involves only products of squared nuisance errors. Consequently, the estimator $\widehat{\tau}_\text{gDR}$ achieves the oracle rate whenever each of these products is of order no larger than the oracle rate. The outcome errors enter only through products with the selection and propensity errors, so a slower outcome model rate can be offset by faster selection and propensity score rates. The oracle rate depends only on the complexity of the true \ac{pcate} and on the choice of second-stage regression.

\subsection{Cross-fitting}
\label{sec:cross-fitting}

Theorem~\ref{thm:gdr-excess-risk-main} requires the nuisance functions $\eta$ to be estimated by sample splitting. This is because Neyman orthogonality removes first-order plug-in bias, but additional overfitting bias may arise when the same observations are used both to estimate the nuisance functions and to construct the pseudo-outcomes. In that case, the residuals used in the pseudo-outcome are in-sample prediction errors which might be too small to provide the intended bias correction.

To avoid efficiency loss, cross-fitting is used, instead of simple sample splitting.
Cross-fitting estimates the nuisance functions on separate training folds, so that, conditional on those folds, they can be treated as fixed when constructing the pseudo-outcomes (see \citet{chernozhukov_doubledebiased_2018,foster_orthogonal_2023}). As an alternative to sample splitting, semiparametric theory typically imposes restrictions on the complexity of the nuisance models, such as a Donsker condition, but these are not appropriate here as they may exclude flexible machine learning algorithms \citep{chernozhukov_doubledebiased_2018}.

In practice, the sample is partitioned into $K$ folds. For each fold, the nuisance functions are estimated using the remaining $K-1$ folds and are then used to construct pseudo-outcomes for the held-out observations. The folds are stratified by $S$, to ensure $\widehat{\omega}$ can be estimated well across all folds. Then, the cross-fitted pseudo-outcomes are pooled and used in the second-stage regression. Full algorithms are provided in Web Appendix~\ref{app:cross-fit-alg}.

\section{Extensions for Incomplete Target Data}
\label{sec:extensions}
The practical utility of the estimators introduced in the previous section rests on which covariates are observed in the trial ($S=1$) and the non-trial population ($S=0$). In many applications, data collection protocols differ across sources: randomized trials often measure rich baseline characteristics, whereas target population datasets (e.g., electronic health records, registries, or censuses) may contain only a restricted subset of covariates. We now present extensions to address this challenge in two specific settings. 

\subsection{Trial-Only Covariates}
\label{sec:prognostic}
RCTs routinely collect additional baseline covariates $X_P$ which, while not required for identification, may be strongly predictive of the outcome $Y$. Incorporating these variables can therefore improve the precision of \ac{pcate} estimates. These variables are typically not available in the non-randomized portion of the data, i.e. we observe $(X_S \cup X_\tau,X_P,A,Y)$ when $S=1$, but only $(X_S \cup X_\tau)$ when $S=0$. We refer to estimators that make use of these additional RCT covariates as `enriched' learners.

If $X_P$ were observed for all individuals in the target population, we could use a two-stage plug-in strategy (similar to that outlined in Section \ref{sec:tlearner}), where we first model $\mu_a(X_S \cup X_\tau,X_P)=\mathbb E[Y|X_S \cup X_\tau,X_P, S=1, A=a]$ in the trial data and then using the trained model, we obtain $\widehat{\mu}_a(X_S \cup X_\tau,X_P)$ across the target population, taking the difference between the two arms and finally regressing the resulting  difference onto $X_\tau$.

Since $X_P$ is not available for $S=0$ observations,  we cannot evaluate $\widehat{\mu}_a(X_S \cup X_\tau, X_P)$ on $S=0$, preventing the direct marginalization over $(X_S,X_P)$ in the target population. 
We therefore first marginalize $X_P$ within the trial by defining
\begin{align}
    g_{a,\text{enr}}(X_S \cup X_\tau) = \mathbb E[\mu_a(X_S \cup X_\tau,X_P)\mid X_S \cup X_\tau, S=1].
\end{align}
We can now identify the pCATE by taking expectations over $(X_S \cup X_\tau)$ in the target population,
$
\tau(X_\tau) = \mathbb{E}[g_{1,\text{enr}}(X_S \cup X_\tau) - g_{0,\text{enr}}(X_S \cup X_\tau) \mid X_\tau],
$
where this outer expectation is taken over the target population, in which $(X_S \cup X_\tau)$ is fully observed. Because randomization implies $A\perp X_P\mid (X_S\cup X_\tau,S=1)$, we have $g_{a,\text{enr}}(X_S\cup X_\tau)=\mu_a(X_S\cup X_\tau)$: the enriched learners target the same nuisance function as the standard learners, but estimate it through a regression that also conditions on $X_P$, which for the gDR-Learner reduces the variance of the residual $Y-\widehat\mu_A$ in the pseudo-outcome. Identification is established in Web Appendix~\ref{app:identification-prognostic}.

In practice, we first estimate $\mu_a(X_S \cup X_\tau, X_P)$ using the trial data and then estimate $g_{a,\mathrm{enr}}(X_S \cup X_\tau)$ by regressing the corresponding cross-fitted predictions $\widehat\mu_a(X_S \cup X_\tau, X_P)$ on $X_S \cup X_\tau$ among trial participants. The resulting $\widehat g_{a,\mathrm{enr}}$ can then be evaluated for all individuals.

We obtain the enriched gT-Learner $\widehat\tau_{\text{gTL,enr}}(X_\tau)$ by first computing, for each unit, the contrast
$
\widehat g_{1,\text{enr}}(X_S \cup X_\tau) - \widehat g_{0,\text{enr}}(X_S \cup X_\tau),
$
and then regressing this contrast on $X_\tau$. 
Similarly, we construct an enriched gDR-Learner using the pseudo-outcome 
\begin{align*} 
\widehat\varphi_{\text{gDR,enr}}(O) & = \frac{\mathbb{I}(S=1)\{A-\widehat\pi_1(X_S)\}} {\widehat\omega(X_S)\widehat\pi_1(X_S)\widehat\pi_0(X_S)} \left\{Y-\widehat\mu_A(X_S \cup X_\tau,X_P)\right\} \\
& ~~+ \widehat g_{1,\text{enr}}(X_S \cup X_\tau) - \widehat g_{0,\text{enr}}(X_S \cup X_\tau). 
\end{align*}
Regressing $\widehat\varphi_{\text{gDR,enr}}(O)$ on $X_\tau$, yields the enriched gDR-Learner, $\widehat\tau_{\text{gDR,enr}}(X_\tau)$.

\subsection{Restricted Selection Covariates}
\label{sec:Estimation-with-Restricted-Target-Covariates}
Target population data often lack covariates required to identify the pCATE, (effect modifiers that also drive trial participation and whose distributions are shifted across the populations). For example, variables affecting selection into a trial might be unmeasured in population registries, leaving $X_S$ partially unobserved outside the trial and preventing direct estimation of the trial selection model $\omega(X_S)$. 

Suppose that an auxiliary external dataset that records both $X_S$ and $X_\tau$ for a representative sample of the target population, such as a biobank or household survey, is available. Let $S'=1$ denote members of the auxiliary sample. We assume that the auxiliary sample is a random sample of the target population drawn independently of trial participation,
\begin{equation}
\label{eq:aux-assump}
    S' \perp (S, X_S),
\end{equation}
which implies $\mathbb P(S'=1\mid X_S,S)=\mathbb P(S'=1)$. Here, $\mathbb P(S'=1)$ is assumed known by design or consistently estimable from the data.

Since the two samples from our target population are not necessarily disjoint, we define an indicator $M = \mathbb I(S=1 \cup S'=1)$, so $M=1$ whenever an individual belongs to either sample. The observed data is then $O = (S, S',X_\tau, M\cdot(X_S\setminus X_\tau),\, S\cdot A,\, S\cdot Y)$ so that $X_\tau$ is observed for every individual in the target population, $X_S$ is observed only when $M=1$, and treatment and outcome are observed only within the trial.

We can then recover $\omega(X_S)$ in the full population under the independence assumption (eq.~\eqref{eq:aux-assump}) and through Bayes' Rule, 
\begin{equation}
\omega(X_S) = \mathbb P(S=1\mid X_S) =\frac{\mathbb P(S'=1)\mathbb P(S=1\mid X_S,M=1)} {1-\{1-\mathbb P(S'=1)\}\mathbb P(S=1\mid X_S,M=1)}.
\label{eq:omega-subsample}
\end{equation}
We obtain $\widehat\omega(X_S)$ by applying this transformation to the estimate of $\mathbb P(S=1|X_S, M=1)$, which we obtain by regressing $S$ onto $X_S$ in the combined trial and auxiliary sample population.

We first consider extensions of the gT-Learner. Once $\widehat\omega(X_S)$ has been recovered from the combined samples, the gT-Learner of Section~\ref{sec:tlearner} can be applied directly. We refer to this estimator as the gT-Learner (Subsample). For each treatment arm $a\in\{0,1\}$, $Y$ is regressed on $X_\tau$ among trial participants with $A=a$, weighting each observation by $1/\{\widehat\omega(X_S)\widehat\pi_a(X_S)\}$. The resulting estimator is
$
\widehat\tau_{\mathrm{gTL,sub}}(X_\tau)=\widehat\mu_{1,\mathrm{sub}}(X_\tau;\widehat\omega,\widehat\pi)-\widehat\mu_{0,\mathrm{sub}}(X_\tau;\widehat\omega,\widehat\pi).
$
Here, $\widehat\omega$ is estimated using $M=1$, $\widehat\pi_a$ and the two weighted outcome regressions estimated using $S=1$. Since $X_S$ is required only to construct weights for trial participants, the gT-Learner (Subsample) remains feasible when $X_S$ is unobserved outside of $M=1$.

An alternative gT-Learner is obtained by marginalizing the arm-specific outcome regressions over $X_S$ within the trial using inverse-selection weighting:
$$
g_{a,\mathrm{res}}(X_\tau)=\mathbb E\left[\frac{\mathbb I(S=1)}{\omega(X_S)}\mu_a(X_S\cup X_\tau)\middle|X_\tau\right].
$$
We estimate $g_{a,\mathrm{res}}(X_\tau)$ by regressing the fitted values $\widehat\mu_a(X_S\cup X_\tau)$ on $X_\tau$ among trial participants ($S=1$), weighting observations by $1/\widehat\omega(X_S)$. The resulting gT-Learner (Restricted) is
$
\widehat\tau_{\mathrm{gTL,res}}(X_\tau)=\widehat g_{1,\mathrm{res}}(X_\tau)-\widehat g_{0,\mathrm{res}}(X_\tau).
$
As shown in Web Appendix~\ref{app:identification-restricted}, this contrast identifies the \ac{pcate}.

We next consider extensions of the gDR-Learner. A direct analogue of the standard gDR-Learner, although infeasible, can be defined by replacing the trial participation model with $\widehat\omega(X_S)$ estimated from pooling the trial and auxiliary samples, while otherwise retaining the standard pseudo-outcome. We refer to this estimator as the gDR-Learner (Subsample), with pseudo-outcome
\begin{align*}
\widehat\varphi_{\mathrm{sub}}(O) & = \frac{\mathbb I(S=1)\{A-\widehat\pi_1(X_S)\}}{\widehat\omega(X_S)\widehat\pi_1(X_S)\widehat\pi_0(X_S)}\left\{Y-\widehat\mu_A(X_S\cup X_\tau)\right\} \\
& ~~ + \widehat\mu_1(X_S\cup X_\tau)-\widehat\mu_0(X_S\cup X_\tau).
\end{align*}
The gDR-Learner (Subsample) would then be obtained by regressing this pseudo-outcome on $X_\tau$ over the full target-population cohort.

Unlike the gT-Learner (Subsample), however, the gDR-Learner (Subsample) is infeasible due to the restricted-observation data structure. Its plug-in component, $\widehat\mu_1(X_S\cup X_\tau)-\widehat\mu_0(X_S\cup X_\tau)$, must be evaluated for every individual in the target population, including those for whom $X_S$ is unobserved. We nevertheless retain the gDR-Learner (Subsample) as a benchmark in the simulation studies. This allows us to distinguish the effect of estimating $\omega(X_S)$ using an auxiliary sample from the additional effect of restricting observation of $X_S$.

To obtain a feasible estimator under restricted observation of $X_S$, we replace the standard pseudo-outcome, eq.~\eqref{eq:pseudo-outcome}, with $\widehat g_{1,\mathrm{res}}(X_\tau)-\widehat g_{0,\mathrm{res}}(X_\tau)$. This gives the restricted pseudo-outcome
\begin{align*}
\widehat\varphi_{\mathrm{res}}(O) & = \frac{\mathbb I(S=1)\{A-\widehat\pi_1(X_S)\}}{\widehat\omega(X_S)\widehat\pi_1(X_S)\widehat\pi_0(X_S)}\left\{Y-\widehat\mu_A(X_S\cup X_\tau)\right\} \\
& ~~ + \widehat g_{1,\mathrm{res}}(X_\tau)-\widehat g_{0,\mathrm{res}}(X_\tau).
\end{align*}
The restricted pseudo-outcomes are well defined for every individual in the target population, unlike in the gDR-Learner (Subsample). The outcome model residuals remain well-defined on trial participants, and the second stage regression can be fitted over the entire target population without requiring $X_S$ to be observed outside $M=1$.

The marginal independence assumption in eq.~\eqref{eq:aux-assump} can be relaxed to the conditional independence assumption $S'\perp(S,X_S)\mid X_\tau$. Under this weaker condition, eq.~\eqref{eq:omega-subsample} holds with $q(X_\tau)=\mathbb P(S'=1\mid X_\tau)$ replacing $\mathbb P(S'=1)$ and $\mathbb P(S=1\mid X_S\cup X_\tau,M=1)$, where $q$ is estimable by regressing $S'$ on $X_\tau$. Alternatively, for the gT-Learners, $1/\omega$ can be replaced by the odds ratio from fitting a classifier of data source (auxiliary vs.\ trial) on $X_S\cup X_\tau$, which avoids estimating $q$. However, the gDR-Learner requires $1/\omega$ itself, so to avoid estimating $q$, we need two additional models. See Web Appendix~\ref{app:aux-relax} for details.

\section{Uncertainty Quantification via Conformal Inference}
\label{sec:conformal}
Confidence intervals for $\tau(X_\tau)$ are non-standard when $X_\tau$ contains continuous covariates. In a fully non-parametric model, $\widehat{\tau}$ will generally be slowly converging, even when the nuisance parameters are known. 
Consequently, unless imposing additional restrictions on the second-stage regression, generic Wald-type inference is not applicable, and bootstrap or asymptotic procedures require additional learner-specific regularity and smoothing assumptions \citep{bickel_asymptotic_1981}. Moreover, a confidence interval for the \ac{pcate} does not quantify the variation in treatment effects experienced by individual patients. We therefore target instead prediction intervals for the individual treatment effect (ITE), $Y^1 - Y^0$, constructed via conformal inference \citep{vovk_algorithmic_2005, tibshirani_conformal_2019, lei_conformal_2021}. These intervals characterize the uncertainty of treatment effects that may arise for a new individual drawn from the target population with covariates $X_\tau$ and thus may consequently be substantially wider than confidence intervals for the \ac{pcate}. The formal conformal coverage guarantee of these intervals is marginal over the distribution of $X_\tau$ in the target population:
\begin{equation}
\label{eq:conformal-coverage}
\mathbb P_{(X,Y^0,Y^1)\sim P_0}\left\{Y^1-Y^0\in\widehat C(X_\tau)\right\}\geq 1-\alpha.
\end{equation}
Thus, the intervals vary with $X_\tau$, although the coverage guarantee is marginal over the target population rather than pointwise conditional. 

For arm $a\in\{0,1\}$, weighted conformal inference corrects the difference between the trial-arm and target covariate distributions using the weights
\begin{equation}
\label{eq:conformal-weights}
w_a(X_S)=\frac{\mathbb P(S=1)}{\omega(X_S)}\frac{\mathbb P(A=a\mid S=1)}{\pi_a(X_S)}\propto\frac{1}{\omega(X_S)\pi_a(X_S)}.
\end{equation}
Hence, the fitted $\widehat\omega$ and $\widehat\pi_a$ used by the gT- and gDR-Learners can also be used to estimate the conformal weights.
For each calibration observation with $S_i=1$ and $A_i=a$, we use the absolute residual $R_i=|Y_i-\widehat m_a(X_{\tau,i})|$. For the gT-Learner, $\widehat m_a(X_\tau)=\widehat\mu_a(X_\tau;\widehat\omega,\widehat\pi)$. For the gDR-Learner, we regress the arm-specific pseudo-outcome
$$
\varphi_a(O)=\frac{\mathbb I(S=1)\mathbb I(A=a)}{\omega(X_S)\pi_a(X_S)}\{Y-\mu_a(X_S\cup X_\tau)\}+\mu_a(X_S\cup X_\tau)
$$
on $X_\tau$ to obtain $\widehat m_a(X_\tau)$. The weighted calibration residuals then yield arm-specific prediction intervals $\widehat C_a(X_\tau)$ for $Y^a$. Since the weights are estimated, rather than known, the resulting finite-sample coverage is approximate rather than guaranteed. Alternative nonconformity scores, such as those based on quantile regression, can better adapt to heteroscedasticity and local variation in the outcome distribution \citep{romano_conformalized_2019}. We do not pursue these here as they would require fitting additional conditional quantile models that are separate from the pseudo-outcome regression used by our proposed learners.

To obtain prediction intervals, we consider the ``naive'' and ``inexact'' procedures of \citet{lei_conformal_2021}. The naive procedure contrasts separately calibrated $1-\alpha/2$ prediction intervals for $Y^1$ and $Y^0$. By the union bound, both potential outcomes are simultaneously covered with probability at least $1-\alpha$, implying at least $1-\alpha$ marginal coverage for their contrast, albeit potentially conservatively. The inexact procedure uses an additional quantile regression step to obtain narrower intervals, but does not provide a formal finite-sample coverage guarantee. Full details are given in Web Appendix~\ref{app:conformal}.

\section{Simulation Study}
\label{sec:simulation}

We conducted simulation studies to evaluate the proposed estimators under the three data availability settings described in the previous sections: (1) the standard setting in which all relevant covariates are observed across the full target population, (2) the covariate setting in which additional outcome-predictive variables are collected only in the trial, and (3) the restricted covariate setting in which components of the selection covariates $X_S$ are unobserved in the target population but available in an external representative subsample. As baseline comparators we include the non-generalized T-Learner \citep{kunzel_metalearners_2019} and DR-Learner \citep{kennedy_towards_2023}, which estimate the \ac{scate} from the trial data alone and therefore do not correct for covariate shift between the trial and the target population.

Across all settings, covariates are drawn independently from a Uniform$(-1, 1)$ distribution, treatment is randomized with equal probability within the trial, and selection into the trial follows a logistic model linear in $X_S$, with the intercept calibrated to achieve the desired expected trial sample size $n_{1}$. Each setting investigates low- and high-dimensional covariate configurations. We vary $\Delta$, the number of covariates in $(X_S\setminus X_\tau)$ that are also treatment-effect modifiers, that is, modifiers that influence trial participation but are not included in the pCATE conditioning set. Increasing $\Delta$ increases the amount of treatment-effect heterogeneity that must be marginalized over when generalizing the trial effect to the target population. Although $\Delta$ does not quantify the lack of generalizability in general, our data-generating process ensures that, holding $d_S$ fixed, increasing $\Delta$ widens the \ac{scate}-\ac{pcate} gap while increasing treatment effect heterogeneity in regions where selection is least likely. Web Appendix~\ref{sec:simulation-details} provides full details of the functional forms. 

Nuisance functions are estimated using XGBoost \citep{chen_xgboost_2016} with 5-fold cross-fitting. Inverse probability weights are normalised to have mean equal to one. The estimated selection probabilities are truncated at their 10th and 90th percentiles. 

For the conformal inference prediction intervals, both the naive and inexact procedures require $X_S$ to be observed at prediction time, ruling out the restricted covariate setting of Section~\ref{sec:Estimation-with-Restricted-Target-Covariates}. We therefore apply conformal inference to simulation setting (1) only.

\subsection{Estimators}
\label{sec:sim-estimators}

Table~\ref{tab:sim_design} lists the learners compared in each setting, and the three settings differ in which data each nuisance function is fitted on and in what each comparison isolates. In all settings, $\widehat\pi_a$ and the outcome regressions $\widehat\mu_a$ are fitted on trial participants ($S=1$), and all second-stage regressions on $X_\tau$ using XGBoost.

In Setting 1, the T- and DR-Learners are fitted on trial participants only, without selection weights, and therefore target the \ac{scate}. Their error against the true \ac{pcate} measures the cost of ignoring covariate shift. The gT- and gDR-Learners additionally fit $\widehat\omega(X_S)$ on the full cohort of $n$ units.

In Setting 2, the standard gT- and gDR-Learners ignore $X_P$, whereas the enriched learners include $X_P$ in $\widehat\mu_a$. The gDR versus gDR (Enr.) comparison therefore isolates the improvement from using the trial-only covariates. 
The gT versus gT (Enr.) comparison instead shows the combined effect of the enrichment and of moving from a weighted regression (eq.~\eqref{eq:ipw-identification}) to an outcome-regression plug-in (eq.~\eqref{eq:g-comp-identification}).

In Setting 3, $X_S\setminus X_\tau$ is observed only in the trial and in an auxiliary sample of size $n_{\text{sub}}$, and all four generalized learners estimate $\widehat\omega(X_S)$ from these pooled samples ($M=1$) through eq.~\eqref{eq:omega-subsample}. For gDR (Subsample), $X_S$ is generated for all $n$ units, and its plug-in term $\widehat\mu_1-\widehat\mu_0$ is evaluated using values outside $M=1$, whereas $\widehat\omega$ is estimated only from the pooled trial and auxiliary samples. The gap between gDR (Subsample) and gDR (Restricted) reflects the price of replacing $\widehat\mu_1-\widehat\mu_0$ with $\widehat g_{1,\mathrm{res}}-\widehat g_{0,\mathrm{res}}$.

\begin{table}[htbp]
\centering
\caption{Simulation design across the three settings. $d_\tau$: number of CATE variables; $d_S$: number of selection modifiers; $d_P$: number of additional trial-only variables. For each setting, the target population has $N=15{,}000$ units, covariates are Uniform$(-1,1)$, selection follows a logistic model linear in $X_S$, and treatment is randomized within the trial.}
\label{tab:sim_design}
\resizebox{\textwidth}{!}{%
\begin{tabular}{llccccccl}
\toprule
\textbf{Setting} & \textbf{Dimension} & $d_\tau$ & $d_S$ & $d_P$ & $\Delta$ & $\tau(X_\tau)$ & $\mu_0(X_\mu)$ & \textbf{Comparators} \\
\midrule

\multirow{2}{*}{1: Standard}
& Low  & 1 & 2  & \multirow{2}{*}{0} & 1
& \multirow{2}{*}{Linear / Non-linear}
& \multirow{2}{*}{Non-linear}
& \multirow{2}{*}{T, DR, gT, gDR} \\
& High & 5 & 10 & & $\{5,\dots,10\}$
& & & \\
\midrule

\multirow{2}{*}{2: Enriched}
& Low  & 1 & 2  & \multirow{2}{*}{$\{0,2,4,6,8\}$} & 1
& \multirow{2}{*}{Linear}
& \multirow{2}{*}{Linear / Non-linear}
& \multirow{2}{*}{T, DR, gT, gDR, gT (Enr.), gDR (Enr.)} \\
& High & 5 & 10 & & 5
& & & \\
\midrule

\multirow{2}{*}{3: Restricted}
& Low  & 2 & 2  & \multirow{2}{*}{0} & $\{0,1,2\}$
& \multirow{2}{*}{Non-linear}
& \multirow{2}{*}{Non-linear}
& \multirow{2}{*}{\shortstack[l]{
T, DR, gT (Subsample)$^\dagger$, gDR (Subsample)$^\dagger$,
\\
gT (Restricted), gDR (Restricted)
}}
\\
& High & 5 & 10 & & $\{5,7,9\}$
& & & \\
\bottomrule
\end{tabular}}
\parbox{\textwidth}{%
\par\smallskip
\footnotesize\raggedright
\textit{Abbreviations:} Enr. = enriched;
Restricted = restricted target covariates.
\par\smallskip
$^\dagger$ In Setting 3, the gT (Subsample) and
gDR (Subsample) comparators use the external subsample
together with the trial to estimate the selection model,
without marginalizing the outcome models. The
gDR (Subsample) additionally requires $X_S$ across
the full target population to construct its pseudo-outcomes
and is an infeasible oracle benchmark.
}
\end{table}
For each simulated dataset and learner, we evaluated \ac{pcate} predictions against the known true $\tau(X_\tau)$ on an independent draw of the full target population ($15{,}000$ units), held fixed across replications, using the Median Squared Error (MedSE), and averaged over $1000$ replications. The MedSE is the median over validation units of the pointwise squared errors. We take the median over validation units because, within a replication, pointwise squared errors can be highly skewed, particularly for estimators involving inverse probability weighting, for which a small number of validation points may have very large errors. The MedSE therefore summarizes typical pointwise accuracy and it is insensitive to errors in the tails of the $X_\tau$ distribution. The median therefore provides a summary of typical performance that is less sensitive to such extreme realizations than the mean squared error.

\subsection{Results}

\begin{figure}
\caption{Setting 1. Median squared error (MedSE) of the generalized (gT, gDR) and trial-only (T, DR) learners for linear and non-linear pCATEs. Left and centre: against $n_1$, with $\Delta=1$ (low-dimensional) and $\Delta=5$ (high-dimensional). Right: against $\Delta=|X_S\setminus X_\tau|$ (high-dimensional), with $n_1=1{,}000$.}
\label{fig:exp1_overlap}
    \centering
    \includegraphics[width = 0.7\linewidth]{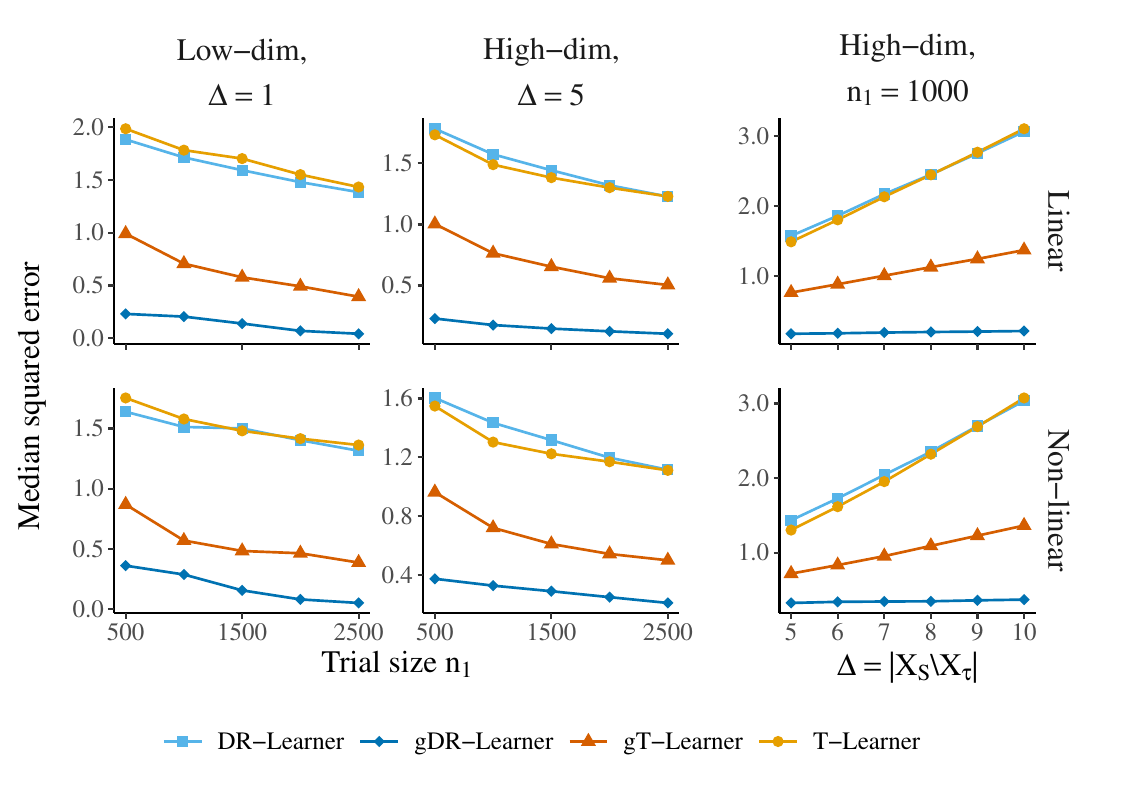}
\end{figure}
The gDR-Learner achieved the lowest MedSE across all settings (Figure~\ref{fig:exp1_overlap}). The gT-Learner consistently outperformed the T-Learner but remained less accurate than gDR. Errors decreased with trial size, the higher errors at small $n_1$ are consistent with the difficulty of estimating nuisance functions and applying estimated \ac{ipsw} with fewer trial participants. As $\Delta$ increased, the accuracy of the trial-only learners deteriorated substantially, whereas gDR's MedSE increased only modestly in both linear and non-linear settings. The 95\% conformal prediction intervals over-covered the true \ac{pcate} (coverage $0.965$--$1.00$), as expected because they target the more variable \ac{ite} (Web Table~\ref{tab:conformal_coverage}). Against the \ac{ite}, naive intervals remained conservative (coverage $0.954$--$0.995$), whereas the narrower inexact intervals undercovered: coverage was $0.87$--$0.89$ in low-dimensional settings and $0.92$--$0.95$ in high-dimensional settings (Web Table~\ref{tab:conformal_coverage_ite}; Web Figure~\ref{fig:conformal-interval-widths}). Thus, the reduction in interval width came at a cost in coverage.

\begin{figure}
\caption{Setting 2: MedSE (log scale) of the standard and enriched (Enr.) learners, by dimension and outcome model. Top: against the number of trial-only covariates $d_P$, with $n_1=1{,}500$. Bottom: against $n_1$, with $d_P=4$. $\Delta=1$ (low-dimensional), $\Delta=5$ (high-dimensional).}
    \centering
    \includegraphics[width = \linewidth]{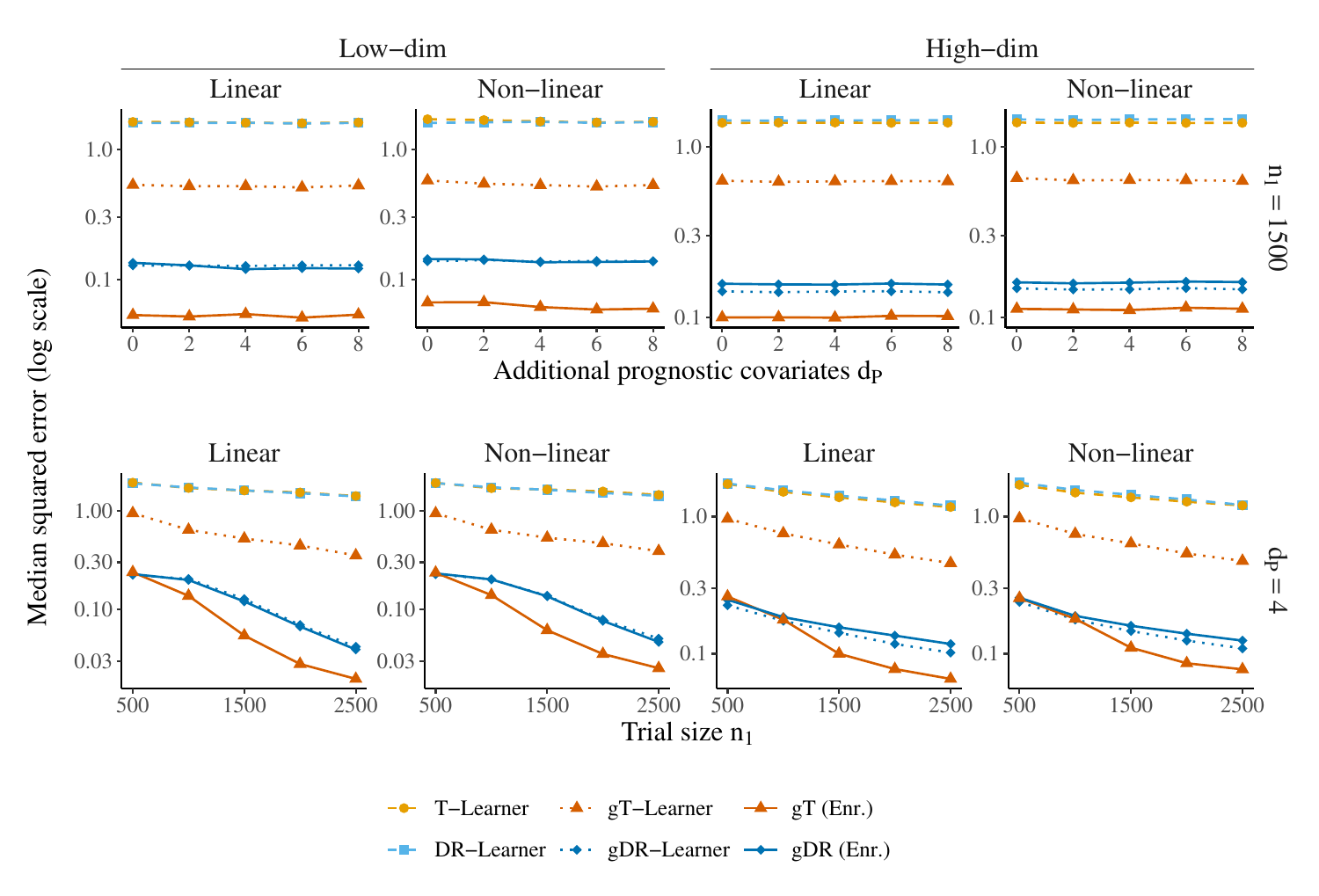}
    \label{fig:exp2_prog}
\end{figure}
In Setting 2, gT (Enr.) achieved the lowest MedSE across all scenarios and values of $d_P$ at $n_1=1{,}500$ (Figure~\ref{fig:exp2_prog}, top row). Compared to the previous setting, this suggests that richer outcome modelling can improve the accuracy of the gT-Learner. gDR (Enr.) yielded small improvements at larger $d_P$ in low-dimensional settings but slightly increased MedSE in high-dimensional settings. Across trial sizes (bottom row), standard gDR performed best at $n_1=500$, while gT (Enr.) generally performed best at larger sizes. These results show no consistent accuracy gain from gDR (Enr.), and suggest that the extra marginalization step can offset the benefit of additional predictors in finite samples.

\begin{figure}
    \caption{Setting 3. MedSE of the trial-only, Subsample and Restricted learners for non-linear pCATE, with $\Delta=1$ (low-dimensional) and $\Delta=9$ (high-dimensional). Left: against external subsample size $n_{\text{sub}}$ (log scale), with $n_1=2{,}000$. Right: against $n_1$, with $n_{\text{sub}}=5{,}000$. gDR (Subsample) is an infeasible benchmark.}
        \centering
    \includegraphics[width = \linewidth]{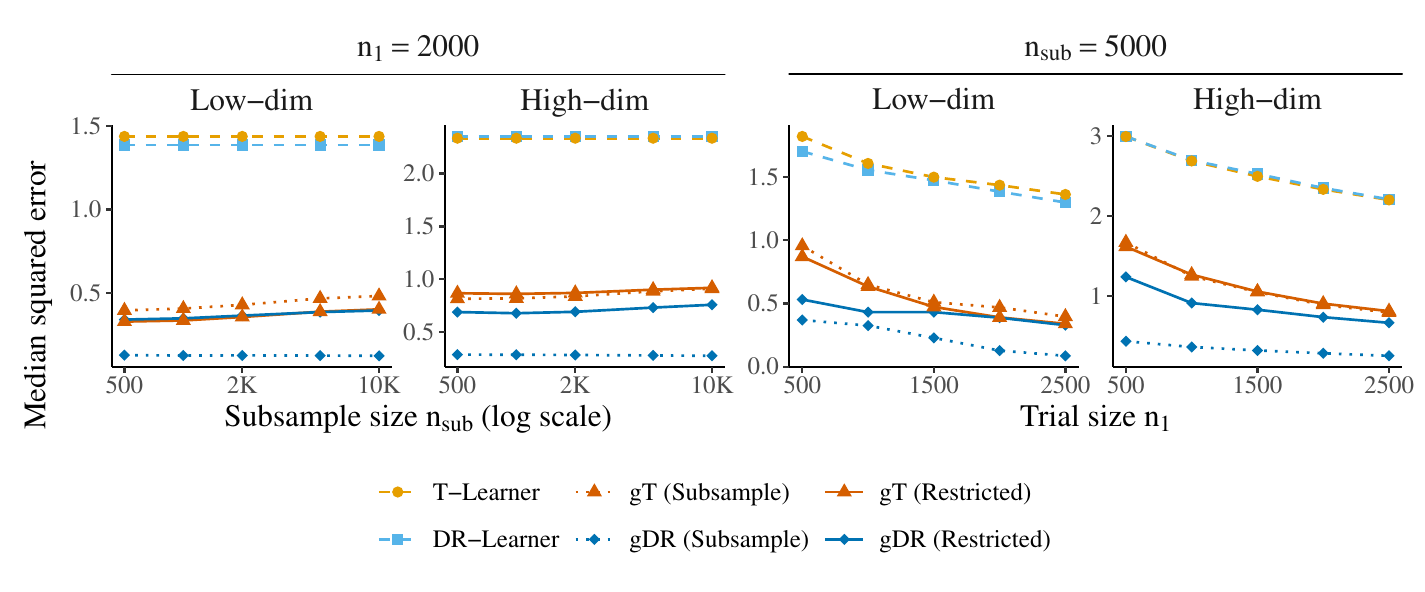}
    \label{fig:exp3_trial}
\end{figure}
In Setting 3, gT (Restricted) had lower MedSE than gT (Subsample) in low-dimensional settings, but generally slightly higher MedSE in high-dimensional settings (Figure~\ref{fig:exp3_trial}). The feasible gDR (Restricted) had higher MedSE than the infeasible gDR (Subsample) benchmark in both dimensions, showing a substantial finite-sample cost of marginalizing unobserved selection covariates. Among the feasible learners, gDR (Restricted) attained the lowest MedSE at every trial size. Increasing trial size reduced errors substantially, whereas increasing the external subsample from $500$ to $10{,}000$ yielded little improvement. This suggests that enlarging the subsample alone does not resolve the difficulty of estimating the additional marginalization regressions from trial data.

\section{Case Study: the Coronary Artery Surgery Study}
\label{sec:cass}

We illustrate the proposed methods by re-analysing the Coronary Artery Surgery Study (CASS) \citep{killip_coronary_1985}, applying the gT- and gDR-Learners and corresponding conformal prediction intervals in a setting with missing baseline covariates. CASS includes a randomized trial of 780 patients and a registry of 1,319 eligible non-randomized patients. Our primary estimand is the \ac{pcate} conditional on left ventricular ejection fraction (\ac{ef}) and history of myocardial infarction (\ac{mi}). Secondary analyses by age and sex are in Web Appendix~\ref{app:cass}.

\subsection{Methods}

We estimate the \ac{pcate} using gT- and gDR-Learners with random forest nuisance models and second-stage regression \citep{breiman_random_2001}, aggregated over multiple cross-fitting splits for stability. Parametric equivalents using spline based linear models following \citet{robertson_regression-based_2023} are included for comparison, alongside parametric 95\% confidence intervals and uniform bands. Baseline \ac{ef} is missing for $\sim$25\% of registry and $\sim$6\% of trial patients. We apply MICE with random forest imputation for non-parametric learners and regression-based imputation for parametric learners \citep{buuren_mice_2011}, averaging point estimates across imputations. Full specifications, diagnostics, and covariate balance are in Web Appendices~\ref{app:cass-baseline}--\ref{app:conformal-cass}.

\subsection{Results}
\begin{figure}[h]
\caption{CASS: non-parametric gDR- and gT-Learner estimates of the pCATE against ejection fraction, by history of MI. Curves are pointwise medians over 50 cross-fitting replications. Horizontal lines mark the mean over the ejection fraction grid.}
    \label{fig:np-ejec_frac}
    \centering
    \includegraphics[width=0.7\linewidth]{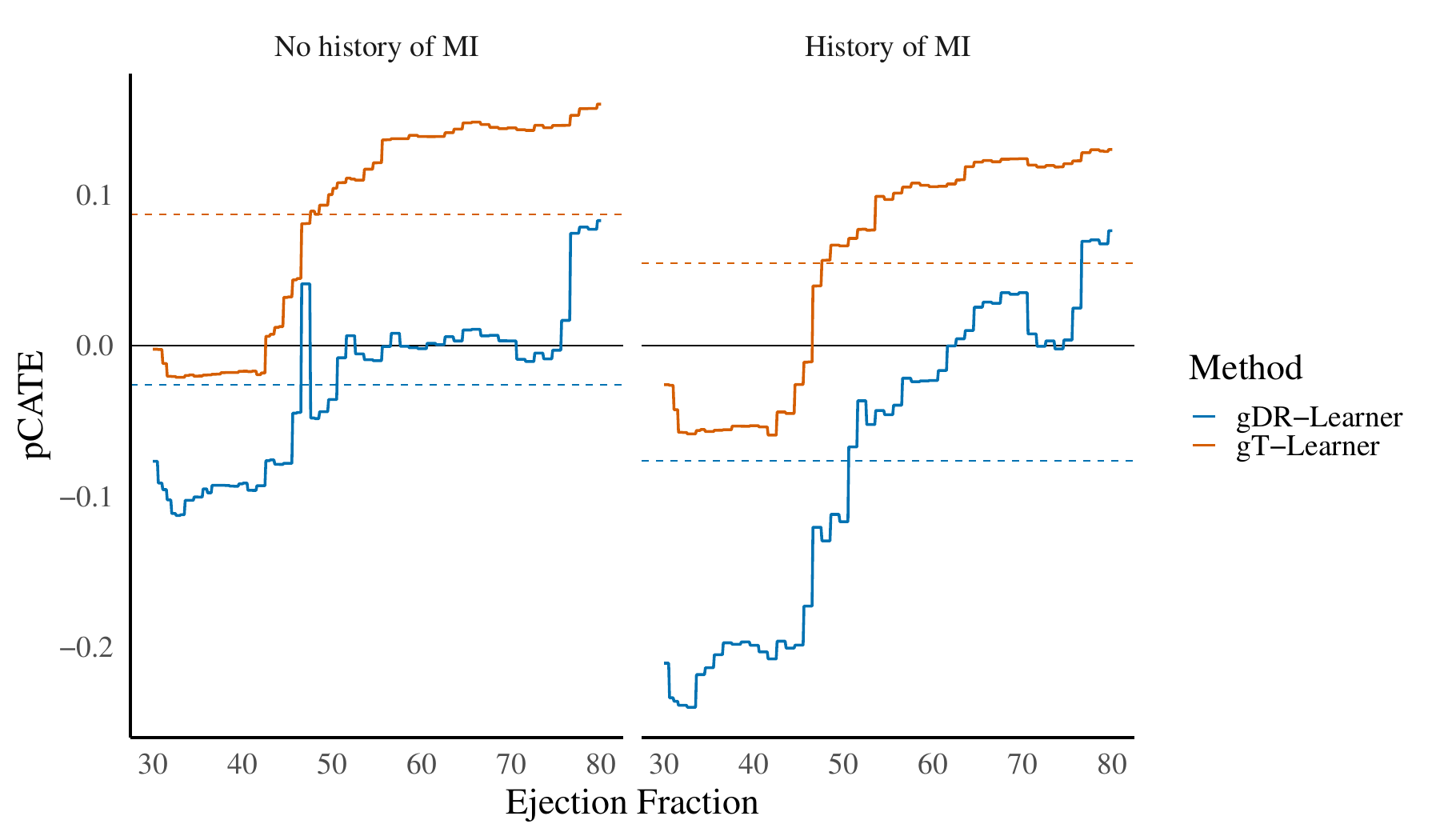}    
\end{figure}
Figure~\ref{fig:np-ejec_frac} displays estimates of the \ac{pcate} as a function of \ac{ef}, stratified by history of \ac{mi}, obtained using the non-parametric gT- and gDR-Learners. These estimates were aggregated using the median over repeated refits of each learner to mitigate variability due to sample splitting \citep{jacob_cross-fitting_2020}. Estimates from both methods suggest that patients with lower \ac{ef} values tend to benefit more from surgery relative to medical therapy. Among patients with a history of \ac{mi}, estimated treatment effects are negative for \ac{ef} values below approximately 50\%, indicating a relative benefit of surgical intervention in this subgroup. At higher values of \ac{ef}, the gT-Learner yields higher \ac{pcate} estimates than the gDR-Learner, which may reflect first-order bias from the estimated inverse probability weights. Among patients with prior \ac{mi}, both parametric and non-parametric approaches showed increasing \ac{pcate} with \ac{ef}, although the parametric models suggested greater benefit from surgery at low \ac{ef}. Among patients without prior \ac{mi}, the parametric gDR-Learner produced larger positive estimates at low \ac{ef} than the non-parametric learners, indicating greater sensitivity to model choice. 
Conformal prediction intervals are wide, reflecting that they target the individual treatment effect, which is more variable than the \ac{pcate}. 
Web Figure~\ref{fig:np-ejec_frac-subgroup} shows analogous results by age and \ac{mi} history, with the gDR-Learner remaining stable across subgroups while the gT-Learner produces larger estimates, particularly in smaller subgroups. The parametric confidence intervals (Web Appendix~\ref{app:cass-parametric}) were substantially narrower than the conformal prediction intervals, although they target different quantities and rely on different assumptions, and thus are not directly comparable. 

\section{Discussion}
\label{sec:discussion}

In this paper, we developed a generalized T-Learner for estimating population conditional average treatment effects from nested randomized trials under covariate shift, and studied a generalized DR-Learner previously proposed by \citet{robertson_regression-based_2023}, which had been presented using parametric models. For the gDR-Learner, we establish excess risk bounds showing that, when combined with cross-fitting, nuisance estimation errors enter through higher-order products rather than as first-order errors, allowing slower convergence of individual nuisance estimators provided the relevant products converge sufficiently fast relative to the oracle rate. We further extend the framework to settings in which important covariates are only partially observed in the target population. In this work we assume there is access to a sample $S'$ of the target population that satisfies a marginal independence assumption (with respect to trial inclusion and selection variables), eq.~\eqref{eq:aux-assump}. We outline the methods under a weaker assumption in the appendix, leaving empirical comparisons under this weaker condition to future work. The methods presented here are for nested trial designs, the same principles can apply more broadly to transportability problems involving covariate shift between source and target populations. 

Like other weighting-based estimators, the proposed approaches depend on adequate overlap between the trial and target populations. When the probability of trial participation is small, inverse probability weights may become unstable, leading to highly variable pseudo-outcomes. Approaches such as the EP-Learner \citep{van_der_laan_combining_2024}, which constructs bounded pseudo-outcomes through infinite-dimensional targeted minimum loss-based estimation provide a promising direction for improving robustness.

The inexact conformal procedure gave narrower intervals but undercovered the \ac{ite} in low-dimensional settings, because its validity relies on quantile regression and estimated weights. Future work could adopt the debiased conformal approach of \citet{barnatchez2026} for partially observed target covariates, or target approximately conditional coverage \citep{gibbs_conformal_2025}. 

\section*{Supplementary Materials}
Web Appendices, Tables, and Figures referenced in Sections 2--7 are available with this paper. The code used in this paper is publicly available at \url{https://github.com/v-rodriguez/gen-hte}.

\section*{Data Availability}
The Coronary Artery Surgery Study (CASS) data used in this analysis were obtained from the National Heart, Lung, and Blood Institute (NHLBI) Biologic Specimen and Data Repository Information Coordinating Center (BioLINCC) and are available upon request, subject to approval.  



\section*{Acknowledgments}
VR acknowledges the receipt of studentship awards from the Health Data Research UK-The Alan Turing Institute Wellcome PhD Programme in Health Data Science (218529/Z/19/Z). KDO was funded by a Royal Society-Wellcome Trust Sir Henry Dale fellowship (218554/Z/19/Z). 
BL has received research funding from Novo Nordisk A/S. Novo Nordisk had no role in the conception, design, conduct, analysis, or interpretation of the work presented in this manuscript.

The authors acknowledge the use of the UCL Myriad High Performance Computing Facility (Myriad@UCL), and associated support services, in the completion of this work.

\bibliographystyle{unsrtnat}
\bibliography{output}

\setcounter{section}{0}

\newpage
\renewcommand{\thesection}{\Alph{section}}
\renewcommand{\thesubsection}{\Alph{section}.\arabic{subsection}}
\makeatletter
\renewcommand{\proof}[1][Proof]{\reset@font\rm \trivlist \item[\hskip \parindent
  {\reset@font~~\it #1.}]}
\def\appsecprefix@section{Web Appendix~}
\renewcommand{\@seccntformat}[1]{%
  \ifcsname appsecprefix@#1\endcsname\csname appsecprefix@#1\endcsname\fi
  \csname the#1\endcsname\quad
}
\@ifclassloaded{biom}{\@tempswatrue}{\@tempswafalse}
\if@tempswa
  \patchcmd{\NR@sect}
    {\edef\@svsec{\csname the#1\endcsname.\hskip .5em}}
    {\edef\@svsec{Web Appendix\noexpand~\csname the#1\endcsname\hskip .5em}}
    {}{\PackageWarning{appendices}{Could not patch biom \string\@sect}}
\fi
\makeatother
\beginsupplement
\setcounter{equation}{0}
\renewcommand{\theequation}{S\arabic{equation}}
\ifdefined\theHequation\renewcommand{\theHequation}{S\arabic{equation}}\fi
\ifdefined\theHsection\renewcommand{\theHsection}{S\Alph{section}}\fi

\setlength{\emergencystretch}{2em}
\setcounter{section}{0}


\section{Identification in the Standard Setting}
\label{app:identification}

Here we provide proofs for the identification of the \ac{pcate} using observed data from the RCT.

G-computation identification (eq.~\eqref{eq:g-comp-identification} in main text):
\begin{equation*}
    \mathbb{E}[Y^1 - Y^0 \mid X_\tau]
    = \mathbb{E}\bigl[
        \mathbb{E}[Y \mid X_S \cup X_\tau, S=1, A=1]
        - \mathbb{E}[Y \mid X_S \cup X_\tau, S=1, A=0]
      \bigm| X_\tau
    \bigr].
\end{equation*}

\begin{proof}
\begin{align*}
\mathbb{E}[Y^1 - Y^0 \mid X_\tau]
&= \mathbb{E}\bigl[
     \mathbb{E}[Y^1 - Y^0 \mid X_S \cup X_\tau]
   \bigm| X_\tau
   \bigr]
   \tag{Tower property} \\
&= \mathbb{E}\bigl[
     \mathbb{E}[Y^1 \mid X_S \cup X_\tau]
   - \mathbb{E}[Y^0 \mid X_S \cup X_\tau]
   \bigm| X_\tau
   \bigr]
   \tag{Linearity of expectation} \\
&= \mathbb{E}\bigl[
     \mathbb{E}[Y^1 \mid X_S \cup X_\tau, S=1] \\
&\qquad - \mathbb{E}[Y^0 \mid X_S \cup X_\tau, S=1]
   \bigm| X_\tau
   \bigr]
   \tag{Assumption~\ref{assump:mean_exchangeability_populations}} \\
&= \mathbb{E}\bigl[
     \mathbb{E}[Y^1 \mid X_S \cup X_\tau, S=1, A=1] \\
&\qquad - \mathbb{E}[Y^0 \mid X_S \cup X_\tau, S=1, A=0]
   \bigm| X_\tau
   \bigr]
   \tag{Assumption~\ref{assump:mean_exchangeability_trial}} \\
&= \mathbb{E}\bigl[
     \mathbb{E}[Y \mid X_S \cup X_\tau, S=1, A=1] \\
&\qquad - \mathbb{E}[Y \mid X_S \cup X_\tau, S=1, A=0]
   \bigm| X_\tau
   \bigr].
   \tag{Consistency}
\end{align*}
\end{proof}

IPW identification (eq.~\eqref{eq:ipw-identification} in main text):
\begin{equation*}
    \mathbb{E}[Y^1 - Y^0 \mid X_\tau]
    = \mathbb{E}\left[
        \frac{\mathbb{I} (S=1)\mathbb{I} (A=1)}{\omega(X_S)\pi_1(X_S)}Y
        - \frac{\mathbb{I} (S=1)\mathbb{I} (A=0)}{\omega(X_S)\pi_0(X_S)}Y
      \middle| X_\tau
    \right].
\end{equation*}

\begin{proof}
It suffices to show that, for each $a \in \{0,1\}$,
\begin{equation*}
\mathbb{E}\left[\frac{\mathbb{I} (S=1)\mathbb{I} (A=a)}{\omega(X_S)\pi_a(X_S)}Y \middle| X_\tau\right] = \mathbb{E}[Y^a \mid X_\tau].
\end{equation*}
This result follows from positivity, as $\pi_a(X_S)$ and $\omega(X_S)$ are never 0, and so we can write
\begin{align*}
\lefteqn{\mathbb{E}\left[\frac{\mathbb{I} (S=1)\mathbb{I} (A=a)}{\omega(X_S)\pi_a(X_S)}Y \middle| X_\tau\right]} \\*
\qquad
&= \mathbb{E}\left[
     \mathbb{E}\left[\frac{\mathbb{I} (S=1)\mathbb{I} (A=a)}{\omega(X_S)\pi_a(X_S)}Y \middle| X_S \cup X_\tau\right]
   \middle| X_\tau\right]
   \tag{Tower property} \\
&= \mathbb{E}\left[
     \mathbb{E}\left[\frac{\mathbb{I} (S=1)\mathbb{I} (A=a)}{\omega(X_S)\pi_a(X_S)}Y^a \middle| X_S \cup X_\tau\right]
   \middle| X_\tau\right]
   \tag{Consistency} \\
&= \mathbb{E}\left[
     \mathbb{E}\left[\frac{\mathbb{I} (S=1)\mathbb{I} (A=a)}{\omega(X_S)\pi_a(X_S)} \middle| X_S \cup X_\tau\right]
     \mathbb{E}[Y^a \mid X_S \cup X_\tau]
   \middle| X_\tau\right]
   \tag{Assumptions~\ref{assump:mean_exchangeability_trial}, \ref{assump:mean_exchangeability_populations}} \\
&= \mathbb{E}\bigl[\mathbb{E}[Y^a \mid X_S \cup X_\tau] \bigm| X_\tau\bigr] \\
&= \mathbb{E}[Y^a \mid X_\tau].
   \tag{Tower property}
\end{align*}
The fourth line follows because $\mathbb{E}\left[\{\mathbb{I} (A=a)\mathbb{I} (S=1)\}/\{\pi_a(X_S)\omega(X_S)\} \mid X_S \cup X_\tau\right] = 1$.

\end{proof}

\section{Derivation of the \ac{eif} of the \ac{pcate} risk}
\label{app:eif}

We derive the orthogonal loss stated in eq.~\eqref{eq:orthogonal-risk}. The argument follows the derivation given by \citet{pryce_causal_2025} for the \ac{cate} risk under missing outcome data.

Let $P$ denote the true observed data distribution, let $f(x)$ be the density of $X$ in the target population, and write $x_\tau$ and $x_S$ for the effect-modifier and selection coordinates of $x$, abbreviating $\mu_a(x) = \mu_a(x_S \cup x_\tau)$. Fix a candidate function $\tau$. The estimand of interest is the risk
\begin{equation*}
    L(\tau; P) = \mathbb{E}_P\left[\tau(X_\tau)^2 - 2\tau(X_\tau)\left(\mathbb{E}_P[Y^1 \mid X_\tau] - \mathbb{E}_P[Y^0 \mid X_\tau]\right)\right].
\end{equation*}
Under Assumptions~\ref{assump:mean_exchangeability_trial}--\ref{assump:mean_exchangeability_populations} we have $\mathbb{E}_P[Y^a \mid X_\tau] = \mathbb{E}_P[\mu_a(X_S \cup X_\tau) \mid X_\tau]$, so by the tower property the risk depends on $P$ only through $f$ and the two conditional means. Writing it as an integral,
\begin{equation}
\label{eq:app-risk-integral}
    L(\tau; P) = \int f(x)\bigl[\tau(x_\tau)^2 - 2\tau(x_\tau)\bigl\{\mu_1(x) - \mu_0(x)\bigr\}\bigr] \,\mathrm{d}x.
\end{equation}

To obtain the \ac{eif} of \eqref{eq:app-risk-integral} we perturb $P$ along the one-dimensional mixture
\begin{equation*}
    P_t = t\widetilde{P} + (1-t)P,
\end{equation*}
where $\widetilde{P}$ is a point mass at $\tilde{o} = (\tilde{x}, \tilde{s}, \tilde{a}, \tilde{y})$, whose support is contained in that of $P$. The perturbed covariate density is
\begin{equation*}
    f_{P_t}(x) = t\mathbb{I}(x=\tilde{x}) + (1-t)f(x),
\end{equation*}
and the \ac{eif} is the Gateaux derivative $\left. \mathrm{d}L(\tau; P_t)/\mathrm{d}t \right|_{t=0}$.

Applying the product rule to \eqref{eq:app-risk-integral} splits this derivative into a contribution from the covariate density and a contribution from the two conditional means,
\begin{align}
    \left. \frac{\mathrm{d}L(\tau; P_t)}{\mathrm{d}t} \right|_{t=0}
    = {} & \int \bigl\{\mathbb{I}(x=\tilde{x}) - f(x)\bigr\}\bigl[\tau(x_\tau)^2 - 2\tau(x_\tau)\bigl\{\mu_1(x) - \mu_0(x)\bigr\}\bigr]\,\mathrm{d}x \notag \\
    & - 2\int f(x)\,\tau(x_\tau)
    \left.\frac{\mathrm{d}}{\mathrm{d}t}\bigl\{\mu_{1,t}(x) - \mu_{0,t}(x)\bigr\}\right|_{t=0} \mathrm{d}x. \label{eq:app-mean-term}
\end{align}
The first term integrates directly to
\begin{equation}
\label{eq:app-density-result}
    \tau(\tilde{x}_\tau)^2 - 2\tau(\tilde{x}_\tau)\left\{\mu_1(\tilde{x}) - \mu_0(\tilde{x})\right\} - L(\tau; P).
\end{equation}

For the second term, we use the Gateaux derivative of a conditional mean under point-mass contamination,
\begin{equation}
\label{eq:app-cond-mean-deriv}
    \left. \frac{\mathrm{d}}{\mathrm{d}t} \mu_{a,t}(x) \right|_{t=0}
    = \frac{\mathbb{I}(x=\tilde{x})\,\mathbb{I}(\tilde{s}=1)\,\mathbb{I}(\tilde{a}=a)}
           {f(x)\,\mathbb{P}(S=1, A=a \mid X=x)}
      \bigl\{\tilde{y} - \mu_a(x)\bigr\}.
\end{equation}
The denominator factorizes to $\mathbb{P}(S=1 \mid X = x) = \omega(x_S)$ and $\mathbb{P}(A = a \mid X = x, S=1) = \pi_a(x_S)$, giving
\begin{equation*}
    \mathbb{P}(S=1, A=a \mid X = x) = \omega(x_S)\pi_a(x_S).
\end{equation*}

Substituting \eqref{eq:app-cond-mean-deriv} into \eqref{eq:app-mean-term},
\begin{equation}
\label{eq:app-mean-result}
    -2\tau(\tilde{x}_\tau)\,\mathbb{I}(\tilde{s}=1)\left[\frac{\mathbb{I}(\tilde{a}=1)}{\omega(\tilde{x}_S)\pi_1(\tilde{x}_S)} - \frac{\mathbb{I}(\tilde{a}=0)}{\omega(\tilde{x}_S)\pi_0(\tilde{x}_S)}\right]\left\{\tilde{y} - \mu_{\tilde{a}}(\tilde{x})\right\}.
\end{equation}
Since treatment is binary, $\pi_0 = 1 - \pi_1$, and the bracket simplifies:
\begin{equation*}
    \frac{\mathbb{I}(\tilde{a}=1)}{\pi_1(\tilde{x}_S)} - \frac{\mathbb{I}(\tilde{a}=0)}{\pi_0(\tilde{x}_S)} = \frac{\tilde{a} - \pi_1(\tilde{x}_S)}{\pi_1(\tilde{x}_S)\pi_0(\tilde{x}_S)}.
\end{equation*}

Adding \eqref{eq:app-density-result} and \eqref{eq:app-mean-result}, and dropping tildes, gives
\begin{align*}
\psi_\tau(O) = {} & \tau(X_\tau)^2 - 2\tau(X_\tau)\bigl\{\mu_1(X_S \cup X_\tau) - \mu_0(X_S \cup X_\tau)\bigr\} \\
& - 2\tau(X_\tau)\frac{\mathbb{I}(S=1)\{A-\pi_1(X_S)\}}{\omega(X_S)\pi_1(X_S)\pi_0(X_S)}\bigl\{Y-\mu_A(X_S \cup X_\tau)\bigr\} - L(\tau; P).
\end{align*}
Define
\begin{align}
\varphi_{\mathrm{gDR}}(O)
= {} & \frac{\mathbb{I}(S=1)\{A-\pi_1(X_S)\}}{\omega(X_S)\pi_1(X_S)\pi_0(X_S)}
       \bigl\{Y-\mu_A(X_S\cup X_\tau)\bigr\} \notag \\
& + \mu_1(X_S\cup X_\tau)-\mu_0(X_S\cup X_\tau).
\label{eq:app-pseudo-outcome}
\end{align}
Then
\begin{equation*}
\psi_\tau(O) = \tau(X_\tau)^2 - 2\tau(X_\tau)\varphi_{\mathrm{gDR}}(O) - L(\tau;P).
\end{equation*}
The corresponding orthogonal criterion is therefore
\begin{equation}
L_{\mathrm{orth}}(\tau,\eta)
= \mathbb{E}\bigl[\tau(X_\tau)^2 - 2\tau(X_\tau)\varphi_{\mathrm{gDR}}(O)\bigr].
\label{eq:app-orthogonal-risk}
\end{equation}
At the true nuisance functions, $L_{\mathrm{orth}}(\tau,\eta)=L(\tau;P)$, and the criterion is Neyman orthogonal with respect to $\eta$.

Completing the square,
\begin{equation*}
L_{\mathrm{orth}}(\tau,\eta)
= \mathbb{E}\bigl[\{\varphi_{\mathrm{gDR}}(O)-\tau(X_\tau)\}^2\bigr]
- \mathbb{E}\bigl[\varphi_{\mathrm{gDR}}(O)^2\bigr].
\end{equation*}
Since the final term does not depend on $\tau$,
$$
\operatorname*{arg\,min}_\tau L_{\mathrm{orth}}(\tau,\eta)
= \operatorname*{arg\,min}_\tau
  \mathbb{E}\bigl[\{\varphi_{\mathrm{gDR}}(O)-\tau(X_\tau)\}^2\bigr].
$$
Hence,
$$
\tau_{\mathrm{gDR}}(X_\tau) = \mathbb{E}[\varphi_{\mathrm{gDR}}(O) \mid X_\tau],
$$
so estimation reduces to regressing the pseudo-outcome $\varphi_{\mathrm{gDR}}(O)$ on $X_\tau$.

\section{Identification in the Prognostic Setting}
\label{app:identification-prognostic}

\begin{proposition}[Identification of $g_{a,\text{enr}}$ with prognostic variables]
\label{prop:prog-identification}
Suppose Assumptions~\ref{assump:mean_exchangeability_trial}--\ref{assump:mean_exchangeability_populations} hold, with Assumptions~\ref{assump:mean_exchangeability_trial} and \ref{assump:treatment_positivity} understood to hold with $X_P$ added to the conditioning set, as guaranteed by randomization. Define
\begin{equation*}
g_{a,\text{enr}}(X_S \cup X_\tau)
= \mathbb{E}[\mu_a(X_S \cup X_\tau, X_P) \mid X_S \cup X_\tau, S=1].
\end{equation*}
Then $g_{a,\text{enr}}$ is identified from trial data alone, and satisfies
\begin{equation}
\label{eq:g-prog-id}
g_{a,\text{enr}}(X_S \cup X_\tau) = \mathbb{E}[Y^a \mid X_S \cup X_\tau],
\end{equation}
so that the pCATE is identified by
\begin{equation*}
\tau(X_\tau)
= \mathbb{E}[g_{1,\text{enr}}(X_S \cup X_\tau) - g_{0,\text{enr}}(X_S \cup X_\tau) \mid X_\tau].
\end{equation*}
\end{proposition}

\begin{proof}
Starting from the definition of $g_{a,\text{enr}}$,
\begin{align*}
g_{a,\text{enr}}(X_S \cup X_\tau)
&= \mathbb{E}\bigl[
    \mathbb{E}[Y \mid X_S \cup X_\tau, X_P, S=1, A=a]
    \bigm| X_S \cup X_\tau, S=1
  \bigr] \\
&= \mathbb{E}\bigl[
    \mathbb{E}[Y^a \mid X_S \cup X_\tau, X_P, S=1, A=a]
    \bigm| X_S \cup X_\tau, S=1
  \bigr]
  \tag{Consistency} \\
&= \mathbb{E}\bigl[
    \mathbb{E}[Y^a \mid X_S \cup X_\tau, X_P, S=1]
    \bigm| X_S \cup X_\tau, S=1
  \bigr]
  \tag{Assumption~\ref{assump:mean_exchangeability_trial}} \\
&= \mathbb{E}[Y^a \mid X_S \cup X_\tau, S=1]
  \tag{Tower property} \\
&= \mathbb{E}[Y^a \mid X_S \cup X_\tau],
  \tag{Assumption~\ref{assump:mean_exchangeability_populations}}
\end{align*}
where the third line applies Assumption~\ref{assump:mean_exchangeability_trial} conditional on $X_P$, and the last line uses the definition of $X_S$ and $X_\tau$ together with Assumption~\ref{assump:mean_exchangeability_populations}. This establishes \eqref{eq:g-prog-id}.
The final claim then follows,
\begin{align*}
\tau(X_\tau)
&= \mathbb{E}[Y^1 - Y^0 \mid X_\tau] \\
&= \mathbb{E}\bigl[
     \mathbb{E}[Y^1 \mid X_S \cup X_\tau] - \mathbb{E}[Y^0 \mid X_S \cup X_\tau]
     \bigm| X_\tau
   \bigr]
  \tag{Tower property, linearity} \\
&= \mathbb{E}\bigl[
     g_{1,\text{enr}}(X_S \cup X_\tau) - g_{0,\text{enr}}(X_S \cup X_\tau)
     \bigm| X_\tau
   \bigr].
  \tag{Eq.~\eqref{eq:g-prog-id}}
\end{align*}
\end{proof}

\section{Identification in the Restricted Covariate Setting}
\label{app:identification-restricted}

\begin{lemma}[Identification of the trial selection probability]
\label{lem:selection_recovery}
Suppose that $M=\mathbb{I}(S=1\cup S'=1)$ and that $\mathbb{P}(S'=1\mid X_S,S=0)=\mathbb{P}(S'=1)$. Then the trial selection probability $\omega(X_S)=\mathbb{P}(S=1\mid X_S)$ is identified from the distribution conditional on $M=1$ as
\begin{equation*}
\omega(X_S) = \frac{\mathbb{P}(S'=1)\,\mathbb{P}(S=1\mid X_S,M=1)}{1-\{1-\mathbb{P}(S'=1)\}\,\mathbb{P}(S=1\mid X_S,M=1)}.
\end{equation*}
\end{lemma}

\begin{proof}
Since $S=1$ implies $M=1$,
\begin{equation*}
\mathbb{P}(S=1\mid X_S,M=1) = \frac{\mathbb{P}(S=1\mid X_S)}{\mathbb{P}(M=1\mid X_S)}.
\end{equation*}
Moreover,
\begin{align*}
\mathbb{P}(M=1\mid X_S)
&= \mathbb{P}(S=1\mid X_S) + \mathbb{P}(S'=1,S=0\mid X_S) \\
&= \mathbb{P}(S=1\mid X_S) + \mathbb{P}(S'=1\mid X_S,S=0)\,\mathbb{P}(S=0\mid X_S) \\
&= \mathbb{P}(S=1\mid X_S) + \mathbb{P}(S'=1)\{1-\mathbb{P}(S=1\mid X_S)\},
\end{align*}
where the final equality follows from the assumption that $\mathbb{P}(S'=1\mid X_S,S=0)=\mathbb{P}(S'=1)$.
Therefore,
\begin{equation*}
\mathbb{P}(S=1\mid X_S,M=1)
= \frac{\mathbb{P}(S=1\mid X_S)}{\mathbb{P}(S'=1) + \{1-\mathbb{P}(S'=1)\}\,\mathbb{P}(S=1\mid X_S)}.
\end{equation*}
Solving for $\mathbb{P}(S=1\mid X_S)$ gives
\begin{equation*}
\mathbb{P}(S=1\mid X_S)
= \frac{\mathbb{P}(S'=1)\,\mathbb{P}(S=1\mid X_S,M=1)}{1-\{1-\mathbb{P}(S'=1)\}\,\mathbb{P}(S=1\mid X_S,M=1)},
\end{equation*}
which gives eq.~\eqref{eq:omega-subsample}.
\end{proof}

\begin{proposition}[Identification of $g_{a,\text{res}}$ with restricted target covariates]
\label{prop:res-identification}

Suppose Assumptions~\ref{assump:mean_exchangeability_trial}--\ref{assump:mean_exchangeability_populations} and the Remark hold. 
Then $\mathbb{E}[Y^a \mid X_\tau]$ is identified from observed data by
\begin{equation}
\label{eq:g-res-id}
g_{a,\text{res}}(X_\tau)
= \mathbb{E}\biggl[
    \frac{\mathbb{I} (S=1)}{\omega(X_S)}\,\mu_a(X_S \cup X_\tau)
    \biggm| X_\tau
  \biggr],
\end{equation}
where $\mu_a(X_S \cup X_\tau) = \mathbb{E}[Y \mid X_S \cup X_\tau, S=1, A=a]$.

\end{proposition}

\begin{proof}
Starting from the right-hand side of \eqref{eq:g-res-id}, and using that
$\mu_a(X_S \cup X_\tau) = \mathbb{E}[Y^a \mid X_S \cup X_\tau]$,
\begin{allowdisplaybreaks}
\begin{align*}
\lefteqn{\mathbb{E}\biggl[\frac{\mathbb{I} (S=1)}{\omega(X_S)}\,\mu_a(X_S\cup X_\tau) \biggm| X_\tau\biggr]} \\*
\qquad
&= \mathbb{E}\biggl[
     \mathbb{E}\biggl[\frac{\mathbb{I} (S=1)}{\omega(X_S)}\,\mathbb{E}[Y^a \mid X_S \cup X_\tau] \biggm| X_S\cup X_\tau\biggr]
   \biggm| X_\tau\biggr]
   \tag{Tower property} \\
&= \mathbb{E}\biggl[
     \frac{\mathbb{P}(S=1 \mid X_S\cup X_\tau)}{\omega(X_S)}\,\mathbb{E}[Y^a \mid X_S \cup X_\tau]
   \biggm| X_\tau\biggr] \\
&= \mathbb{E}\biggl[
     \frac{\omega(X_S)}{\omega(X_S)}\,\mathbb{E}[Y^a \mid X_S \cup X_\tau]
   \biggm| X_\tau\biggr]
   \tag{Definition of $X_S$, $X_\tau$} \\
&= \mathbb{E}\bigl[\mathbb{E}[Y^a \mid X_S \cup X_\tau] \bigm| X_\tau\bigr] \\
&= \mathbb{E}[Y^a \mid X_\tau],
   \tag{Tower property}
\end{align*}
\end{allowdisplaybreaks}
which completes the proof.
\end{proof}

\subsection{Relaxing Auxiliary Sample Independence}
\label{app:aux-relax}

We consider relaxing the marginal independence assumption in eq.~\eqref{eq:aux-assump} to allow the auxiliary selection to depend on the data only through $X_\tau$,
\begin{equation}
\label{eq:aux-cond}
S'\perp(S,X_S)\mid X_\tau, \qquad q(X_\tau)=\mathbb P(S'=1\mid X_\tau)>0.
\end{equation}

\begin{lemma}[Selection probability under conditional independence]
\label{lem:selection_recovery_cond}
Under eq.~\eqref{eq:aux-cond},
\begin{equation}
\label{eq:omega-subsample-cond}
\omega(X_S)=\frac{q(X_\tau)\,\mathbb P(S=1\mid X_S\cup X_\tau,M=1)}{1-\{1-q(X_\tau)\}\,\mathbb P(S=1\mid X_S\cup X_\tau,M=1)}.
\end{equation}
\end{lemma}
\begin{proof}
The proof of Lemma~\ref{lem:selection_recovery} applies with every probability conditioned on $X_S\cup X_\tau$ rather than on $X_S$. By the Remark following Assumption~\ref{assump:identification}, $\mathbb P(S=1\mid X_S\cup X_\tau)=\omega(X_S)$, and eq.~\eqref{eq:aux-cond} gives $\mathbb P(S'=1\mid X_S\cup X_\tau,S=0)=q(X_\tau)$, so that $\mathbb P(M=1\mid X_S\cup X_\tau)=\omega(X_S)+q(X_\tau)\{1-\omega(X_S)\}$. The remaining steps are unchanged.
\end{proof}

Unless known by design, $q(X_\tau)$ can be estimated by regressing $S'$ on $X_\tau$ in the full cohort.
Estimating $q$ can be avoided for the gT-Learner by using the density ratio between the auxiliary and trial samples,
\begin{align}
\label{eq:aux-density-ratio}
h(X_S\cup X_\tau)
 &=\frac{p(X_S\cup X_\tau\mid S'=1)}{p(X_S\cup X_\tau\mid S=1)}
=\frac{p(X_S\mid X_\tau)}{p(X_S\mid X_\tau,S=1)}\,\frac{p(X_\tau\mid S'=1)}{p(X_\tau\mid S=1)} \notag \\
& =\frac{\mathbb P(S=1\mid X_\tau)}{\omega(X_S)}\,\frac{p(X_\tau\mid S'=1)}{p(X_\tau\mid S=1)},
\end{align}
where the second equality uses $p(X_S\mid X_\tau,S'=1)=p(X_S\mid X_\tau)$, implied by eq.~\eqref{eq:aux-cond}, and the third follows from Bayes' rule. To estimate $h$, we construct a classifier that predicts whether a record comes from the trial or the auxiliary sample. To define its label $G$, we stack the two samples: each unit with $S'=1$ contributes a record with $G=1$ and each unit with $S=1$ a record with $G=0$, so that a unit with $S=S'=1$ contributes one record of each. Then $p(X_S\cup X_\tau\mid G=1)=p(X_S\cup X_\tau\mid S'=1)$ and $p(X_S\cup X_\tau\mid G=0)=p(X_S\cup X_\tau\mid S=1)$, and Bayes' rule gives
\begin{equation*}
\frac{\mathbb P(G=1\mid X_S\cup X_\tau)}{\mathbb P(G=0\mid X_S\cup X_\tau)}
=\frac{\mathbb P(G=1)}{\mathbb P(G=0)}\,h(X_S\cup X_\tau),
\end{equation*}
so the odds of a classifier of $G$ on $X_S\cup X_\tau$ estimate $h$ up to the ratio of sample sizes.

By eq.~\eqref{eq:aux-density-ratio}, $h$ differs from $1/\omega$ only by a positive factor depending on $X_\tau$ alone. Multiplying regression weights by such a factor does not change the population target of a regression on $X_\tau$, so $h$ can replace $1/\omega$ in the gT-Learner. 

However, the gDR pseudo-outcome requires $1/\omega$ itself, since its correction term must cancel the plug-in bias on the target-population scale. Taking expectations of eq.~\eqref{eq:aux-density-ratio} over the trial given $X_\tau$ recovers the $X_\tau$ factor, so that
\begin{equation*}
\frac{1}{\omega(X_S)}
=\frac{h(X_S\cup X_\tau)}{\mathbb E\{h(X_S\cup X_\tau)\mid X_\tau,S=1\}\,\mathbb P(S=1\mid X_\tau)}.
\end{equation*}
Using $h$ for the gDR-Learner therefore requires two additional regressions on $X_\tau$ to estimate the denominator.

\section{Learner Error Bounds}
\label{app:error-bounds}

\subsection{T-Learner Error}
\begin{lemma}[T-Learner error decomposition]
Let
\begin{equation*}
\mathcal{B}(X_\tau)=\tau(X_\tau) - \tau^{\mathrm{sCATE}}(X_\tau).
\end{equation*}
Then the mean squared error of the T-Learner satisfies
\begin{equation*}
\mathbb{E}\bigl[\{\tau(X_\tau) - \hat\tau_{\mathrm{TL}}(X_\tau)\}^2\bigr]
\leq 4\bigl(\|\hat\mu_1 - \mu_1\|_2^2 + \|\hat\mu_0 - \mu_0\|_2^2\bigr) + 2\|\mathcal{B}\|_2^2.
\end{equation*}
\end{lemma}

\begin{proof}
By adding and subtracting $\tau^{\mathrm{sCATE}}(X_\tau)$,
\begin{equation*}
\tau(X_\tau) - \hat\tau_{\mathrm{TL}}(X_\tau)
= \tau(X_\tau) - \tau^{\mathrm{sCATE}}(X_\tau) + \tau^{\mathrm{sCATE}}(X_\tau) - \hat\tau_{\mathrm{TL}}(X_\tau).
\end{equation*}
Using $(a+b)^2 \leq 2a^2 + 2b^2$ and taking expectations,
\begin{align*}
\mathbb{E}\bigl[\{\tau(X_\tau) - \hat\tau_{\mathrm{TL}}(X_\tau)\}^2\bigr]
&\leq 2\,\mathbb{E}\bigl[\{\tau(X_\tau) - \tau^{\mathrm{sCATE}}(X_\tau)\}^2\bigr]
   + 2\,\mathbb{E}\bigl[\{\tau^{\mathrm{sCATE}}(X_\tau) - \hat\tau_{\mathrm{TL}}(X_\tau)\}^2\bigr] \\
&= 2\|\mathcal{B}\|_2^2
   + 2\,\mathbb{E}\Bigl[\bigl[\{\mu_1(X_\tau) - \hat\mu_1(X_\tau)\} - \{\mu_0(X_\tau) - \hat\mu_0(X_\tau)\}\bigr]^2\Bigr] \\
&\leq 2\|\mathcal{B}\|_2^2 + 4\bigl(\|\hat\mu_1 - \mu_1\|_2^2 + \|\hat\mu_0 - \mu_0\|_2^2\bigr).
\end{align*}
\end{proof}

\subsection{gT-Learner Error}
\begin{lemma}[gT-Learner error decomposition]
\label{lem:gtl-error}
Under Assumptions~\ref{assump:mean_exchangeability_trial}--\ref{assump:mean_exchangeability_populations}, bounded outcomes ($|Y| \leq M$ a.s.),
and uniform overlap ($\omega(X_S) \geq \delta > 0$ and
$\pi_a(X_S) \geq \delta' > 0$ a.s.),
\begin{align*}
\mathbb{E}\bigl[\{\tau(X_\tau)-\hat\tau_{\mathrm{gTL}}(X_\tau)\}^2\bigr]
&\leq 4\sum_{a=0}^1\bigl\|\hat\mu_a(X_\tau)-\mu_a(X_\tau;\hat\omega,\hat\pi)\bigr\|_2^2 \\
&\quad + C_{\delta,\delta',M}\bigl(\|\hat\omega-\omega\|_2^2+\|\hat\pi-\pi\|_2^2\bigr),
\end{align*}
where
\begin{equation*}
\mu_a(x_\tau;\hat\omega,\hat\pi)
= \mathbb{E}\left[\frac{\mathbb{I} (S=1)\mathbb{I} (A=a)}{\hat\omega(X_S)\hat\pi_a(X_S)}Y \Bigm| X_\tau=x_\tau\right],
\qquad
\|\hat\pi-\pi\|_2^2 = \sum_{a=0}^1\|\hat\pi_a-\pi_a\|_2^2,
\end{equation*}
and $C_{\delta,\delta',M} = 16M^2/\bigl\{\delta^2\delta'^2\min(\delta^2,\delta'^2)\bigr\}$.
\end{lemma}

\begin{proof}
For $a\in\{0,1\}$, define
\begin{equation*}
\mu_a(x_\tau;\omega,\pi)
= \mathbb{E}\left[\frac{\mathbb{I} (S=1)\mathbb{I} (A=a)}{\omega(X_S)\pi_a(X_S)}Y \Bigm| X_\tau=x_\tau\right].
\end{equation*}
Under Assumptions~\ref{assump:mean_exchangeability_trial}--\ref{assump:mean_exchangeability_populations},
\begin{equation*}
\mu_a(x_\tau;\omega,\pi) = \mu_a(x_\tau),
\qquad
\tau(x_\tau) = \mu_1(x_\tau)-\mu_0(x_\tau).
\end{equation*}

By sample splitting, we condition on the sample used to estimate $\hat\omega$ and $\hat\pi$, so that these nuisance estimates can be treated as fixed functions in the second-stage analysis.

We decompose
\begin{align*}
\tau(X_\tau)-\hat\tau_{\mathrm{gTL}}(X_\tau)
&= \bigl[\mu_1(X_\tau;\omega,\pi)-\mu_1(X_\tau;\hat\omega,\hat\pi)\bigr]
   - \bigl[\mu_0(X_\tau;\omega,\pi)-\mu_0(X_\tau;\hat\omega,\hat\pi)\bigr] \\
&\quad + \bigl[\mu_1(X_\tau;\hat\omega,\hat\pi)-\hat\mu_1(X_\tau)\bigr]
   - \bigl[\mu_0(X_\tau;\hat\omega,\hat\pi)-\hat\mu_0(X_\tau)\bigr].
\end{align*}
Therefore,
\begin{align}
\mathbb{E}\bigl[\{\tau(X_\tau)-\hat\tau_{\mathrm{gTL}}(X_\tau)\}^2\bigr]
&\leq 4\sum_{a=0}^1\bigl\|\hat\mu_a(X_\tau)-\mu_a(X_\tau;\hat\omega,\hat\pi)\bigr\|_2^2 \notag \\
&\quad + 4\sum_{a=0}^1\bigl\|\mu_a(X_\tau;\hat\omega,\hat\pi)-\mu_a(X_\tau;\omega,\pi)\bigr\|_2^2.
\label{eq:gtl-decomp}
\end{align}

For each $a\in\{0,1\}$,
\begin{equation}
\frac{1}{\omega\pi_a}-\frac{1}{\hat\omega\hat\pi_a}
= \frac{\hat\omega-\omega}{\omega\hat\omega\pi_a}
+ \frac{\hat\pi_a-\pi_a}{\hat\omega\pi_a\hat\pi_a}.
\label{eq:weight-decomp}
\end{equation}
Define
\begin{equation*}
Z_a = \mathbb{I} (S=1)\mathbb{I} (A=a)Y
\left\{
\frac{\hat\omega-\omega}{\omega\hat\omega\pi_a}
+ \frac{\hat\pi_a-\pi_a}{\hat\omega\pi_a\hat\pi_a}
\right\}.
\end{equation*}
Then
\begin{equation*}
\mu_a(X_\tau;\omega,\pi)-\mu_a(X_\tau;\hat\omega,\hat\pi) = \mathbb{E}[Z_a\mid X_\tau],
\end{equation*}
and hence
\begin{equation*}
\bigl\|\mu_a(X_\tau;\omega,\pi)-\mu_a(X_\tau;\hat\omega,\hat\pi)\bigr\|_2^2 \leq \mathbb{E}[Z_a^2].
\end{equation*}

Assume that the estimated probabilities are truncated so that $\hat\omega(X_S)\geq\delta$ and $\hat\pi_a(X_S)\geq\delta'$ almost surely. Since $|Y|\leq M$,
\begin{align}
\mathbb{E}[Z_a^2]
&\leq 2M^2\,\mathbb{E}\left[
\frac{(\hat\omega-\omega)^2}{\omega^2\hat\omega^2\pi_a^2}
+ \frac{(\hat\pi_a-\pi_a)^2}{\hat\omega^2\pi_a^2\hat\pi_a^2}
\right] \notag \\
&\leq 2M^2\left\{
\frac{\|\hat\omega-\omega\|_2^2}{\delta^4\delta'^2}
+ \frac{\|\hat\pi_a-\pi_a\|_2^2}{\delta^2\delta'^4}
\right\}.
\label{eq:arm-bound}
\end{align}
Summing over $a\in\{0,1\}$,
\begin{align*}
4\sum_{a=0}^1\bigl\|\mu_a(X_\tau;\omega,\pi)-\mu_a(X_\tau;\hat\omega,\hat\pi)\bigr\|_2^2
&\leq \frac{16M^2}{\delta^4\delta'^2}\|\hat\omega-\omega\|_2^2
   + \frac{8M^2}{\delta^2\delta'^4}\sum_{a=0}^1\|\hat\pi_a-\pi_a\|_2^2 \\
&\leq 16M^2\max\left\{\frac{1}{\delta^4\delta'^2}, \frac{1}{\delta^2\delta'^4}\right\}
   \bigl(\|\hat\omega-\omega\|_2^2+\|\hat\pi-\pi\|_2^2\bigr).
\end{align*}
Since
\begin{equation*}
\max\left\{\frac{1}{\delta^4\delta'^2}, \frac{1}{\delta^2\delta'^4}\right\}
= \frac{1}{\delta^2\delta'^2\min(\delta^2,\delta'^2)},
\end{equation*}
substitution into \eqref{eq:gtl-decomp} gives
\begin{align*}
\mathbb{E}\bigl[\{\tau(X_\tau)-\hat\tau_{\mathrm{gTL}}(X_\tau)\}^2\bigr]
&\leq 4\sum_{a=0}^1\bigl\|\hat\mu_a(X_\tau)-\mu_a(X_\tau;\hat\omega,\hat\pi)\bigr\|_2^2 \\
&\quad + C_{\delta,\delta',M}\bigl(\|\hat\omega-\omega\|_2^2+\|\hat\pi-\pi\|_2^2\bigr),
\end{align*}
where $C_{\delta,\delta',M} = 16M^2/\bigl\{\delta^2\delta'^2\min(\delta^2,\delta'^2)\bigr\}$.
\end{proof}

\subsection{gDR-Learner Excess Risk}

Define the population risk
$L(\tau, \eta) = \mathbb{E}[(\varphi(O;\eta) - \tau(X_\tau))^2]$, where
$\varphi(O;\eta) = \varphi_1(O;\eta) - \varphi_0(O;\eta)$ and $\varphi_a$ is as
defined in the pseudo-outcome construction above. Let
$\eta_0 = (\pi_0, \omega_0, \mu_{1,0}, \mu_{0,0})\in \mathcal{H}$ denote the true nuisance
parameter vector and $\hat{\eta} = (\hat{\pi}, \hat{\omega}, \hat{\mu}_1,
\hat{\mu}_0)$ its estimate. 

We define the excess risk as the difference
\begin{equation}\label{ex-risk}
    L(\hat{\tau},\eta_0)-L(\tau^*,\eta_0)
\end{equation}
in MSE risk $L(\cdot,\eta_0)$ when evaluated at $\hat{\tau}$ and at $\tau^*$, where $\tau^*$ is defined as the minimiser of the population risk, $L(\tau^*,\eta_0) = \inf_{\tau}L(\tau,\eta_0)$.

When the nuisance functions are known, an upper bound for this excess risk can be written in terms of the convergence rate of the CATE estimates obtained from gDR-learner's second stage pseudo-outcome regression:
$$L(\hat{\tau},\eta_0)-L(\tau^*,\eta_0) \leq \text{Rate}_{\tau}(\eta_0). $$
In practice, the gDR-learner does not know the true nuisance parameters $\eta_0$, but estimates them, $\hat{\eta}$, with this upper bound taking the form:
$$L(\hat{\tau},\hat{\eta})-L(\tau^*,\hat{\eta}) \leq \text{Rate}_{\tau}(\hat{\eta}). $$

Below, we use the criteria and arguments outlined by \citet{foster_orthogonal_2023,pryce_causal_2025} to write the right hand side of this bound in terms of the known nuisance functions, rather than the estimated nuisance functions.

\begin{definition}[Algorithm and Rates]
\label{def:rates}
\leavevmode
\begin{enumerate}[label=(\alph*)]
    \item \textit{Nuisance algorithms and rates. For the gDR-learner, as $\eta=(\pi,\mu_0,\mu_1,\omega)$, we define the first stage algorithms and their corresponding rate functions for each component of $\eta$.}

    \smallskip
    \textit{The algorithm} $\mathrm{Alg}_{\pi}$ \textit{outputs a predictor of $\pi$, $\hat{\pi}$, for which}
    \begin{equation*}
        \|\hat{\pi}-\pi_0\|_2 \leq \mathrm{Rate}_{\pi}(\delta) \quad \text{with probability at least } 1-\delta/5
    \end{equation*}

    \textit{The algorithm} $\mathrm{Alg}_{\omega}$ \textit{outputs a predictor of $\omega$, $\hat{\omega}$, for which}
    \begin{equation*}
        \|\hat{\omega}-\omega_0\|_2 \leq \mathrm{Rate}_{\omega}(\delta) \quad \text{with probability at least } 1-\delta/5
    \end{equation*}

    \textit{The algorithm} $\mathrm{Alg}_{\mu_0}$ \textit{outputs a predictor of $\mu_0$, $\hat{\mu}_0$, for which}
    \begin{equation*}
        \|\hat{\mu}_0-\mu_{0,0}\|_2 \leq \mathrm{Rate}_{\mu_0}(\delta) \quad \text{with probability at least } 1-\delta/5
    \end{equation*}

    \textit{The algorithm} $\mathrm{Alg}_{\mu_1}$ \textit{outputs a predictor of $\mu_1$, $\hat{\mu}_1$, for which}
    \begin{equation*}
        \|\hat{\mu}_1-\mu_{1,0}\|_2 \leq \mathrm{Rate}_{\mu_1}(\delta) \quad \text{with probability at least } 1-\delta/5
    \end{equation*}

    \item \textit{Target algorithm and rate. Let $\tau^*\in{\mathcal{T}}$ and define the second stage algorithm} $\mathrm{Alg}(\eta)$\textit{, with $\eta=(\pi,\mu_0,\mu_1,\omega)$, to be a predictor of $\tau$, $\hat{\tau}$, for which}
    \begin{equation*}
        L(\hat{\tau},\eta) - L(\tau^*,\eta) \leq \mathrm{Rate}_{\tau}(\delta;\eta) \quad \text{with probability at least } 1-\delta/5
    \end{equation*}

\end{enumerate}
\noindent\textbf{Notation:} When writing these rates we drop $\delta$ from the notation, writing $\mathrm{Rate}_{\pi}$, $\mathrm{Rate}_{\mu_0}$, $\mathrm{Rate}_{\mu_1}$, $\mathrm{Rate}_{\omega}$, $\mathrm{Rate}_{\tau}(\eta)$.
\end{definition}

\begin{definition}[Directional Derivative]
\label{def:derivative}
Let $\mathcal{F}$ be a vector space of functions. For a functional $F : \mathcal{F} \to \mathbb{R}$, we define the derivative operator $D_f F(f)[h] = \frac{d}{dt} F(f + th) \big|_{t=0}$ for a pair of functions $f, h \in \mathcal{F}$. Likewise, we define $D^k_f F(f)[h_1, \dots, h_k] = \frac{\partial^k}{\partial t_1 \dots \partial t_k} F(f + t_1 h_1 + \dots + t_k h_k) \big|_{t_1 = \dots = t_k = 0}$. When considering a functional in two arguments, e.g. $L(\tau, \eta)$, we write $D_\tau L(\tau, \eta)$ and $D_\eta L(\tau, \eta)$ to highlight which component the argument is taken with respect to.
\end{definition}

\renewcommand{\theassumption}{\Alph{assumption}}

\begin{assumption}[Neyman orthogonal loss function] \label{assump:orthogonal}
The population risk $L(\tau^*,\eta_0)$ is Neyman orthogonal:
$$D_{\eta_j}D_\tau L(\tau^*,\eta_0)[\tau-\tau^*, \eta_j - \eta_{j,0}] = 0,$$
for each $\eta_j \in \{\pi,\omega,\mu_1,\mu_0\}$.

\end{assumption}

\begin{assumption}[First order optimality]
For an element $x\in \mathcal{X}$, we define the star hull as in (\ref{starhull}) and adopt the shorthand star($\mathcal{X}$) $:=$ star($\mathcal{X}$,0).
\begin{equation} \label{starhull}
    \text{star}(\mathcal{X},x) = \bigl\{t x+(1-t) x' \bigm| x'\in \mathcal{X},\ t\in [0,1]\bigr\}.
\end{equation}
\label{assump:optimality}
    The minimiser for the population risk satisfies the first-order optimality condition:
\begin{equation*}
   D_{\tau}L(\tau^*,\eta_0)[\tau - \tau^*] = 0 \quad \text{for all } \tau\in \text{star}(\mathcal{T},\tau^*).
\end{equation*}

\end{assumption}

\begin{assumption}[Strong convexity in prediction]
\label{assump:convexity}
    The population risk $L$ is strongly convex with respect to the prediction: For all $\tau \in \mathcal{T}$ and $\eta \in \mathcal{H}$, 
\begin{equation*}
    D^2_{\tau}L(\bar{\tau},\eta)[\tau-\tau^*, \tau-\tau^*] \geq \lambda\|\tau - \tau^*\|_2^2 \quad \text{for all } \bar{\tau}\in \text{star}(\mathcal{T},\tau^*).
\end{equation*}
\end{assumption}

\begin{assumption}[Higher order smoothness]
    \label{assump:smoothness}

    There exist constants $\beta_1,~\beta_2>0$ such that the following derivative bounds hold: 
\begin{enumerate}[label=(\alph*)]
    \item Second-order smoothness with respect to target. For all $\tau \in \mathcal{T}$ and all $\bar{\tau} \in \text{star}(\mathcal{T},\tau^*)$,
    \begin{equation*}
       D^2_{\tau}L(\bar{\tau},\eta_0)[\tau-\tau^*,\tau-\tau^*]\leq \beta_1\|\tau-\tau^*\|_2^2.
    \end{equation*}
    \item Higher-order smoothness. For all $\tau \in \text{star}(\mathcal{T},\tau^*)$, $\eta\in \mathcal{H}$, and $\bar{\eta}\in \text{star}(\mathcal H,\eta_0)$,
    \begin{equation*}
       \bigl|D_\eta^2D_{\tau}L(\tau^*,\bar{\eta})[\tau-\tau^*,\eta-\eta_0,\eta-\eta_0]\bigr|\leq \beta_2\|\tau-\tau^*\|_2 \, \|\eta-\eta_0\|_2^2.
    \end{equation*}
    \end{enumerate}
\end{assumption}

\setcounter{theorem}{\numexpr\getrefnumber{thm:gdr-excess-risk-main}-1\relax}
\ifdefined\theHtheorem\renewcommand{\theHtheorem}{W\arabic{theorem}}\fi
\begin{theorem}[Excess risk bound for the gDR-Learner]
\label{thm:gdr-excess-risk}
Suppose Assumptions~\ref{assump:orthogonal}--\ref{assump:smoothness} hold and
that there exists $\tau^* \in \operatorname{arg\,min}_{\tau \in \mathcal{T}} L(\tau, \eta_0)$.
Then the gDR-Learner produces an estimate $\hat{\tau}$ such that,
with probability at least $1 - \delta$ for positive constants
$\delta_1, \ldots, \delta_7$ with $\sum_{j=1}^{7} \delta_j < \lambda/2$,
\begin{align}
\label{eq:gdr-excess-risk}
L(\hat{\tau}, \eta_0) - L(\tau^*, \eta_0)
&\leq \frac{\beta_1}{\lambda - 2\sum_{j=1}^{7}\delta_j}
\biggl(
\mathrm{Rate}_\tau(\hat{\eta})
+ \frac{\beta_{2,1}}{\delta_1}\,\mathrm{Rate}_\pi^4
+ \frac{\beta_{2,2}}{\delta_2}\,\mathrm{Rate}_\pi^2\,\mathrm{Rate}_{\mu_1}^2 \notag \\
&\qquad
+ \frac{\beta_{2,3}}{\delta_3}\,\mathrm{Rate}_\pi^2\,\mathrm{Rate}_{\mu_0}^2
+ \frac{\beta_{2,4}}{\delta_4}\,\mathrm{Rate}_\pi^2\,\mathrm{Rate}_\omega^2
+ \frac{\beta_{2,5}}{\delta_5}\,\mathrm{Rate}_\omega^2\,\mathrm{Rate}_{\mu_1}^2 \notag \\
&\qquad
+ \frac{\beta_{2,6}}{\delta_6}\,\mathrm{Rate}_\omega^2\,\mathrm{Rate}_{\mu_0}^2
+ \frac{\beta_{2,7}}{\delta_7}\,\mathrm{Rate}_\omega^4
\biggr),
\end{align}
where $\lambda$ and $\beta_1$ are the strong convexity and smoothness
constants from Assumptions~\ref{assump:convexity}
and~\ref{assump:smoothness}(a) respectively; $\mathrm{Rate}_\tau(\hat{\eta})$
is the convergence rate of the second-stage regression under estimated
nuisances; and $\mathrm{Rate}_\pi = \|\hat{\pi} - \pi_0\|_2$,
$\mathrm{Rate}_\omega = \|\hat{\omega} - \omega_0\|_2$,
$\mathrm{Rate}_{\mu_a} = \|\hat{\mu}_a - \mu_{a,0}\|_2$ are the
nuisance estimation rates.
The constants $\beta_{2,j}$ are defined as
\begin{align*}
\beta_{2,1}
&= \mathbb{E}\left[
    \left(\frac{SA}{\hat{\omega}\hat{\pi}^3}(Y - \hat{\mu}_1)
    - \frac{S(1-A)}{\hat{\omega}(1-\hat{\pi})^3}(Y - \hat{\mu}_0)\right)^2
\right], \\[6pt]
\beta_{2,2}
&= \mathbb{E}\left[\left(
    \frac{SA}{\hat{\omega}\hat{\pi}^2}
\right)^2\right],
\qquad
\beta_{2,3}
= \mathbb{E}\left[\left(
    \frac{S(1-A)}{\hat{\omega}(1-\hat{\pi})^2}
\right)^2\right], \\[6pt]
\beta_{2,4}
&= \mathbb{E}\left[
    \left(\frac{SA}{\hat{\pi}^2\hat{\omega}^2}(Y - \hat{\mu}_1)
    + \frac{S(1-A)}{(1-\hat{\pi})^2\hat{\omega}^2}(Y - \hat{\mu}_0)\right)^2
\right], \\[6pt]
\beta_{2,5}
&= \mathbb{E}\left[\left(
    \frac{SA}{\hat{\pi}\hat{\omega}^2}
\right)^2\right],
\qquad
\beta_{2,6}
= \mathbb{E}\left[\left(
    \frac{S(1-A)}{(1-\hat{\pi})\hat{\omega}^2}
\right)^2\right], \\[6pt]
\beta_{2,7}
&= \mathbb{E}\left[
    \left(\frac{SA}{\hat{\pi}\hat{\omega}^3}(Y - \hat{\mu}_1)
    - \frac{S(1-A)}{(1-\hat{\pi})\hat{\omega}^3}(Y - \hat{\mu}_0)\right)^2
\right].
\end{align*}
\end{theorem}

\subsection{Proof of the gDR-Learner excess risk bound}

We verify the four assumptions of \citet{foster_orthogonal_2023} for the gDR-Learner pseudo-outcome and then follow the arguments from Theorem~1 of \citet{pryce_causal_2025} to determine the excess risk bounds.

Throughout, write $\pi \equiv \pi_0 = \mathbb{P}(A=1\mid X_S,S=1)$, so that
$1-\pi = \mathbb{P}(A=0\mid X_S,S=1)$. For notational ease, we rewrite the pseudo-outcome for the gDR-Learner as
\begin{equation}
\label{eq:pseudo-outcome-proof}
\varphi(O;\eta)
= \frac{SA}{\pi\omega}(Y-\mu_1)
- \frac{S(1-A)}{(1-\pi)\omega}(Y-\mu_0)
+ \mu_1 - \mu_0,
\end{equation}
with nuisance vector $\eta = (\pi,\omega,\mu_1,\mu_0)$ and define the population risk
$L(\tau,\eta) = \mathbb{E}[\{\varphi(O;\eta) - \tau(X_\tau)\}^2]$.
The vector of true nuisance parameters is denoted by $\eta_0=(\pi_0, \omega_0, \mu_{1,0}, \mu_{0,0})$.

\begin{lemma}[Conditional moment condition]
\label{lem:moment}
Under Assumptions~\ref{assump:mean_exchangeability_trial}--\ref{assump:mean_exchangeability_populations}, for each $a\in\{0,1\}$,
\begin{equation*}
\mathbb{E}\left[
\frac{SI(A=a)}{\pi_{a,0}(X_S)\omega_0(X_S)}
\bigl\{Y - \mu_{a,0}(X_S \cup X_\tau)\bigr\}
\Bigm| X_S \cup X_\tau
\right] = 0.
\end{equation*}
\end{lemma}

\begin{proof}
By the tower property, conditioning further on $(X_S \cup X_\tau, S)$,
\begin{align*}
\lefteqn{\mathbb{E}\bigl[\mathbb{I} (A=a)\mathbb{I} (S=1)\{Y-\mu_{a,0}\}\bigm| X_S \cup X_\tau\bigr]} \\*
\qquad
&= \mathbb{E}\bigl[
  \mathbb{I} (S=1)\,\mathbb{E}[\mathbb{I} (A=a)\{Y-\mu_{a,0}\}\mid X_S \cup X_\tau,S=1]
  \bigm| X_S \cup X_\tau\bigr].
\end{align*}
When $S=1$, by Assumption~\ref{assump:mean_exchangeability_trial} (conditional exchangeability in the trial),
\begin{align*}
\mathbb{E}\bigl[\mathbb{I} (A=a)\{Y-\mu_{a,0}\}\bigm| X_S \cup X_\tau,S=1\bigr]
&= \pi_{a,0}(X_S)\,\mathbb{E}[Y-\mu_{a,0}\mid X_S \cup X_\tau,S=1,A=a] \\
&= \pi_{a,0}(X_S)\cdot 0,
\end{align*}
where the last equality uses the definition
$\mu_{a,0}(X_S \cup X_\tau) = \mathbb{E}[Y\mid X_S \cup X_\tau,S=1,A=a]$.
Since $\pi_{a,0}(X_S)\omega_0(X_S) > 0$ by the positivity assumptions,
dividing gives the result.
\end{proof}

\begin{proof}[Proof of Assumption~\ref{assump:orthogonal} (Neyman orthogonal loss function)]
We verify that
\begin{equation*}
D_{\eta_j}D_\tau L(\tau^*,\eta_0)[\tau-\tau^*, \eta_j - \eta_{j,0}] = 0
\end{equation*}
for each $\eta_j \in \{\pi,\omega,\mu_1,\mu_0\}$, where
\begin{align*}
D_\tau L(\tau^*,\eta_0)[\tau-\tau^*]
&= \left.\frac{\partial}{\partial t}\,\mathbb{E}\Bigl[\bigl\{\varphi(O;\eta_0) - t\,\tau(X_\tau) - (1-t)\,\tau^*(X_\tau)\bigr\}^2\Bigr]\right|_{t=0} \\
&= -2\,\mathbb{E}\bigl[\{\varphi(O;\eta_0)-\tau^*(X_\tau)\}\{\tau(X_\tau)-\tau^*(X_\tau)\}\bigr].
\end{align*}

\begin{enumerate}
\item \textit{Orthogonality with respect to $\pi$.}
The directional derivative of $\varphi$ in the direction $\pi - \pi_0$ is
\begin{align*}
D_\pi\varphi(O;\eta_0)[\pi-\pi_0]
&= \frac{\partial}{\partial t}\biggl\{\frac{SA}{\omega_0\{\pi_0+t(\pi-\pi_0)\}}(Y-\mu_{1,0}) \\
&\qquad - \frac{S(1-A)}{\omega_0\{1-\pi_0-t(\pi-\pi_0)\}}(Y-\mu_{0,0}) + \mu_{1,0} - \mu_{0,0}\biggr\}\biggr|_{t=0} \\
&= -\biggl\{\frac{SA}{\omega_0\pi_0^2}(Y-\mu_{1,0}) + \frac{S(1-A)}{\omega_0(1-\pi_0)^2}(Y-\mu_{0,0})\biggr\}(\pi-\pi_0),
\end{align*}
so that
\begin{allowdisplaybreaks}
\begin{align*}
\lefteqn{D_\pi D_\tau L(\tau^*,\eta_0)[\tau-\tau^*,\pi-\pi_0]} \\*
\qquad
&= 2\,\mathbb{E}\biggl[
     \biggl\{\frac{SA}{\omega_0\pi_0^2}(Y-\mu_{1,0}) + \frac{S(1-A)}{\omega_0(1-\pi_0)^2}(Y-\mu_{0,0})\biggr\}
     (\pi-\pi_0)\{\tau(X_\tau)-\tau^*(X_\tau)\}
   \biggr] \\
&= 2\,\mathbb{E}\biggl[
     \mathbb{E}\biggl[
       \biggl\{\frac{SA}{\omega_0\pi_0^2}(Y-\mu_{1,0}) + \frac{S(1-A)}{\omega_0(1-\pi_0)^2}(Y-\mu_{0,0})\biggr\} \\
&\qquad\qquad \times (\pi-\pi_0)\{\tau(X_\tau)-\tau^*(X_\tau)\}
     \biggm| X_S \cup X_\tau, S\biggr]
   \biggr] \\
&= 2\,\mathbb{E}\biggl[
     \mathbb{E}\biggl[
       \frac{SA}{\omega_0\pi_0^2}(Y-\mu_{1,0}) + \frac{S(1-A)}{\omega_0(1-\pi_0)^2}(Y-\mu_{0,0})
     \biggm| X_S \cup X_\tau\biggr] \\
&\qquad\qquad \times (\pi-\pi_0)\{\tau(X_\tau)-\tau^*(X_\tau)\}
   \biggr] \\
&= 2\,\mathbb{E}\biggl[
     \biggl(\frac{1}{\omega_0\pi_0^2}\bigl\{\mathbb{E}[SAY\mid X_S \cup X_\tau] - \mathbb{E}[SA\mid X_S \cup X_\tau]\,\mu_{1,0}\bigr\} \\
&\qquad\qquad + \frac{1}{\omega_0(1-\pi_0)^2}\bigl\{\mathbb{E}[S(1-A)Y\mid X_S \cup X_\tau] - \mathbb{E}[S(1-A)\mid X_S \cup X_\tau]\,\mu_{0,0}\bigr\}\biggr) \\
&\qquad\qquad \times (\pi-\pi_0)\{\tau(X_\tau)-\tau^*(X_\tau)\}
   \biggr] \\
&= 2\,\mathbb{E}\biggl[
     \biggl(\frac{\omega_0\pi_0\mu_{1,0} - \omega_0\pi_0\mu_{1,0}}{\omega_0\pi_0^2}
     + \frac{\omega_0(1-\pi_0)\mu_{0,0} - \omega_0(1-\pi_0)\mu_{0,0}}{\omega_0(1-\pi_0)^2}\biggr) \\
&\qquad\qquad \times (\pi-\pi_0)\{\tau(X_\tau)-\tau^*(X_\tau)\}
   \biggr] \\
&= 0.
\end{align*}
\end{allowdisplaybreaks}

\item \textit{Orthogonality with respect to $\mu_1$.}
The directional derivative of $\varphi$ in the direction $\mu_1 - \mu_{1,0}$ is
\begin{align*}
D_{\mu_1}\varphi(O;\eta_0)[\mu_1 - \mu_{1,0}]
&= \frac{\partial}{\partial t}\biggl\{\frac{AS}{\omega_0\pi_0}\bigl\{Y-\mu_{1,0} - t(\mu_{1}-\mu_{1,0})\bigr\}
   - \frac{(1-A)S}{\omega_0(1-\pi_0)}(Y-\mu_{0,0}) \\
&\qquad + \mu_{1,0} + t(\mu_1 - \mu_{1,0}) - \mu_{0,0}\biggr\}\biggr|_{t=0} \\
&= -\biggl\{\frac{AS}{\omega_0\pi_0}(\mu_1-\mu_{1,0}) - \mu_1 + \mu_{1,0}\biggr\},
\end{align*}
so then
\begin{allowdisplaybreaks}
\begin{align*}
\lefteqn{D_{\mu_1}D_\tau L(\tau^*,\eta_0)[\tau-\tau^*,\mu_1 - \mu_{1,0}]} \\*
\qquad
&= 2\,\mathbb{E}\biggl[
     \biggl\{\frac{AS}{\omega_0\pi_0}(\mu_1-\mu_{1,0}) - \mu_1 + \mu_{1,0}\biggr\}
     \{\tau(X_\tau)-\tau^*(X_\tau)\}
   \biggr] \\
&= 2\,\mathbb{E}\biggl[
     \mathbb{E}\biggl[
       \biggl\{\frac{AS}{\omega_0\pi_0}(\mu_1-\mu_{1,0}) - \mu_1 + \mu_{1,0}\biggr\}
       \{\tau(X_\tau)-\tau^*(X_\tau)\}
     \biggm| X_S \cup X_\tau\biggr]
   \biggr] \\
&= 2\,\mathbb{E}\biggl[
     \mathbb{E}\biggl[
       \frac{AS}{\omega_0\pi_0}(\mu_1-\mu_{1,0}) - \mu_1 + \mu_{1,0}
     \biggm| X_S \cup X_\tau\biggr]
     \{\tau(X_\tau)-\tau^*(X_\tau)\}
   \biggr] \\
&= 2\,\mathbb{E}\biggl[
     \biggl\{\frac{\mathbb{E}[AS\mid X_S \cup X_\tau]}{\omega_0\pi_0}(\mu_1-\mu_{1,0}) - \mu_1 + \mu_{1,0}\biggr\}
     \{\tau(X_\tau)-\tau^*(X_\tau)\}
   \biggr] \\
&= 2\,\mathbb{E}\bigl[
     (\mu_1-\mu_{1,0} - \mu_1 + \mu_{1,0})
     \{\tau(X_\tau)-\tau^*(X_\tau)\}
   \bigr] \\
&= 0.
\end{align*}
\end{allowdisplaybreaks}

\item \textit{Orthogonality with respect to $\mu_0$.}
This condition can be verified using the same arguments as for $\mu_1$.

\item \textit{Orthogonality with respect to $\omega$.}
The directional derivative of $\varphi$ in the direction $\omega - \omega_0$ is
\begin{align*}
D_{\omega}\varphi(O;\eta_0)[\omega - \omega_0]
&= \frac{\partial}{\partial t}\biggl\{\frac{SA}{\{\omega_0 + t(\omega-\omega_0)\}\pi_0}(Y-\mu_{1,0}) \\
&\qquad - \frac{S(1-A)}{\{\omega_0 + t(\omega-\omega_0)\}(1-\pi_0)}(Y-\mu_{0,0}) + \mu_{1,0} - \mu_{0,0}\biggr\}\biggr|_{t=0} \\
&= -\biggl\{\frac{SA}{\omega_0^2\pi_0}(Y-\mu_{1,0})-\frac{S(1-A)}{\omega_0^2(1-\pi_0)}(Y-\mu_{0,0})\biggr\}(\omega-\omega_0),
\end{align*}
so then
\begin{allowdisplaybreaks}
\begin{align*}
\lefteqn{D_{\omega}D_\tau L(\tau^*,\eta_0)[\tau-\tau^*,\omega-\omega_0]} \\*
\qquad
&= 2\,\mathbb{E}\biggl[
     \biggl\{\frac{SA}{\omega_0^2\pi_0}(Y-\mu_{1,0})-\frac{S(1-A)}{\omega_0^2(1-\pi_0)}(Y-\mu_{0,0})\biggr\}
     (\omega-\omega_0)\{\tau(X_\tau)-\tau^*(X_\tau)\}
   \biggr] \\
&= 2\,\mathbb{E}\biggl[
     \mathbb{E}\biggl[
       \biggl\{\frac{SA}{\omega_0^2\pi_0}(Y-\mu_{1,0})-\frac{S(1-A)}{\omega_0^2(1-\pi_0)}(Y-\mu_{0,0})\biggr\} \\
&\qquad\qquad \times (\omega-\omega_0)\{\tau(X_\tau)-\tau^*(X_\tau)\}
     \biggm| X_S \cup X_\tau\biggr]
   \biggr] \\
&= 2\,\mathbb{E}\biggl[
     \mathbb{E}\biggl[
       \frac{SA}{\omega_0^2\pi_0}(Y-\mu_{1,0})-\frac{S(1-A)}{\omega_0^2(1-\pi_0)}(Y-\mu_{0,0})
     \biggm| X_S \cup X_\tau\biggr] \\
&\qquad\qquad \times (\omega-\omega_0)\{\tau(X_\tau)-\tau^*(X_\tau)\}
   \biggr] \\
&= 2\,\mathbb{E}\biggl[
     \biggl(\frac{1}{\omega_0^2\pi_0}\bigl\{\mathbb{E}[SAY\mid X_S \cup X_\tau] - \mathbb{E}[SA\mid X_S \cup X_\tau]\,\mu_{1,0}\bigr\} \\
&\qquad\qquad - \frac{1}{\omega_0^2(1-\pi_0)}\bigl\{\mathbb{E}[S(1-A)Y\mid X_S \cup X_\tau] - \mathbb{E}[S(1-A)\mid X_S \cup X_\tau]\,\mu_{0,0}\bigr\}\biggr) \\
&\qquad\qquad \times (\omega-\omega_0)\{\tau(X_\tau)-\tau^*(X_\tau)\}
   \biggr] \\
&= 2\,\mathbb{E}\biggl[
     \biggl(\frac{\mu_{1,0}-\mu_{1,0}}{\omega_0^2\pi_0} - \frac{\mu_{0,0}- \mu_{0,0}}{\omega_0^2(1-\pi_0)}\biggr)
     (\omega-\omega_0)\{\tau(X_\tau)-\tau^*(X_\tau)\}
   \biggr] \\
&= 0.
\end{align*}
\end{allowdisplaybreaks}
\end{enumerate}
\end{proof}

\begin{proof}[Proof of Assumption~\ref{assump:optimality} (First order optimality)]
We compute the directional derivative of $L(\cdot,\eta_0)$ at $\tau^*$ in the direction $\tau - \tau^*$:
\begin{align*}
D_{\tau}L(\tau^*,\eta_0)[\tau - \tau^*]
&= -2\,\E\bigl[\{\varphi(O;\eta_0)-\tau^*(X_\tau)\}\{\tau(X_\tau)-\tau^*(X_\tau)\}\bigr] \\
&= -2\,\E\Bigl[\E\bigl[\{\varphi(O;\eta_0)-\tau^*(X_\tau)\}\{\tau(X_\tau)-\tau^*(X_\tau)\}\bigm| X_\tau\bigr]\Bigr] \\
&= -2\,\E\Bigl[\bigl\{\E[\varphi(O;\eta_0)\mid X_\tau]-\tau^*(X_\tau)\bigr\}\{\tau(X_\tau)-\tau^*(X_\tau)\}\Bigr] \\
&= 0,
\end{align*}
since the pseudo-outcome satisfies $\E[\varphi(O;\eta_0)\mid X_\tau] = \E[Y^1 - Y^0 \mid X_\tau] = \tau^*(X_\tau)$,
which was shown in Web Appendix~\ref{app:identification}. Hence $D_\tau L(\tau^*,\eta_0)[\tau-\tau^*] = 0$ for all $\tau \in \text{star}(\mathcal{T},\tau^*)$.
\end{proof}

\begin{proof}[Proof of Assumption~\ref{assump:convexity} (Strong convexity in prediction)]
\begin{align*}
D^2_{\tau}L(\bar{\tau},\eta)[\tau-\tau^*,\tau-\tau^*]
&= \left.\frac{\partial^2}{\partial t_1 \partial t_2}\,\E\Bigl[\bigl\{\varphi(O;\eta)-\bar{\tau} - t_1(\tau - \tau^*) - t_2(\tau - \tau^*)\bigr\}^2\Bigr]\right|_{t_1=t_2=0} \\
&= \left.\frac{\partial}{\partial t_2}\Bigl(-2\,\E\bigl[\{\varphi(O;\eta)-\bar{\tau} - t_2(\tau - \tau^*)\}(\tau-\tau^*)\bigr]\Bigr)\right|_{t_2=0} \\
&= 2\,\E[(\tau-\tau^*)^2].
\end{align*}
Therefore, for some $\lambda>0$,
$$ 2\,\E[(\tau-\tau^*)^2] \geq \lambda \|\tau-\tau^*\|_2^2,$$
hence Assumption~\ref{assump:convexity} holds with $\lambda = 2$.
\end{proof}

\begin{proof}[Proof of Assumption~\ref{assump:smoothness}(a) (Second-order smoothness)]
From the strong convexity calculation in the proof of Assumption~\ref{assump:convexity}, $D^2_\tau L = 2\,\mathbb{E}[(\tau-\tau^*)^2] \leq\beta_1\|\tau-\tau^*\|_2^2$ for any $\beta_1\geq 2$.
\end{proof}

\begin{proof}[Proof of Assumption~\ref{assump:smoothness}(b) (Higher-order smoothness)]
We need to find
\begin{align*}
\lefteqn{D_{\eta}^2D_{\tau}L(\tau^*,\bar{\eta})[\tau-\tau^*,\eta-\eta_0,\eta-\eta_0]} \\*
\qquad
&= \mathbb{E}\bigl[
    (\eta-\eta_0)^{\top}\,
    \nabla^2_{\eta\eta}\nabla_{\tau}l(\tau^*,\bar{\eta})\,
    (\eta-\eta_0)\,
    (\tau-\tau^*)\bigr],
\end{align*}
where $\eta-\eta_0 = (\pi-\pi_0,\ \omega-\omega_0,\ \mu_1-\mu_{1,0},\ \mu_0-\mu_{0,0})^{\top}$ and
\begin{equation*}
\nabla_{\eta\eta}^2\nabla_{\tau}l(\tau^*,\bar{\eta}) = {}
\begin{pmatrix}
\nabla^2_{\pi\pi}\nabla_{\tau}l & \nabla^2_{\pi\omega}\nabla_{\tau}l & \nabla^2_{\pi\mu_1}\nabla_{\tau}l & \nabla^2_{\pi\mu_0}\nabla_{\tau}l \\
\nabla^2_{\omega\pi}\nabla_{\tau}l & \nabla^2_{\omega\omega}\nabla_{\tau}l & \nabla^2_{\omega\mu_1}\nabla_{\tau}l & \nabla^2_{\omega\mu_0}\nabla_{\tau}l \\
\nabla^2_{\mu_1\pi}\nabla_{\tau}l & \nabla^2_{\mu_1\omega}\nabla_{\tau}l & \nabla^2_{\mu_1\mu_1}\nabla_{\tau}l & \nabla^2_{\mu_1\mu_0}\nabla_{\tau}l \\
\nabla^2_{\mu_0\pi}\nabla_{\tau}l & \nabla^2_{\mu_0\omega}\nabla_{\tau}l & \nabla^2_{\mu_0\mu_1}\nabla_{\tau}l & \nabla^2_{\mu_0\mu_0}\nabla_{\tau}l
\end{pmatrix}.
\end{equation*}
Since the sample loss is $l(\tau,\eta) = \{\varphi(O;\eta)-\tau(X_\tau)\}^2$, we have
$\nabla_\tau l(\tau^*,\eta) = -2\{\varphi(O;\eta)-\tau^*(X_\tau)\}$, and hence
$\nabla_{\eta_j\eta_k}^2\nabla_\tau l = -2\,\partial^2_{\eta_j\eta_k}\varphi$.

Writing $\bar\pi,\bar\omega,\bar\mu_1,\bar\mu_0$ for the components of $\bar{\eta}$, the second-order partial derivatives of $\varphi$ with respect to
$\eta=(\pi,\omega,\mu_1,\mu_0)$ evaluated at $\bar{\eta}$ are
\begin{align*}
\frac{\partial^2\varphi}{\partial\pi^2}\bigg|_{\bar{\eta}}
&= \frac{2SA}{\bar\omega\bar\pi^3}(Y-\bar\mu_1)
  -\frac{2S(1-A)}{\bar\omega(1-\bar\pi)^3}(Y-\bar\mu_0), \\
\frac{\partial^2\varphi}{\partial\omega^2}\bigg|_{\bar{\eta}}
&= \frac{2SA}{\bar\pi\bar\omega^3}(Y-\bar\mu_1)
  -\frac{2S(1-A)}{(1-\bar\pi)\bar\omega^3}(Y-\bar\mu_0), \\
\frac{\partial^2\varphi}{\partial\pi\,\partial\omega}\bigg|_{\bar{\eta}}
&= \frac{SA}{\bar\pi^2\bar\omega^2}(Y-\bar\mu_1)
  +\frac{S(1-A)}{(1-\bar\pi)^2\bar\omega^2}(Y-\bar\mu_0), \\
\frac{\partial^2\varphi}{\partial\pi\,\partial\mu_1}\bigg|_{\bar{\eta}}
&= \frac{SA}{\bar\omega\bar\pi^2}, \qquad
\frac{\partial^2\varphi}{\partial\pi\,\partial\mu_0}\bigg|_{\bar{\eta}}
= \frac{S(1-A)}{\bar\omega(1-\bar\pi)^2}, \\
\frac{\partial^2\varphi}{\partial\omega\,\partial\mu_1}\bigg|_{\bar{\eta}}
&= \frac{SA}{\bar\pi\bar\omega^2}, \qquad
\frac{\partial^2\varphi}{\partial\omega\,\partial\mu_0}\bigg|_{\bar{\eta}}
= -\frac{S(1-A)}{(1-\bar\pi)\bar\omega^2},
\end{align*}
and all second derivatives involving only $\mu_1$ or $\mu_0$ are zero.

Substituting the respective derivatives and collecting all seven non-zero
contributions from the quadratic form gives
\begin{allowdisplaybreaks}
\begin{align*}
\lefteqn{D_{\eta}^2D_{\tau}L(\tau^*,\bar{\eta})[\tau-\tau^*,\eta-\eta_0,\eta-\eta_0]} \\*
\qquad
&= -4\,\mathbb{E}\biggl[
      \biggl\{\frac{SA}{\bar\omega\bar\pi^3}(Y-\bar\mu_1)
           -\frac{S(1-A)}{\bar\omega(1-\bar\pi)^3}(Y-\bar\mu_0)\biggr\}
      (\pi-\pi_0)^2(\tau-\tau^*)\biggr] \\
&\quad -4\,\mathbb{E}\biggl[
      \frac{SA}{\bar\omega\bar\pi^2}
      (\pi-\pi_0)(\mu_1-\mu_{1,0})(\tau-\tau^*)\biggr] \\
&\quad -4\,\mathbb{E}\biggl[
      \frac{S(1-A)}{\bar\omega(1-\bar\pi)^2}
      (\pi-\pi_0)(\mu_0-\mu_{0,0})(\tau-\tau^*)\biggr] \\
&\quad -4\,\mathbb{E}\biggl[
      \biggl\{\frac{SA}{\bar\pi^2\bar\omega^2}(Y-\bar\mu_1)
           +\frac{S(1-A)}{(1-\bar\pi)^2\bar\omega^2}(Y-\bar\mu_0)\biggr\}
      (\pi-\pi_0)(\omega-\omega_0)(\tau-\tau^*)\biggr] \\
&\quad -4\,\mathbb{E}\biggl[
      \frac{SA}{\bar\pi\bar\omega^2}
      (\omega-\omega_0)(\mu_1-\mu_{1,0})(\tau-\tau^*)\biggr] \\
&\quad +4\,\mathbb{E}\biggl[
      \frac{S(1-A)}{(1-\bar\pi)\bar\omega^2}
      (\omega-\omega_0)(\mu_0-\mu_{0,0})(\tau-\tau^*)\biggr] \\
&\quad -4\,\mathbb{E}\biggl[
      \biggl\{\frac{SA}{\bar\pi\bar\omega^3}(Y-\bar\mu_1)
           -\frac{S(1-A)}{(1-\bar\pi)\bar\omega^3}(Y-\bar\mu_0)\biggr\}
      (\omega-\omega_0)^2(\tau-\tau^*)\biggr] \\
&= -4\,\E\bigl[C_1(\pi-\pi_0)^2(\tau-\tau^*)\bigr]
   - 4\,\E\bigl[C_2(\pi-\pi_0)(\mu_1-\mu_{1,0})(\tau-\tau^*)\bigr] \\
&\quad - 4\,\E\bigl[C_3(\pi-\pi_0)(\mu_0-\mu_{0,0})(\tau-\tau^*)\bigr]
   - 4\,\E\bigl[C_4(\pi-\pi_0)(\omega-\omega_0)(\tau-\tau^*)\bigr] \\
&\quad - 4\,\E\bigl[C_5(\omega-\omega_0)(\mu_1-\mu_{1,0})(\tau-\tau^*)\bigr]
   + 4\,\E\bigl[C_6(\omega-\omega_0)(\mu_0-\mu_{0,0})(\tau-\tau^*)\bigr] \\
&\quad - 4\,\E\bigl[C_7(\omega-\omega_0)^2(\tau-\tau^*)\bigr],
\end{align*}
\end{allowdisplaybreaks}
where
\begin{align*}
C_{1} &:= \frac{SA}{\bar\omega\bar\pi^3}(Y-\bar\mu_1)
       -\frac{S(1-A)}{\bar\omega(1-\bar\pi)^3}(Y-\bar\mu_0), \\
C_{2} &:= \frac{SA}{\bar\omega\bar\pi^2}, \qquad
C_{3} := \frac{S(1-A)}{\bar\omega(1-\bar\pi)^2}, \\
C_{4} &:= \frac{SA}{\bar\pi^2\bar\omega^2}(Y-\bar\mu_1)
       +\frac{S(1-A)}{(1-\bar\pi)^2\bar\omega^2}(Y-\bar\mu_0), \\
C_{5} &:= \frac{SA}{\bar\pi\bar\omega^2}, \qquad
C_{6} := \frac{S(1-A)}{(1-\bar\pi)\bar\omega^2}, \\
C_{7} &:= \frac{SA}{\bar\pi\bar\omega^3}(Y-\bar\mu_1)
       -\frac{S(1-A)}{(1-\bar\pi)\bar\omega^3}(Y-\bar\mu_0).
\end{align*}
Using the triangle inequality,
\begin{allowdisplaybreaks}
\begin{align*}
\lefteqn{\bigl|D_{\eta}^2D_{\tau}L(\tau^*,\bar{\eta})[\tau-\tau^*,\eta-\eta_0,\eta-\eta_0]\bigr|} \\*
\qquad
&\leq 4\Bigl(\E\bigl[|C_1(\pi-\pi_0)^2|\,|\tau-\tau^*|\bigr]
      + \E\bigl[|C_2(\pi-\pi_0)(\mu_1-\mu_{1,0})|\,|\tau-\tau^*|\bigr] \\
&\qquad + \E\bigl[|C_3(\pi-\pi_0)(\mu_0-\mu_{0,0})|\,|\tau-\tau^*|\bigr]
      + \E\bigl[|C_4(\pi-\pi_0)(\omega-\omega_0)|\,|\tau-\tau^*|\bigr] \\
&\qquad + \E\bigl[|C_5(\omega-\omega_0)(\mu_1-\mu_{1,0})|\,|\tau-\tau^*|\bigr]
      + \E\bigl[|C_6(\omega-\omega_0)(\mu_0-\mu_{0,0})|\,|\tau-\tau^*|\bigr] \\
&\qquad + \E\bigl[|C_7(\omega-\omega_0)^2|\,|\tau-\tau^*|\bigr]\Bigr).
\end{align*}
Applying Cauchy--Schwarz to each term to extract $\|\tau-\tau^*\|_2$, keeping $C_j$ inside the expectation,
\begin{align*}
\lefteqn{\bigl|D_{\eta}^2D_{\tau}L(\tau^*,\bar{\eta})[\tau-\tau^*,\eta-\eta_0,\eta-\eta_0]\bigr|} \\*
\qquad
&\leq 4\Bigl(
      \E\bigl[C_1^2(\pi-\pi_0)^4\bigr]^{1/2}
      + \E\bigl[C_2^2(\pi-\pi_0)^2(\mu_1-\mu_{1,0})^2\bigr]^{1/2} \\
&\qquad + \E\bigl[C_3^2(\pi-\pi_0)^2(\mu_0-\mu_{0,0})^2\bigr]^{1/2}
      + \E\bigl[C_4^2(\pi-\pi_0)^2(\omega-\omega_0)^2\bigr]^{1/2} \\
&\qquad + \E\bigl[C_5^2(\omega-\omega_0)^2(\mu_1-\mu_{1,0})^2\bigr]^{1/2}
      + \E\bigl[C_6^2(\omega-\omega_0)^2(\mu_0-\mu_{0,0})^2\bigr]^{1/2} \\
&\qquad + \E\bigl[C_7^2(\omega-\omega_0)^4\bigr]^{1/2}
      \Bigr)\|\tau-\tau^*\|_2.
\end{align*}
Applying Cauchy--Schwarz to separate $C_j$ from the nuisance products,
\begin{align*}
\lefteqn{\bigl|D_{\eta}^2D_{\tau}L(\tau^*,\bar{\eta})[\tau-\tau^*,\eta-\eta_0,\eta-\eta_0]\bigr|} \\*
\qquad
&\leq 4\Bigl(
      \E\bigl[C_1^4\bigr]^{1/4}\|\pi-\pi_0\|_8^2
      + \E\bigl[\{C_2(\pi-\pi_0)\}^4\bigr]^{1/4}\|\mu_1-\mu_{1,0}\|_4 \\
&\qquad + \E\bigl[\{C_3(\pi-\pi_0)\}^4\bigr]^{1/4}\|\mu_0-\mu_{0,0}\|_4
      + \E\bigl[\{C_4(\pi-\pi_0)\}^4\bigr]^{1/4}\|\omega-\omega_0\|_4 \\
&\qquad + \E\bigl[\{C_5(\omega-\omega_0)\}^4\bigr]^{1/4}\|\mu_1-\mu_{1,0}\|_4
      + \E\bigl[\{C_6(\omega-\omega_0)\}^4\bigr]^{1/4}\|\mu_0-\mu_{0,0}\|_4 \\
&\qquad + \E\bigl[C_7^4\bigr]^{1/4}\|\omega-\omega_0\|_8^2
      \Bigr)\|\tau-\tau^*\|_2.
\end{align*}
Applying Cauchy--Schwarz once more to separate the joint $C_j$--nuisance factors,
\begin{align*}
\lefteqn{\bigl|D_{\eta}^2D_{\tau}L(\tau^*,\bar{\eta})[\tau-\tau^*,\eta-\eta_0,\eta-\eta_0]\bigr|} \\*
\qquad
&\leq 4\Bigl(
      \E\bigl[C_1^4\bigr]^{1/4}\|\pi-\pi_0\|_8^2
      + \E\bigl[C_2^8\bigr]^{1/8}\|\pi-\pi_0\|_8\|\mu_1-\mu_{1,0}\|_4 \\
&\qquad + \E\bigl[C_3^8\bigr]^{1/8}\|\pi-\pi_0\|_8\|\mu_0-\mu_{0,0}\|_4
      + \E\bigl[C_4^8\bigr]^{1/8}\|\pi-\pi_0\|_8\|\omega-\omega_0\|_4 \\
&\qquad + \E\bigl[C_5^8\bigr]^{1/8}\|\omega-\omega_0\|_8\|\mu_1-\mu_{1,0}\|_4
      + \E\bigl[C_6^8\bigr]^{1/8}\|\omega-\omega_0\|_8\|\mu_0-\mu_{0,0}\|_4 \\
&\qquad + \E\bigl[C_7^4\bigr]^{1/4}\|\omega-\omega_0\|_8^2
      \Bigr)\|\tau-\tau^*\|_2 \\
&\leq 4\Bigl(
      \E\bigl[C_1^2\bigr]^{1/2}\|\pi-\pi_0\|_2^2
      + \E\bigl[C_2^2\bigr]^{1/2}\|\pi-\pi_0\|_2\|\mu_1-\mu_{1,0}\|_2 \\
&\qquad + \E\bigl[C_3^2\bigr]^{1/2}\|\pi-\pi_0\|_2\|\mu_0-\mu_{0,0}\|_2
      + \E\bigl[C_4^2\bigr]^{1/2}\|\pi-\pi_0\|_2\|\omega-\omega_0\|_2 \\
&\qquad + \E\bigl[C_5^2\bigr]^{1/2}\|\omega-\omega_0\|_2\|\mu_1-\mu_{1,0}\|_2
      + \E\bigl[C_6^2\bigr]^{1/2}\|\omega-\omega_0\|_2\|\mu_0-\mu_{0,0}\|_2 \\
&\qquad + \E\bigl[C_7^2\bigr]^{1/2}\|\omega-\omega_0\|_2^2
      \Bigr)\|\tau-\tau^*\|_2.
\end{align*}
\end{allowdisplaybreaks}
Setting $\|\eta-\eta_0\|_2:=\max_j\|\eta_j-\eta_{j,0}\|_2$, so that each component $L_2$ norm is bounded by $\|\eta-\eta_0\|_2$, gives the desired result,
\begin{equation*}
\bigl|D_{\eta}^2D_{\tau}L(\tau^*,\bar{\eta})[\tau-\tau^*,\eta-\eta_0,\eta-\eta_0]\bigr|
\leq \beta_2\|\tau-\tau^*\|_2 \, \|\eta-\eta_0\|_2^2,
\end{equation*}
where $\beta_2 = 4(\sqrt{\beta_{2,1}}+\sqrt{\beta_{2,2}}+\sqrt{\beta_{2,3}}+\sqrt{\beta_{2,4}} + \sqrt{\beta_{2,5}} + \sqrt{\beta_{2,6}}+\sqrt{\beta_{2,7}})$.
\end{proof}

\begin{proof}[Proof of Theorem~\ref{thm:gdr-excess-risk} (gDR-Learner excess risk)]
We abbreviate $R_{\mathcal{T}} := \mathrm{Rate}_{\tau}(\hat{\eta})$ throughout. We prove the bound on $\|\hat{\tau} - \tau^*\|_2^2$ first, then derive the excess risk bound at $\eta_0$.

By a second-order Taylor expansion of $L(\cdot, \hat{\eta})$ around $\tau^*$, there exists $\bar{\tau} \in \mathrm{star}(\mathcal{T}, \tau^*)$ such that
\begin{align*}
L(\hat{\tau}, \hat{\eta}) - L(\tau^*, \hat{\eta}) - D_\tau L(\tau^*, \hat{\eta})[\hat{\tau} - \tau^*]
&= \frac{1}{2} D^2_\tau L(\bar{\tau}, \hat{\eta})[\hat{\tau} - \tau^*, \hat{\tau} - \tau^*] \\
&\geq \frac{\lambda}{2} \|\hat{\tau} - \tau^*\|_2^2,
\end{align*}
where the inequality applies Assumption~\ref{assump:convexity} (strong convexity).
Rearranging and using the target algorithm rate $L(\hat{\tau}, \hat{\eta}) - L(\tau^*, \hat{\eta}) \leq R_{\mathcal{T}}$, we conclude that
\begin{equation}
\frac{\lambda}{2} \|\hat{\tau} - \tau^*\|_2^2
\leq R_{\mathcal{T}} - D_\tau L(\tau^*, \hat{\eta})[\hat{\tau} - \tau^*]. \label{eq:phase1}
\end{equation}

We expand $-D_\tau L(\tau^*, \hat{\eta})[\hat{\tau} - \tau^*]$ in $\eta$ around $\eta_0$. By Assumption~\ref{assump:smoothness}, there exists $\bar{\eta} \in \mathrm{star}(\mathcal{H}, \eta_0)$ such that
\begin{align*}
-D_\tau L(\tau^*, \hat{\eta})[\hat{\tau} - \tau^*]
&= -D_\tau L(\tau^*, \eta_0)[\hat{\tau} - \tau^*]
   - D_\eta D_\tau L(\tau^*, \eta_0)[\hat{\tau} - \tau^*, \hat{\eta} - \eta_0] \\
&\quad - \frac{1}{2} D^2_\eta D_\tau L(\tau^*, \bar{\eta})[\hat{\tau} - \tau^*, \hat{\eta} - \eta_0, \hat{\eta} - \eta_0].
\end{align*}
Applying Assumption~\ref{assump:orthogonal} (Neyman orthogonality), the first-order cross term vanishes for each component $\eta_j \in \{\pi, \omega, \mu_1, \mu_0\}$, and applying the explicit higher-order smoothness bound established in the proof of Assumption~\ref{assump:smoothness}(b) to the remainder gives
\begin{align*}
-D_\tau L(\tau^*, \hat{\eta})[\hat{\tau} - \tau^*]
&= -D_\tau L(\tau^*, \eta_0)[\hat{\tau} - \tau^*] \\
&\quad - \frac{1}{2} D^2_\eta D_\tau L(\tau^*, \bar{\eta})[\hat{\tau} - \tau^*, \hat{\eta} - \eta_0, \hat{\eta} - \eta_0] \\
&\leq -D_\tau L(\tau^*, \eta_0)[\hat{\tau} - \tau^*] \\
&\quad + 2\E[C_1^2]^{1/2}\, \|\hat{\tau} - \tau^*\|_2\, \|\hat{\pi} - \pi_0\|_2^2 \\
&\quad + 2\E[C_2^2]^{1/2}\, \|\hat{\tau} - \tau^*\|_2\, \|\hat{\pi} - \pi_0\|_2 \|\hat{\mu}_1 - \mu_{1,0}\|_2 \\
&\quad + 2\E[C_3^2]^{1/2}\, \|\hat{\tau} - \tau^*\|_2\, \|\hat{\pi} - \pi_0\|_2 \|\hat{\mu}_0 - \mu_{0,0}\|_2 \\
&\quad + 2\E[C_4^2]^{1/2}\, \|\hat{\tau} - \tau^*\|_2\, \|\hat{\pi} - \pi_0\|_2 \|\hat{\omega} - \omega_0\|_2 \\
&\quad + 2\E[C_5^2]^{1/2}\, \|\hat{\tau} - \tau^*\|_2\, \|\hat{\omega} - \omega_0\|_2 \|\hat{\mu}_1 - \mu_{1,0}\|_2 \\
&\quad + 2\E[C_6^2]^{1/2}\, \|\hat{\tau} - \tau^*\|_2\, \|\hat{\omega} - \omega_0\|_2 \|\hat{\mu}_0 - \mu_{0,0}\|_2 \\
&\quad + 2\E[C_7^2]^{1/2}\, \|\hat{\tau} - \tau^*\|_2\, \|\hat{\omega} - \omega_0\|_2^2.
\end{align*}
Writing $\beta_{2,j} = \E[C_j^2]$ and applying Young's inequality $2\sqrt{\beta_{2,j}}\, \|\hat{\tau} - \tau^*\|_2 \, v_j \leq \delta_j \|\hat{\tau} - \tau^*\|_2^2 + (\beta_{2,j}/\delta_j) v_j^2$ to each of the seven terms separately with parameters $\delta_1, \ldots, \delta_7 > 0$, and combining with \eqref{eq:phase1}, gives
\begin{align}
\biggl(\frac{\lambda}{2} - \sum_{i=1}^{7} \delta_i\biggr) \|\hat{\tau} - \tau^*\|_2^2
&\leq - D_\tau L(\tau^*, \eta_0)[\hat{\tau} - \tau^*] + R_{\mathcal{T}}
   + \frac{\beta_{2,1}}{\delta_1} \|\hat{\pi} - \pi_0\|_2^4 \notag \\
&\quad + \frac{\beta_{2,2}}{\delta_2} \|\hat{\pi} - \pi_0\|_2^2 \|\hat{\mu}_1 - \mu_{1,0}\|_2^2
   + \frac{\beta_{2,3}}{\delta_3} \|\hat{\pi} - \pi_0\|_2^2 \|\hat{\mu}_0 - \mu_{0,0}\|_2^2 \notag \\
&\quad + \frac{\beta_{2,4}}{\delta_4} \|\hat{\pi} - \pi_0\|_2^2 \|\hat{\omega} - \omega_0\|_2^2
   + \frac{\beta_{2,5}}{\delta_5} \|\hat{\omega} - \omega_0\|_2^2 \|\hat{\mu}_1 - \mu_{1,0}\|_2^2 \notag \\
&\quad + \frac{\beta_{2,6}}{\delta_6} \|\hat{\omega} - \omega_0\|_2^2 \|\hat{\mu}_0 - \mu_{0,0}\|_2^2
   + \frac{\beta_{2,7}}{\delta_7} \|\hat{\omega} - \omega_0\|_2^4. \label{eq:phase2}
\end{align}
We require $\sum_{i=1}^{7} \delta_i < \lambda/2$ so that the left-hand coefficient is strictly positive. Applying Assumption~\ref{assump:optimality} (first-order optimality), $D_\tau L(\tau^*, \eta_0)[\hat{\tau} - \tau^*] \geq 0$, so the negative first-order term may be discarded. Dividing through by $\lambda/2 - \sum_{i=1}^7 \delta_i > 0$ yields
\begin{align}
\|\hat{\tau} - \tau^*\|_2^2
&\leq \frac{1}{\lambda/2 - \sum_{i=1}^{7} \delta_i} \Biggl(
    R_{\mathcal{T}}
    + \frac{\beta_{2,1}}{\delta_1} \|\hat{\pi} - \pi_0\|_2^4
    + \frac{\beta_{2,2}}{\delta_2} \|\hat{\pi} - \pi_0\|_2^2 \|\hat{\mu}_1 - \mu_{1,0}\|_2^2 \notag \\
&\qquad + \frac{\beta_{2,3}}{\delta_3} \|\hat{\pi} - \pi_0\|_2^2 \|\hat{\mu}_0 - \mu_{0,0}\|_2^2
    + \frac{\beta_{2,4}}{\delta_4} \|\hat{\pi} - \pi_0\|_2^2 \|\hat{\omega} - \omega_0\|_2^2 \notag \\
&\qquad + \frac{\beta_{2,5}}{\delta_5} \|\hat{\omega} - \omega_0\|_2^2 \|\hat{\mu}_1 - \mu_{1,0}\|_2^2
    + \frac{\beta_{2,6}}{\delta_6} \|\hat{\omega} - \omega_0\|_2^2 \|\hat{\mu}_0 - \mu_{0,0}\|_2^2
    + \frac{\beta_{2,7}}{\delta_7} \|\hat{\omega} - \omega_0\|_2^4
\Biggr). \label{eq:theta-bound}
\end{align}
To bound $L(\hat{\tau}, \eta_0) - L(\tau^*, \eta_0)$, we perform a further Taylor expansion of $L(\cdot, \eta_0)$ around $\tau^*$. There exists $\bar{\tau} \in \mathrm{star}(\mathcal{T}, \tau^*)$ such that
\begin{align}
L(\hat{\tau}, \eta_0) - L(\tau^*, \eta_0)
&= D_\tau L(\tau^*, \eta_0)[\hat{\tau} - \tau^*] + \frac{1}{2} D^2_\tau L(\bar{\tau}, \eta_0)[\hat{\tau} - \tau^*, \hat{\tau} - \tau^*] \notag \\
&\leq D_\tau L(\tau^*, \eta_0)[\hat{\tau} - \tau^*] + \frac{\beta_1}{2} \|\hat{\tau} - \tau^*\|_2^2, \label{eq:phase3}
\end{align}
where the inequality applies the second-order smoothness bound from Assumption~\ref{assump:smoothness}(a).
Noting that
\begin{equation*}
\frac{\beta_1}{2}\|\hat\tau - \tau^*\|_2^2
= \frac{\beta_1}{\lambda/2 - \sum_i\delta_i} \biggl(\frac{\lambda}{2} - \sum_i\delta_i\biggr)\frac{1}{2}\|\hat\tau-\tau^*\|_2^2,
\end{equation*}
we substitute the bound \eqref{eq:phase2} into \eqref{eq:phase3}:
\begin{allowdisplaybreaks}
\begin{align*}
L(\hat{\tau}, \eta_0) - L(\tau^*, \eta_0)
&\leq D_\tau L(\tau^*, \eta_0)[\hat{\tau} - \tau^*] \\
&\quad + \frac{\beta_1}{\lambda - 2\sum_{i=1}^7 \delta_i} \Biggl(
    -D_\tau L(\tau^*, \eta_0)[\hat{\tau} - \tau^*] + R_{\mathcal{T}}
    + \frac{\beta_{2,1}}{\delta_1} \|\hat{\pi} - \pi_0\|_2^4 \\
&\qquad + \frac{\beta_{2,2}}{\delta_2} \|\hat{\pi} - \pi_0\|_2^2 \|\hat{\mu}_1 - \mu_{1,0}\|_2^2
    + \frac{\beta_{2,3}}{\delta_3} \|\hat{\pi} - \pi_0\|_2^2 \|\hat{\mu}_0 - \mu_{0,0}\|_2^2 \\
&\qquad + \frac{\beta_{2,4}}{\delta_4} \|\hat{\pi} - \pi_0\|_2^2 \|\hat{\omega} - \omega_0\|_2^2
    + \frac{\beta_{2,5}}{\delta_5} \|\hat{\omega} - \omega_0\|_2^2 \|\hat{\mu}_1 - \mu_{1,0}\|_2^2 \\
&\qquad + \frac{\beta_{2,6}}{\delta_6} \|\hat{\omega} - \omega_0\|_2^2 \|\hat{\mu}_0 - \mu_{0,0}\|_2^2
    + \frac{\beta_{2,7}}{\delta_7} \|\hat{\omega} - \omega_0\|_2^4
\Biggr) \\
&= \biggl(1 - \frac{\beta_1}{\lambda - 2\sum_{i=1}^7 \delta_i}\biggr) D_\tau L(\tau^*, \eta_0)[\hat{\tau} - \tau^*] \\
&\quad + \frac{\beta_1}{\lambda - 2\sum_{i=1}^7 \delta_i} \Biggl(
    R_{\mathcal{T}}
    + \frac{\beta_{2,1}}{\delta_1} \|\hat{\pi} - \pi_0\|_2^4
    + \frac{\beta_{2,2}}{\delta_2} \|\hat{\pi} - \pi_0\|_2^2 \|\hat{\mu}_1 - \mu_{1,0}\|_2^2 \\
&\qquad + \frac{\beta_{2,3}}{\delta_3} \|\hat{\pi} - \pi_0\|_2^2 \|\hat{\mu}_0 - \mu_{0,0}\|_2^2
    + \frac{\beta_{2,4}}{\delta_4} \|\hat{\pi} - \pi_0\|_2^2 \|\hat{\omega} - \omega_0\|_2^2 \\
&\qquad + \frac{\beta_{2,5}}{\delta_5} \|\hat{\omega} - \omega_0\|_2^2 \|\hat{\mu}_1 - \mu_{1,0}\|_2^2
    + \frac{\beta_{2,6}}{\delta_6} \|\hat{\omega} - \omega_0\|_2^2 \|\hat{\mu}_0 - \mu_{0,0}\|_2^2 \\
&\qquad + \frac{\beta_{2,7}}{\delta_7} \|\hat{\omega} - \omega_0\|_2^4
\Biggr).
\end{align*}
\end{allowdisplaybreaks}
Since $\beta_1 \geq \lambda - 2\sum_i\delta_i$ (which holds because $\beta_1 \geq 2 = \lambda$ from Assumptions~\ref{assump:convexity} and~\ref{assump:smoothness}(a), and $\lambda - 2\sum_i\delta_i < \lambda$), the coefficient on the first-order term satisfies $1 - \beta_1/(\lambda - 2\sum_i\delta_i) \leq 0$. Combined with Assumption~\ref{assump:optimality}, which gives $D_\tau L(\tau^*, \eta_0)[\hat\tau - \tau^*] \geq 0$, this term is non-positive, giving the final excess risk bound
\begin{align}
L(\hat{\tau}, \eta_0) - L(\tau^*, \eta_0)
&\leq \frac{\beta_1}{\lambda - 2\sum_{i=1}^{7} \delta_i} \Biggl(
    R_{\mathcal{T}}
    + \frac{\beta_{2,1}}{\delta_1} \|\hat{\pi} - \pi_0\|_2^4
    + \frac{\beta_{2,2}}{\delta_2} \|\hat{\pi} - \pi_0\|_2^2 \|\hat{\mu}_1 - \mu_{1,0}\|_2^2 \notag \\
&\qquad + \frac{\beta_{2,3}}{\delta_3} \|\hat{\pi} - \pi_0\|_2^2 \|\hat{\mu}_0 - \mu_{0,0}\|_2^2
    + \frac{\beta_{2,4}}{\delta_4} \|\hat{\pi} - \pi_0\|_2^2 \|\hat{\omega} - \omega_0\|_2^2 \notag \\
&\qquad + \frac{\beta_{2,5}}{\delta_5} \|\hat{\omega} - \omega_0\|_2^2 \|\hat{\mu}_1 - \mu_{1,0}\|_2^2
    + \frac{\beta_{2,6}}{\delta_6} \|\hat{\omega} - \omega_0\|_2^2 \|\hat{\mu}_0 - \mu_{0,0}\|_2^2 \notag \\
&\qquad + \frac{\beta_{2,7}}{\delta_7} \|\hat{\omega} - \omega_0\|_2^4
\Biggr), \label{eq:excess-risk}
\end{align}
where the $\delta_i > 0$ are parameters subject to $\sum_{i=1}^7 \delta_i < \lambda/2$.
Applying Definition~\ref{def:rates}(a), $\|\hat{\pi}-\pi_0\|_2 \leq \mathrm{Rate}_\pi$,
$\|\hat{\omega}-\omega_0\|_2 \leq \mathrm{Rate}_\omega$,
$\|\hat{\mu}_1-\mu_{1,0}\|_2 \leq \mathrm{Rate}_{\mu_1}$, and
$\|\hat{\mu}_0-\mu_{0,0}\|_2 \leq \mathrm{Rate}_{\mu_0}$, each with probability at least $1-\delta/5$,
and recalling $R_{\mathcal{T}} = \mathrm{Rate}_\tau(\hat{\eta})$, which holds with probability at least $1-\delta/5$ by Definition~\ref{def:rates}(b).
The bound \eqref{eq:excess-risk} requires all five of these guarantees to hold simultaneously; since each fails with probability at most $\delta/5$, a union bound gives that they hold jointly with probability at least $1 - 5\cdot(\delta/5) = 1-\delta$.
Substituting into \eqref{eq:excess-risk} on this joint event gives the bound stated in Theorem~\ref{thm:gdr-excess-risk}.
\end{proof}

\section{Cross-fitted Estimators}
\label{app:cross-fit-alg}
Here we describe the algorithms with cross-fitting for the gT-Learner, and gDR-Learner, for the standard settings, with extensions to the enriched and restricted learners.

\begin{breakablealgorithm}
\caption{Cross-Fitted Generalized T-Learner (gT-Learner)}
\begin{algorithmic}[1]
\renewcommand{\baselinestretch}{1}\selectfont

\State \textbf{Input:} Data $\mathcal D$, test features $X_{\tau,\mathrm{test}}$, number of folds $K$, and estimation setting.
\State Partition $\{1,\ldots,n\}$ into $K$ disjoint folds $\mathcal I_1,\ldots,\mathcal I_K$, stratified by $S$.

\For{$k=1$ to $K$, with $\mathcal I_{-k}=\{1,\ldots,n\}\setminus\mathcal I_k$}

    \State Estimate $\hat\pi_a^{(k)}(X_S)$ using trial units in $\mathcal I_{-k}$.
    \State Estimate $\hat\omega^{(k)}(X_S)$ using $\mathcal I_{-k}$; in the Subsample or Restricted setting, recover $\hat\omega^{(k)}$ from $\mathbb P(S=1\mid X_S,M=1)$.

    \If{Standard or Subsample setting}
        \State For $a\in\{0,1\}$, regress $Y$ on $X_\tau$ among trial units with $A=a$, weighted by
        $1/\{\hat\omega^{(k)}(X_S)\hat\pi_a^{(k)}(X_S)\}$, to obtain $\hat\mu_a^{(k)}(X_\tau)$.
        \State Set $\hat\tau^{(k)}(X_\tau)=\hat\mu_1^{(k)}(X_\tau)-\hat\mu_0^{(k)}(X_\tau)$.

    \ElsIf{Prognostic setting}
        \State For $a\in\{0,1\}$, estimate $\hat\mu_a^{(k)}(X_S\cup X_\tau,X_P)$ in the trial.
        \State Regress its cross-fitted predictions on $X_S\cup X_\tau$ among trial participants to obtain $\hat g_{a,\mathrm{enr}}^{(k)}$.
        \State Regress $\hat g_{1,\mathrm{enr}}^{(k)}-\hat g_{0,\mathrm{enr}}^{(k)}$ on $X_\tau$ over the target population to obtain $\hat\tau^{(k)}_{\mathrm{gTL,enr}}$.

    \ElsIf{Restricted setting}
        \State For $a\in\{0,1\}$, estimate $\hat\mu_a^{(k)}(X_S\cup X_\tau)$ in the trial.
        \State Regress its cross-fitted predictions on $X_\tau$ among trial participants, weighted by $1/\hat\omega^{(k)}(X_S)$, to obtain $\hat g_{a,\mathrm{res}}^{(k)}$.
        \State Set $\hat\tau^{(k)}_{\mathrm{gTL,res}}(X_\tau)=
        \hat g_{1,\mathrm{res}}^{(k)}(X_\tau)-\hat g_{0,\mathrm{res}}^{(k)}(X_\tau)$.
    \EndIf

    \For{$i\in\mathcal I_k$}
        \State Evaluate the corresponding fold-specific estimator at $X_{\tau,i}$ to obtain $\hat\tau_i$.
    \EndFor
\EndFor

\State Form $\hat\tau_{\mathrm{gTL}}(\cdot)=K^{-1}\sum_{k=1}^K\hat\tau^{(k)}(\cdot)$.
\State Predict $\hat\tau_{\mathrm{test}}=\hat\tau_{\mathrm{gTL}}(X_{\tau,\mathrm{test}})$.

\end{algorithmic}
\end{breakablealgorithm}

\clearpage

\begin{breakablealgorithm}
\caption{Cross-Fitted Generalized DR-Learner (gDR-Learner)}
\begin{algorithmic}[1]
\renewcommand{\baselinestretch}{1}\selectfont

\State \textbf{Input:} Data $\mathcal{D}$, test features $X_{\tau,\mathrm{test}}$, number of folds $K$, and estimation setting.
\State Partition $\{1,\dots,n\}$ into $K$ disjoint folds $\mathcal{I}_1,\dots,\mathcal{I}_K$, stratified by $S$.

\For{$k=1$ to $K$, with $\mathcal{I}_{-k}=\{1,\dots,n\}\setminus\mathcal{I}_k$}

    \State Estimate $\hat{\pi}_a^{(k)}(X_S)$ using trial units in $\mathcal{I}_{-k}$.
    \State Estimate $\hat{\omega}^{(k)}(X_S)$ using $\mathcal{I}_{-k}$; in the Subsample or Restricted setting, recover $\hat{\omega}^{(k)}$ from $\hat P(S=1\mid X_S,M=1)$ using eq.~\eqref{eq:omega-subsample}.

    \If{Standard or Subsample setting}
        \State For $a\in\{0,1\}$, estimate
        $\hat{\mu}_a^{(k)}(X_S\cup X_\tau)$ using trial units with $A=a$.

    \ElsIf{Prognostic setting}
        \State For $a\in\{0,1\}$, estimate
        $\hat{\mu}_a^{(k)}(X_S\cup X_\tau,X_P)$ using trial units with $A=a$.
        \State Regress its cross-fitted predictions on $X_S\cup X_\tau$ among trial participants to obtain
        $\hat g_{a,\mathrm{enr}}^{(k)}(X_S\cup X_\tau)$.

    \ElsIf{Restricted setting}
        \State For $a\in\{0,1\}$, estimate
        $\hat{\mu}_a^{(k)}(X_S\cup X_\tau)$ using trial units with $A=a$.
        \State Regress its cross-fitted predictions on $X_\tau$ among trial participants,
        weighted by $1/\hat{\omega}^{(k)}(X_S)$, to obtain
        $\hat g_{a,\mathrm{res}}^{(k)}(X_\tau)$.
    \EndIf

    \For{$i\in\mathcal{I}_k$}
        \State For trial participants, calculate the residual correction
        $$
        \widehat R_i^{(k)}
        =
        \frac{
        A_i-\hat{\pi}_1^{(k)}(X_{S,i})
        }{
        \hat{\omega}^{(k)}(X_{S,i})
        \hat{\pi}_1^{(k)}(X_{S,i})
        \hat{\pi}_0^{(k)}(X_{S,i})
        }
        \bigl\{Y_i-\hat{\mu}_{A_i}^{(k)}\bigr\},
        $$
        and set $\widehat R_i^{(k)}=0$ when $S_i=0$.

        \If{Standard or Subsample setting}
            \State Set
            $$
            \hat{\varphi}_i
            =
            \widehat R_i^{(k)}
            +\hat{\mu}_1^{(k)}(X_{S,i}\cup X_{\tau,i})
            -\hat{\mu}_0^{(k)}(X_{S,i}\cup X_{\tau,i}).
            $$

        \ElsIf{Prognostic setting}
            \State Set
            $$
            \hat{\varphi}_i
            =
            \widehat R_i^{(k)}
            +\hat g_{1,\mathrm{enr}}^{(k)}(X_{S,i}\cup X_{\tau,i})
            -\hat g_{0,\mathrm{enr}}^{(k)}(X_{S,i}\cup X_{\tau,i}).
            $$

        \ElsIf{Restricted setting}
            \State Set
            $$
            \hat{\varphi}_i
            =
            \widehat R_i^{(k)}
            +\hat g_{1,\mathrm{res}}^{(k)}(X_{\tau,i})
            -\hat g_{0,\mathrm{res}}^{(k)}(X_{\tau,i}).
            $$
        \EndIf
    \EndFor
\EndFor

\State Regress the cross-fitted pseudo-outcomes $\hat{\varphi}$ on $X_\tau$
to obtain $\hat{\tau}_{\mathrm{gDR}}(\cdot)$.
\State Predict
$\hat{\tau}_{\mathrm{test}}
=\hat{\tau}_{\mathrm{gDR}}(X_{\tau,\mathrm{test}})$.

\end{algorithmic}
\end{breakablealgorithm}

\section{Conformal Prediction Intervals}
\label{app:conformal}

This appendix provides full algorithmic details for the conformal inference procedures summarized in Section~\ref{sec:conformal}. Throughout, $\alpha \in (0,1)$ denotes the nominal miscoverage level and we use $R_i$ for conformity scores.

\subsection{Split Conformal Prediction}

Split conformal prediction provides finite-sample, distribution-free prediction intervals under exchangeability \citep{vovk_algorithmic_2005}. Given a regression model $\hat{g}$ trained on a training set $D_\mathrm{tr}$, one computes conformity scores $R_i = |Y_i - \hat{g}(X_i)|$ on a held-out calibration set $D_\mathrm{ca} = \{(X_i,Y_i)\}_{i \in \mathcal{I}_\mathrm{ca}}$. The prediction interval at a test point $x$ is
\begin{equation}
  \hat{C}(x) = \bigl[\hat{g}(x) - \hat{q},\hat{g}(x) + \hat{q}\bigr],
\end{equation}
where $\hat{q}$ is the $\lceil(1-\alpha)(1 + |\mathcal{I}_\mathrm{ca}|^{-1})\rceil$-th smallest conformity score. This yields marginal coverage $\mathbb{P}(Y \in \hat{C}(X)) \geq 1 - \alpha$ whenever calibration and test points are exchangeable.

\subsection{Weighted Conformal Inference for Counterfactuals}
\label{app:conformal-weighted}

Under covariate shift, calibration and test points are no longer exchangeable and the standard procedure fails. \citet{tibshirani_conformal_2019} restore validity by replacing the uniform empirical distribution over calibration scores with a likelihood-ratio-weighted distribution.

In our setting, a separate weighted conformal procedure is run for each treatment arm $a \in \{0,1\}$. The calibration set for arm $a$ consists of trial participants assigned to that arm, $\mathcal{I}^a_\mathrm{ca} = \{i \in \mathcal{I}_\mathrm{ca} : S_i = 1, A_i = a\}$. For each $i \in \mathcal{I}^a_\mathrm{ca}$, the conformity score is the absolute residual
\begin{equation}
  R_i = |Y_i - \hat{g}_a(X_i)|,
\end{equation}
where $\hat{g}_a$ is the fitted potential outcome model: for the gT-Learner, $\hat{g}_a(x) = \hat{\mu}_a(x)$; for the gDR-Learner, $\hat{g}_a(x) = \hat{\mathbb{E}}[\hat{\varphi}_a \mid X = x]$. The likelihood ratio weight, derived in eq.~\eqref{eq:conformal-weights}, is estimated as $W_i = \hat{w}_a(X_i) = \hat{\mathbb{P}}(S=1) / [\hat{\omega}(X_{S,i})\hat{\pi}_a(X_{S,i})]$, using the same nuisance models fitted for the gT- or gDR-Learner.

For a given test point $x$, define normalized weights
\begin{equation}
  \hat{p}_i(x) = \frac{W_i}{\displaystyle\sum_{j \in \mathcal{I}^a_\mathrm{ca}} W_j + \hat{w}_a(x)}, \qquad \hat{p}_\infty(x) = \frac{\hat{w}_a(x)}{\displaystyle\sum_{j \in \mathcal{I}^a_\mathrm{ca}} W_j + \hat{w}_a(x)}.
\end{equation}
The $(1-\alpha)$-th quantile of the weighted empirical distribution
\begin{equation}
  \hat{P}_a(x) = \sum_{i \in \mathcal{I}^a_\mathrm{ca}} \hat{p}_i(x)\delta_{R_i} + \hat{p}_\infty(x)\delta_\infty
\end{equation}
defines the half-width $\hat{q}_a(x)$. The weighted conformal interval for $Y^a$ is then
\begin{equation}
  \hat{C}_a(x) = \bigl[\hat{Y}^L(a,x),\hat{Y}^R(a,x)\bigr] = \bigl[\hat{g}_a(x) - \hat{q}_a(x),\hat{g}_a(x) + \hat{q}_a(x)\bigr],
\end{equation}
which achieves approximate marginal coverage $\mathbb{P}_{(X,Y^a)\sim P_0}(Y^a \in \hat{C}_a(X)) \geq 1 - \alpha$ over the target population distribution \citep{tibshirani_conformal_2019}.

\subsection{Naive Conformal Intervals}

Given weighted conformal intervals $\hat{C}_0(x)$ and $\hat{C}_1(x)$ at level $1-\alpha$, a prediction set for the \ac{ite} $Y^1 - Y^0$ is obtained by contrasting the two intervals:
\begin{equation}
  \hat{C}_\mathrm{ITE}(x) = \bigl[\hat{Y}^L(1;x) - \hat{Y}^R(0;x), \hat{Y}^R(1;x) - \hat{Y}^L(0;x)\bigr].
\end{equation}
If each counterfactual interval satisfies marginal coverage at level $1-\alpha/2$, then by a union bound,
\begin{equation}
  \mathbb{P}\left(Y^1 - Y^0 \in \hat{C}_\mathrm{ITE}(X)\right) \geq 1 - \alpha,
\end{equation}
regardless of the joint distribution of $(Y^0, Y^1)$ \citep{lei_conformal_2021}. Because the interval is constructed as if the two potential outcomes were independent, it is conservative whenever $Y^0$ and $Y^1$ are positively correlated within individuals.

\subsection{Inexact Conformal Inference}

\begin{figure}[t]
    \centering
    \includegraphics[width=\linewidth]{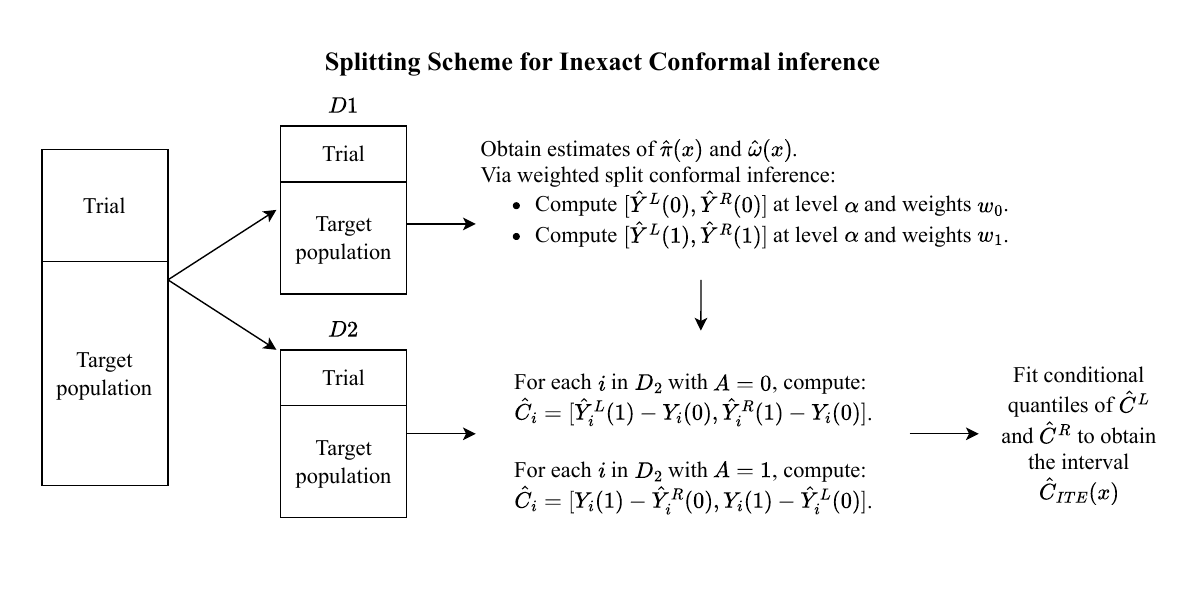}
    \caption{Sample splitting for the inexact conformal procedure. Weighted conformal intervals for $Y^0$ and $Y^1$ are fitted on $D_1$ and used to form pseudo-ITE intervals for trial units in $D_2$. Quantile regression on $D_2$ then gives $\tilde C_{\mathrm{ITE}}(x)$ for all units, including non-trial units.}
    \label{fig:inexact-splitting}
\end{figure}

The inexact procedure exploits the within-individual dependence between $Y^0$ and $Y^1$ to produce narrower intervals at the cost of forgoing the formal coverage guarantee \citep{lei_conformal_2021}. The data are partitioned into two folds $D_1$ and $D_2$ (Web Figure~\ref{fig:inexact-splitting}).

\textbf{Step 1.} On $D_1$, fit nuisance models $\hat{\mu}_a$, $\hat{\pi}_a$, $\hat{\omega}$ and run the weighted split conformal procedure of Web Appendix~\ref{app:conformal-weighted} to obtain counterfactual intervals $\hat{C}_0(\cdot)$ and $\hat{C}_1(\cdot)$ at level $1-\alpha$.

\textbf{Step 2.} For each trial unit $i \in D_2$, define the pseudo-\ac{ite} interval
\begin{equation}
  \hat{C}_i = \hat{C}_\mathrm{ITE}(X_i;A_i,Y_i) = \begin{cases} Y_i - \hat{C}_0(X_i), & A_i = 1, \\[4pt] \hat{C}_1(X_i) - Y_i, & A_i = 0. \end{cases}
\end{equation}
Conditional on $D_1$, the dataset $\{(X_i, \hat{C}_i)\}_{i \in D_2, S_i=1}$ is i.i.d.\ and each $\hat{C}_i$ covers the \ac{ite} with probability at least $1-\alpha$.

\textbf{Step 3.} Fit a quantile regression model $\tilde{C}_\mathrm{ITE}(x)$ to predict the left and right endpoints of $\hat{C}_i$ given $X_i$, using quantile levels $\gamma$ and $1-\gamma$ respectively for a chosen $\gamma$. The resulting interval generalizes to non-trial units ($S_i = 0$) for whom neither potential outcome is observed.

The intervals $\tilde{C}_\mathrm{ITE}(x)$ do not carry the marginal coverage guarantee in eq.~\eqref{eq:conformal-coverage}; their validity depends on the quantile regression model correctly capturing the conditional distribution of the pseudo-\ac{ite} interval widths.

\section{Simulation Design Details}
\label{sec:simulation-details}
For a target population of size $N$, we generate selection covariates $X_S\in[-1,1]^{d_S}$, treatment effect modifiers $X_\tau\in[-1,1]^{d_\tau}$, and additional prognostic covariates $X_P\in[-1,1]^{d_P}$. Their components are independent Uniform$(-1,1)$ variables apart from the prescribed overlap between $X_S$ and $X_\tau$. The outcome model uses $X_\mu=(X_S,X_P)$, where $d_\mu=d_S+d_P$. We set the first $d_S-\Delta$ components of $X_\tau$ equal to the corresponding components of $X_S$, where $\Delta=|X_S\setminus X_\tau|$, and denote the remaining selection covariates by $W=X_S\setminus X_\tau$.
The pCATE is
\begin{equation*}
\tau(X_\tau)=1+\frac{1}{\sqrt{d_\tau}}\sum_{j=1}^{d_\tau}h(X_{\tau,j}),
\end{equation*}
where $h(x)=x$ in the linear setting and $h(x)=x+\sin(5x)$ in the non-linear setting. For $\Delta>0$, write $W=(W_1,\ldots,W_\Delta)$ and define the selection score contributed by $W$ and its standardised tail transformation as
\begin{equation*}
U=\frac{1}{\sqrt{d_S}}\sum_{j=1}^{\Delta}W_j,
\qquad
q(u)=-u^2\mathbb I(u<0),
\qquad
r(u)=\frac{q(u)-\mathbb E\{q(U)\}}{\operatorname{sd}\{q(U)\}}.
\end{equation*}
The conditional treatment effect given $(X_\tau,W)$ is then
\begin{equation*}
\tau(X_\tau,W)=\tau(X_\tau)+3U+3r(U).
\end{equation*}
The additional terms introduce treatment-effect heterogeneity associated with trial selection, with nonlinear variation concentrated in the lower tail of the selection score. Both terms are set to zero when $\Delta=0$.

The linear outcome model is
\begin{equation*}
\mu_0^{\mathrm{lin}}(X_\mu)=\frac{1}{\sqrt{d_\mu}}\sum_{j=1}^{d_\mu}X_{\mu,j}.
\end{equation*}
The non-linear outcome model is:
\begin{equation*}
\mu_0^{\mathrm{nonlin}}(X_\mu)=\mu_0^{\mathrm{lin}}(X_\mu)
+\frac{1}{\sqrt{d_\mu}}\sum_{j=1}^{d_\mu}\sin(4\pi X_{\mu,j})
+\frac{1}{\sqrt{P}}\sum_{1\leq j<k\leq d_\mu}\sin(2\pi X_{\mu,j}X_{\mu,k}),
\end{equation*}
where $P=d_\mu(d_\mu-1)/2$ is the number of covariate pairs (the interaction term is omitted when $d_\mu=1$).
The potential outcomes are $Y(a)=\mu_0(X_\mu)+a\,\tau(X_\tau,W)+\varepsilon_a$ for $a\in\{0,1\}$, where $\varepsilon_0,\varepsilon_1\sim\mathcal N(0,1)$ are independent of each other and of the covariates, so that the \ac{ite} is $Y(1)-Y(0)=\tau(X_\tau,W)+\varepsilon_1-\varepsilon_0$.
Let $S\in\{0,1\}$ denote trial participation. Selection follows
$$
\mathbb{P}(S=1\mid X_S)=\operatorname{expit}\left(\alpha+\frac{1}{\sqrt{d_S}}\sum_{j=1}^{d_S}X_{S,j}\right),
$$
where $\alpha$ is chosen numerically so that $\mathbb{P}(S=1)=n_{1}/N$ \citep{robertson_using_2022}.
Treatment is randomized within the trial, $A\mid S=1\sim\operatorname{Bernoulli}(0.5)$, and recorded as $A=0$ when $S=0$. The outcome is
$$
Y=(1-A)Y(0)+AY(1)=\mu_0(X_\mu)+A\tau(X_\tau,W)+\varepsilon_A.
$$
Conformal coverage of the \ac{ite} is assessed against $\tau(X_\tau,W)+\varepsilon_1-\varepsilon_0$, using the realized noise of the observed arm and an independent draw for the other arm.
In Setting 3, a representative subsample of size $n_{\text{sub}}$ is drawn uniformly without replacement from the target population. Table~\ref{tab:sim_design} summarizes the setting-specific parameters.

\subsection{Simulation Implementation}
\label{app:simulation-implementation}
We use 5-fold cross-fitting. All outcome regressions, including the weighted regressions of the T- and gT-Learners, intermediate regressions of the enriched and restricted learners, and second-stage regressions on $X_\tau$, use  regularized gradient-boosted trees (XGBoost; maximum depth 2, learning rate 0.03, minimum child weight 100, $L_2$ penalty 10, row subsampling 0.8). The number of boosting rounds is selected by 5-fold internal cross-validation, with a maximum of 1000 rounds and early stopping after 30 rounds without improvement. Weighted regressions rescale their weights to have mean one. The treatment and trial participation models use XGBoost with 20 boosting rounds and otherwise default settings. Treatment-model predictors match the first-stage outcome predictors of each learner. Estimated participation probabilities are truncated at their 10th and 90th percentiles within each prediction set used for weighting or residual correction. The residual weights of the DR- and gDR-Learner pseudo-outcomes are stabilized by H\'ajek normalization.

\section{Additional Simulation Results}
\label{app:sim-results}
\begin{figure}[ht!]
    \centering
        \caption{Setting 1: average width of 95\% naive and inexact conformal prediction intervals, by trial sample size. }
    \includegraphics[width=\linewidth]{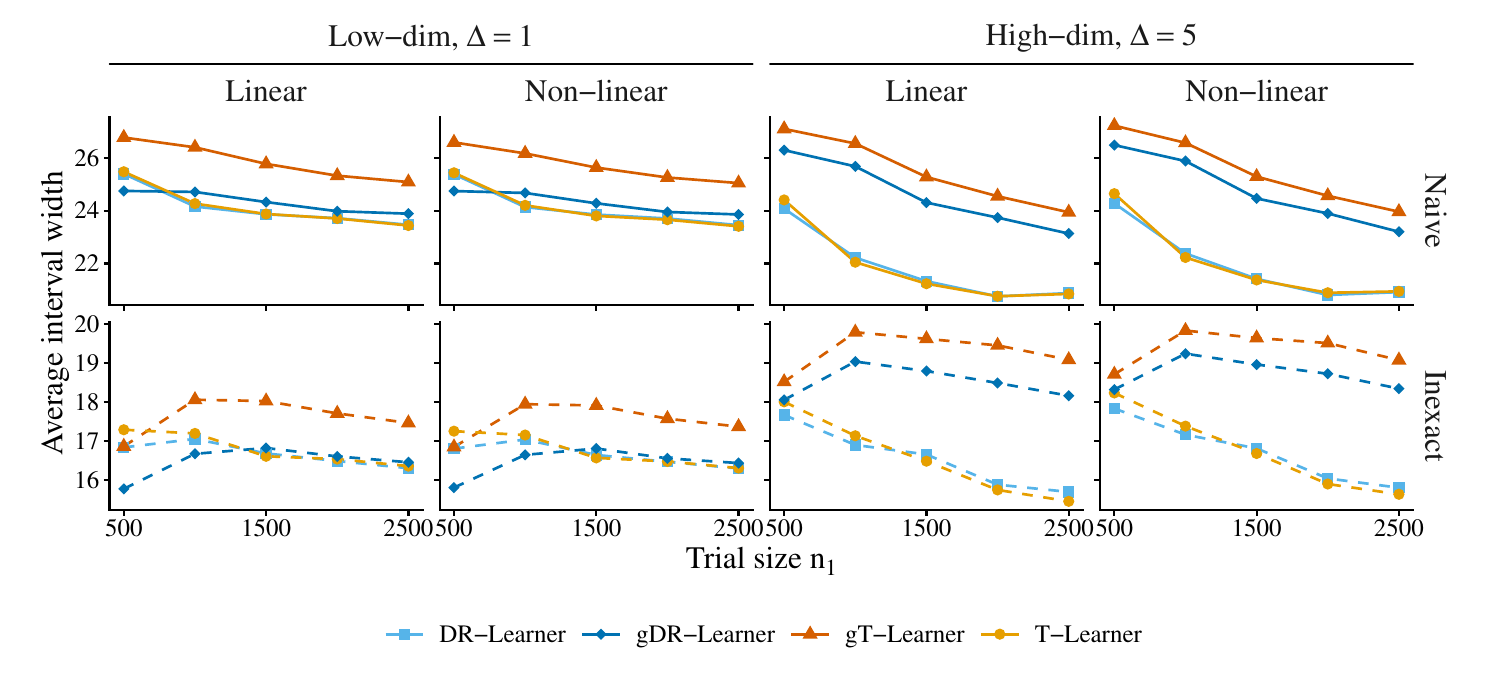}
\label{fig:conformal-interval-widths}
\end{figure}

Here we report supplementary results from the simulation study of Section~\ref{sec:simulation}, focusing on the validity and width of the conformal prediction intervals. Web Tables~\ref{tab:conformal_coverage} and~\ref{tab:conformal_coverage_ite} report empirical coverage of the naive and inexact conformal procedures across all four scenarios and sample sizes, scored against the \ac{pcate} and the \ac{ite}, respectively. Both procedures cover the \ac{pcate} in at least $96.5\%$ of cases throughout, as expected since they target the more variable \ac{ite}. Against the \ac{ite}, the naive procedure remains conservative (coverage $\geq 0.95$), whereas the inexact procedure undercovers: between $0.92$ and $0.95$ in the high-dimensional settings, and between $0.87$ and $0.90$ in the low-dimensional settings. Web Figure~\ref{fig:conformal-interval-widths} contrasts the corresponding interval widths: the inexact procedure yields narrower intervals than the naive procedure in all but one setting (DR-Learner, high-dimensional non-linear setting, $n_1=500$).

\begin{table}[h!]
\centering
\caption{Empirical coverage of the pCATE by 95\% naive and inexact conformal intervals, by trial sample size $n_1$, with $\Delta=1$ (low-dimensional), $\Delta=5$ (high-dimensional).}
\label{tab:conformal_coverage}
\resizebox{\textwidth}{!}{%
\begin{tabular}{llcccccccccc}
\toprule
 & & \multicolumn{5}{c}{Naive Conformal} & \multicolumn{5}{c}{Inexact Conformal} \\
\cmidrule(lr){3-7} \cmidrule(lr){8-12}
Scenario & Method & 500 & 1000 & 1500 & 2000 & 2500 & 500 & 1000 & 1500 & 2000 & 2500 \\
\midrule
Low-Dim: Linear CATE & DR-Learner & 1.000 & 1.000 & 1.000 & 1.000 & 1.000 & 0.972 & 0.972 & 0.970 & 0.969 & 0.968 \\
 & T-Learner & 1.000 & 1.000 & 1.000 & 1.000 & 1.000 & 0.974 & 0.972 & 0.970 & 0.970 & 0.969 \\
 & gDR-Learner & 1.000 & 1.000 & 1.000 & 1.000 & 1.000 & 0.965 & 0.966 & 0.966 & 0.966 & 0.966 \\
 & gT-Learner & 1.000 & 1.000 & 1.000 & 1.000 & 1.000 & 0.968 & 0.969 & 0.968 & 0.968 & 0.968 \\
\midrule
Low-Dim: Non-Linear CATE & DR-Learner & 1.000 & 1.000 & 1.000 & 1.000 & 1.000 & 0.972 & 0.972 & 0.969 & 0.969 & 0.968 \\
 & T-Learner & 1.000 & 1.000 & 1.000 & 1.000 & 1.000 & 0.974 & 0.972 & 0.970 & 0.970 & 0.969 \\
 & gDR-Learner & 1.000 & 1.000 & 1.000 & 1.000 & 1.000 & 0.965 & 0.966 & 0.966 & 0.966 & 0.966 \\
 & gT-Learner & 1.000 & 1.000 & 1.000 & 1.000 & 1.000 & 0.968 & 0.969 & 0.968 & 0.968 & 0.968 \\
\midrule
High-Dim: Linear CATE & DR-Learner & 1.000 & 1.000 & 1.000 & 1.000 & 1.000 & 1.000 & 1.000 & 1.000 & 1.000 & 1.000 \\
 & T-Learner & 1.000 & 1.000 & 1.000 & 1.000 & 1.000 & 1.000 & 1.000 & 1.000 & 1.000 & 1.000 \\
 & gDR-Learner & 1.000 & 1.000 & 1.000 & 1.000 & 1.000 & 1.000 & 1.000 & 1.000 & 1.000 & 1.000 \\
 & gT-Learner & 1.000 & 1.000 & 1.000 & 1.000 & 1.000 & 1.000 & 1.000 & 1.000 & 1.000 & 1.000 \\
\midrule
High-Dim: Non-Linear CATE & DR-Learner & 1.000 & 1.000 & 1.000 & 1.000 & 1.000 & 1.000 & 1.000 & 1.000 & 1.000 & 1.000 \\
 & T-Learner & 1.000 & 1.000 & 1.000 & 1.000 & 1.000 & 1.000 & 1.000 & 1.000 & 1.000 & 1.000 \\
 & gDR-Learner & 1.000 & 1.000 & 1.000 & 1.000 & 1.000 & 1.000 & 1.000 & 1.000 & 1.000 & 1.000 \\
 & gT-Learner & 1.000 & 1.000 & 1.000 & 1.000 & 1.000 & 1.000 & 1.000 & 1.000 & 1.000 & 1.000 \\
\bottomrule
\end{tabular}}
\end{table}

\begin{table}[htbp]
\centering
\caption{Empirical coverage of the ITE by 95\% naive and inexact conformal intervals, by trial sample size $n_1$, with $\Delta=1$ (low-dimensional), $\Delta=5$ (high-dimensional).}
\label{tab:conformal_coverage_ite}
\resizebox{\textwidth}{!}{%
\begin{tabular}{llcccccccccc}
\toprule
 & & \multicolumn{5}{c}{Naive Conformal} & \multicolumn{5}{c}{Inexact Conformal} \\
\cmidrule(lr){3-7} \cmidrule(lr){8-12}
Scenario & Method & 500 & 1000 & 1500 & 2000 & 2500 & 500 & 1000 & 1500 & 2000 & 2500 \\
\midrule
Low-Dim: Linear CATE & DR-Learner & 0.975 & 0.975 & 0.974 & 0.975 & 0.974 & 0.870 & 0.878 & 0.876 & 0.874 & 0.874 \\
 & T-Learner & 0.975 & 0.975 & 0.974 & 0.974 & 0.974 & 0.875 & 0.879 & 0.874 & 0.874 & 0.874 \\
 & gDR-Learner & 0.995 & 0.992 & 0.991 & 0.991 & 0.991 & 0.876 & 0.891 & 0.893 & 0.892 & 0.892 \\
 & gT-Learner & 0.995 & 0.992 & 0.991 & 0.991 & 0.990 & 0.874 & 0.893 & 0.895 & 0.894 & 0.894 \\
\midrule
Low-Dim: Non-Linear CATE & DR-Learner & 0.976 & 0.975 & 0.974 & 0.975 & 0.974 & 0.871 & 0.879 & 0.876 & 0.874 & 0.874 \\
 & T-Learner & 0.976 & 0.976 & 0.974 & 0.975 & 0.974 & 0.876 & 0.880 & 0.875 & 0.875 & 0.875 \\
 & gDR-Learner & 0.995 & 0.992 & 0.991 & 0.991 & 0.991 & 0.877 & 0.891 & 0.894 & 0.892 & 0.892 \\
 & gT-Learner & 0.994 & 0.992 & 0.991 & 0.991 & 0.990 & 0.875 & 0.893 & 0.895 & 0.893 & 0.893 \\
\midrule
High-Dim: Linear CATE & DR-Learner & 0.965 & 0.958 & 0.956 & 0.954 & 0.956 & 0.921 & 0.923 & 0.925 & 0.923 & 0.923 \\
 & T-Learner & 0.964 & 0.957 & 0.956 & 0.954 & 0.955 & 0.922 & 0.924 & 0.924 & 0.922 & 0.922 \\
 & gDR-Learner & 0.986 & 0.977 & 0.974 & 0.973 & 0.971 & 0.938 & 0.944 & 0.946 & 0.947 & 0.947 \\
 & gT-Learner & 0.984 & 0.975 & 0.973 & 0.971 & 0.970 & 0.928 & 0.938 & 0.942 & 0.944 & 0.945 \\
\midrule
High-Dim: Non-Linear CATE & DR-Learner & 0.966 & 0.959 & 0.957 & 0.955 & 0.956 & 0.924 & 0.926 & 0.926 & 0.925 & 0.924 \\
 & T-Learner & 0.966 & 0.959 & 0.957 & 0.955 & 0.956 & 0.925 & 0.926 & 0.927 & 0.924 & 0.924 \\
 & gDR-Learner & 0.986 & 0.977 & 0.974 & 0.973 & 0.971 & 0.939 & 0.946 & 0.947 & 0.948 & 0.948 \\
 & gT-Learner & 0.984 & 0.975 & 0.973 & 0.971 & 0.970 & 0.930 & 0.939 & 0.943 & 0.944 & 0.945 \\
\bottomrule
\end{tabular}}
\end{table}

\section{CASS Analysis}
\label{app:cass}
\subsection{Baseline Covariates}
\label{app:cass-baseline}

Baseline covariates were selected to align with previous analyses of the CASS data \citep{olschewski_analysis_1992, robertson_regression-based_2023} and included demographic variables, clinical characteristics, and measures of disease severity. Summary statistics for all baseline covariates, stratified by randomized trial participation, are reported in Web Table~\ref{tab:summary_stats}.

\begin{figure}
    \centering
    \includegraphics[width=0.8\linewidth]{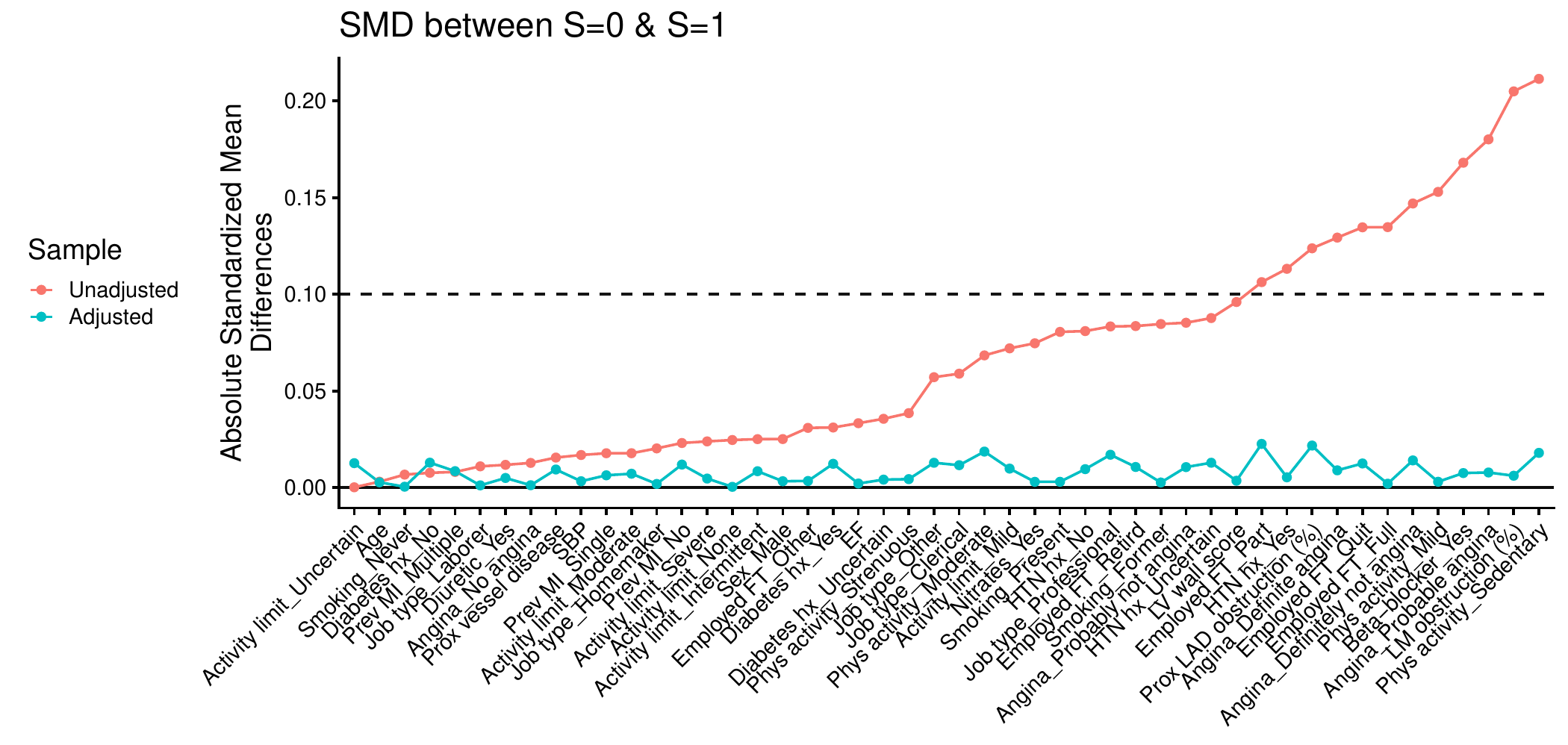}
\caption{CASS: absolute standardized mean differences between trial ($S=1$) and registry ($S=0$) patients, before and after inverse probability of sampling weighting (logistic selection model fitted on complete cases).}
    \label{fig:smd}
\end{figure}

To assess comparability between randomized and non-randomized patients, we examined standardized mean differences (SMDs) before and after re-weighting using inverse probability of sampling weights derived from a logistic regression fitted on complete cases. Web Figure~\ref{fig:smd} shows that prior to re-weighting, several covariates exhibited meaningful imbalance, including angina status and measures of coronary artery obstruction. After weighting, balance was substantially improved across all covariates.

\begin{table}[htbp]
\centering
\small 
\renewcommand{\arraystretch}{0.92} 
\caption{CASS baseline characteristics by trial participation ($S$) and, within the trial, by arm ($A=1$ surgery, $A=0$ medical therapy). Mean (SD) for continuous and $n$ (\%) for binary variables. MI, myocardial infarction; PLAA, proximal left anterior artery; LMA, left main artery.}
\label{tab:summary_stats}
\begin{tabular}{@{}lcccc@{}}
\toprule
\textbf{Variable} & \textbf{S=0} & \textbf{S=1} & \textbf{S=1 \& A=0} & \textbf{S=1 \& A=1} \\ 
\midrule
Number of patients & 1,319 & 780 & 390 & 390 \\
Age & 51 (8) & 51 (7) & 51 (7) & 51 (7) \\
Angina status & 234 (17.7) & 131 (16.8) & 67 (17.2) & 64 (16.4) \\
Regular beta-blocker usage & 719 (54.5) & 338 (43.3) & 166 (42.6) & 172 (44.1) \\
Regular diuretic usage & 205 (15.5) & 119 (15.3) & 52 (13.3) & 67 (17.2) \\
Ejection fraction & 60 (12) & 60 (13) & 60 (13) & 61 (13) \\
Employment status & 893 (67.7) & 527 (67.6) & 250 (64.1) & 277 (71.0) \\
Type of employment & & & & \\
\hspace{3mm}Clerical & 319 (24.2) & 213 (27.3) & 110 (28.2) & 103 (26.4) \\
\hspace{3mm}Homemaker & 62 (4.7) & 32 (4.1) & 19 (4.9) & 13 (3.3) \\
\hspace{3mm}Laborer & 410 (31.1) & 305 (39.1) & 148 (37.9) & 157 (40.3) \\
\hspace{3mm}Professional & 413 (31.3) & 191 (24.5) & 96 (24.6) & 95 (24.4) \\
\hspace{3mm}Other & 115 (8.7) & 39 (5.0) & 17 (4.4) & 22 (5.6) \\
Left ventricular wall score & 7.39 (2.95) & 7.44 (2.92) & 7.41 (2.89) & 7.47 (2.94) \\
Regular nitrates usage & 608 (46.1) & 358 (46.0) & 174 (44.6) & 184 (47.3) \\
History of previous MI & 755 (57.3) & 467 (59.9) & 244 (62.6) & 223 (57.2) \\
Male & 1,196 (90.7) & 704 (90.3) & 351 (90.0) & 353 (90.5) \\
Smoking status & & & & \\
\hspace{3mm}Never & 231 (17.5) & 127 (16.3) & 61 (15.6) & 66 (16.9) \\
\hspace{3mm}Former & 651 (49.4) & 343 (44.0) & 170 (43.6) & 173 (44.4) \\
\hspace{3mm}Present & 437 (33.1) & 310 (39.7) & 159 (40.8) & 151 (38.7) \\
High limitation of activity & 765 (58.0) & 471 (60.4) & 247 (63.3) & 224 (57.4) \\
High level of physical activity & 293 (22.2) & 180 (23.1) & 83 (21.3) & 97 (24.9) \\
History of hypertension & 352 (26.7) & 237 (30.4) & 111 (28.5) & 126 (32.3) \\
History of diabetes & 82 (6.2) & 66 (8.5) & 31 (7.9) & 35 (9.0) \\
PLAA \% obstruction & 40 (39) & 36 (38) & 35 (37) & 37 (38) \\
LMA \% obstruction & 6 (14) & 4 (11) & 3 (10) & 4 (12) \\
Any proximal diseased vessels & 832 (63.3) & 492 (63.2) & 253 (65.0) & 239 (61.4) \\
Systolic blood pressure & 130 (18) & 130 (18) & 131 (19) & 130 (17) \\
\bottomrule
\end{tabular}
\end{table}
\subsection{Non-parametric Learners}
\label{app:cass-np}

The \ac{pcate} was evaluated on a grid of values for the treatment-modifying covariates. For analyses involving \ac{ef}, we considered a grid spanning $[30,80]$ with increments of $0.1$. To reduce sensitivity to a particular sample split, each learner was refitted 50 times using different random splits, and the final point estimates were obtained by taking the median across repetitions.

Both learners used 5 fold cross fitting stratified by S. The outcome regressions were estimated using random forests with 500 trees under default settings. The participation model was fit on all baseline covariates while the treatment model used age, angina, \ac{ef}, systolic blood pressure, PLAA obstruction and left ventricular wall score. Both were trained using probability forests with 200 trees and one third of predictors sampled at each split. Minimum leaf size was chosen from $\{20,50,100,200\}$ by out-of-bag log-loss. The gDR residual weights were stabilized by Hajek normalization. In the second stage we used a regression forest with minimum leaf size chosen from $\{100,200,400,800\}$ by out-of-bag mean squared error.

\subsection{Multiple Imputation}
\label{app:cass-imputation}

Missing baseline covariates were handled using Multiple Imputation by Chained Equations (MICE). For the non-parametric learners, random forest imputation models were used to accommodate non-linearities and interactions among covariates. For the parametric analyses, categorical variables were imputed using logistic regression and continuous variables using normal linear regression.

In total, $M=\M$ imputed datasets were generated for each analytical framework. Within each imputed dataset, the full estimation procedure was applied, yielding imputation-specific estimates $\hat{\tau}(\tilde{x})^{m}$ for $m=1,\ldots,M$. Final point estimates were obtained by averaging across imputations,
\[
\hat{\tau}(\tilde{x}) = \frac{1}{M} \sum_{m=1}^{M} \hat{\tau}(\tilde{x})^{m}.
\]
This approach propagates uncertainty due to missing data into the final estimates while maintaining compatibility between the imputation and analysis models.

\subsection{Parametric Learners}
\label{app:cass-parametric}

Parametric versions of the gT- and gDR-Learners were implemented following \citet{robertson_regression-based_2023}. Outcome models were specified using spline-based regressions to accommodate nonlinear effects of \ac{ef}, with treatment effect heterogeneity modelled through interactions between treatment and selected effect modifiers.

The trial participation model was trained using baseline covariates identified by \citet{olschewski_analysis_1992} (Web Table~\ref{tab:summary_stats}). To allow for heterogeneity, analyses were conducted separately for individuals with and without a history of myocardial infarction (MI). Age and \ac{ef} were modelled using cubic B-splines (order 3) with an interior knot at the median. The treatment propensity score was estimated via logistic regression including severity of angina, heart failure status, and \ac{ef}, consistent with the original treatment assignment protocol.

For the parametric gDR-Learner, pseudo-outcomes were constructed using all baseline covariates. In the second stage, the \ac{pcate} was estimated via ordinary least squares series regression of the pseudo-outcomes on a basis expansion $p(x_T)$ of the treatment-modifying covariates, $\hat\tau(x_T) = p(x_T)\hat\beta$, where $p(x_T)$ consisted of a cubic B-spline basis for \ac{ef} and main effects for age or sex, together with their interactions with \ac{ef}.

For the parametric gT-Learner, inverse probability of sampling weights were used to fit weighted outcome regressions separately under each treatment arm. The outcome models included only the treatment effect modifiers of interest, namely the spline basis for \ac{ef} and its interactions with age or sex.

Standard errors for the parametric gDR-Learner were obtained using the Huber-White sandwich estimator \citep{huber_behavior_1967, white_heteroskedasticity-consistent_1980}. In the parametric specification, the nuisance estimator classes (logistic regression for the propensity and selection models, spline regression for the outcome models) are finite-dimensional and therefore satisfy the Donsker condition. Consequently, the first-stage nuisance estimation error is asymptotically negligible in the influence function expansion, and uncertainty in the nuisance estimates can be ignored when constructing standard errors for the second-stage regression \citep{semenova_debiased_2021}. This contrasts with the non-parametric learners used in the main analysis, where the Donsker condition does not hold and cross-fitting is required (Section~\ref{sec:cross-fitting}).
For the parametric gT-Learner, standard errors were computed using non-parametric bootstrap resampling within each imputed dataset, as recommended by \citet{schomaker_bootstrap_2018}, with $\B$ bootstrap replicates.

Inference was pooled across imputations using Rubin's rules \citep{rubin_multiple_1978}, combining within- and between-imputation variability. Under congenial imputation and analysis models, the resulting variance estimator is asymptotically unbiased and normal \citep{bartlett_bootstrap_2019}. Pointwise $(1-\alpha)\%$ confidence intervals were constructed as
\[
\hat\tau(\tilde x) \pm z_{1-\alpha/2}\hat\sigma(\tilde x).
\]

Uniform confidence bands were obtained for the parametric gDR-Learner using a multiplier (weighted) bootstrap \citep{semenova_debiased_2021, belloni_uniformly_2019, fan_estimation_2022}. Bootstrap weights were drawn from an exponential distribution, and the second-stage regression was refitted under these weights to obtain bootstrap estimates $\hat\tau^{b,m}(\tilde x)$. For each replicate, we computed the maximum absolute $t$-statistic over $\tilde x\in\mathcal Q$, yielding an imputation-specific critical value $\hat{\mathcal C}^m_{1-\alpha}$. These were averaged across imputations to obtain $\bar{\mathcal C}_{1-\alpha}$, and uniform confidence bands were constructed as
\[
\hat\tau(\tilde x) \pm \bar{\mathcal C}_{1-\alpha}\hat\sigma(\tilde x).
\]

We report 95\% pointwise confidence intervals for both parametric learners and uniform confidence bands for the gDR-Learner only. While these parametric intervals are substantially narrower than the conformal prediction sets used for the non-parametric learners, their validity depends on correct specification of the propensity score model and at least one of the outcome or selection models.

\subsection{Conformal Prediction Intervals}
\label{app:conformal-cass}

For the non-parametric learners, uncertainty was quantified using an inexact conformal inference procedure. The data were split into a training set (90\%) used to fit the gT- and gDR-Learners and a calibration set (10\%) used to train quantile regression models targeting the 25th and 75th percentiles of the conformity scores. 

Conformal prediction intervals were constructed using an impute-then-predict strategy based on a single imputed dataset. As shown by \citet{zaffran_conformal_2023}, this approach yields prediction sets that are marginally valid regardless of the imputation model or missingness mechanism. Since the outcome is binary, the resulting counterfactual outcome intervals $\hat{C}_a(x)$ were truncated to lie within $[0,1]$. In instances where the likelihood ratio between treatment groups was extreme, leading to infinite-width intervals, the prediction sets were truncated at the maximum feasible width.

\subsection{Additional Subgroup Analyses}
\label{app:cass-subgroups}

\begin{figure}
    \centering
        \includegraphics[width=0.49\linewidth]{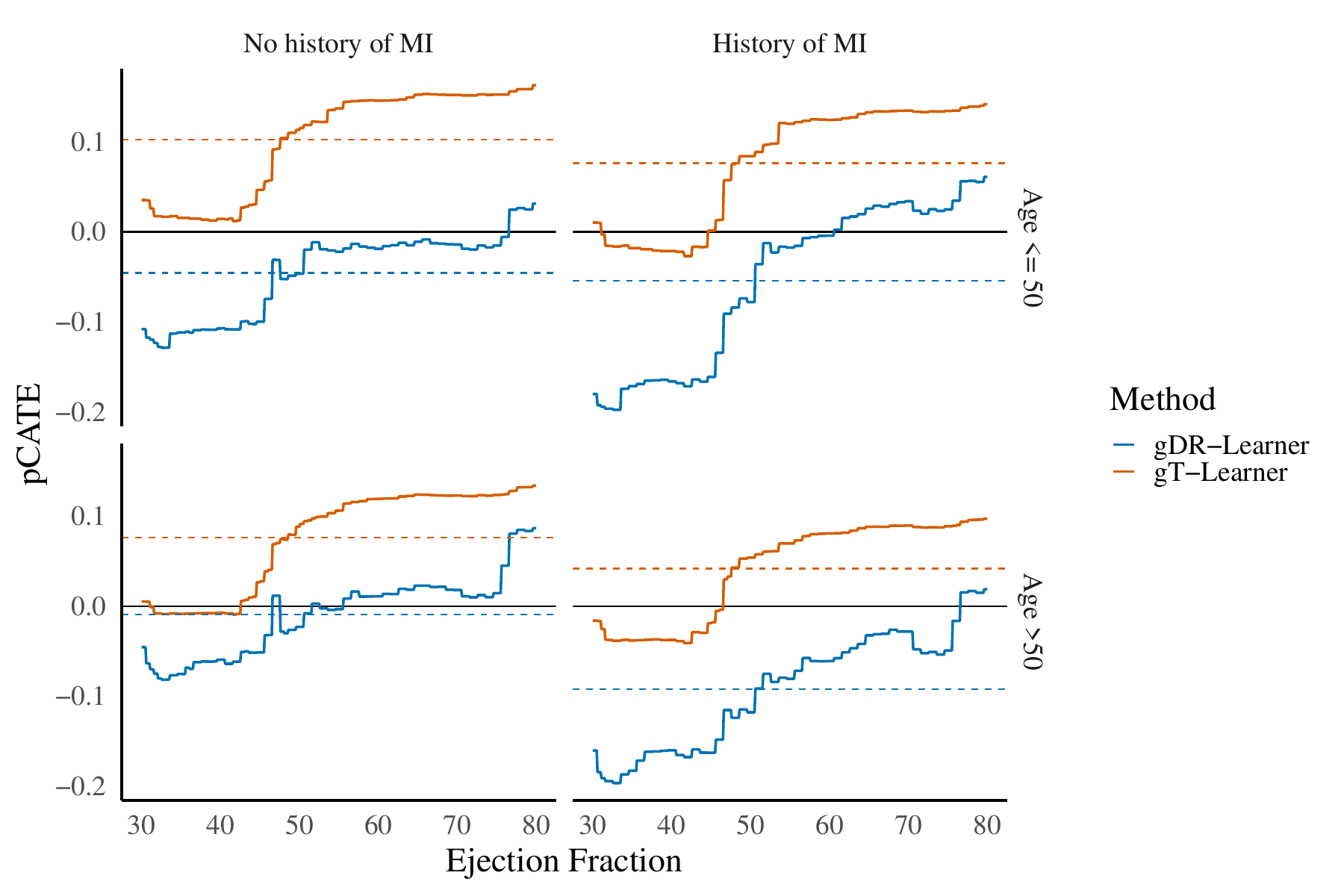}\hfill
        \includegraphics[width=0.49\linewidth]{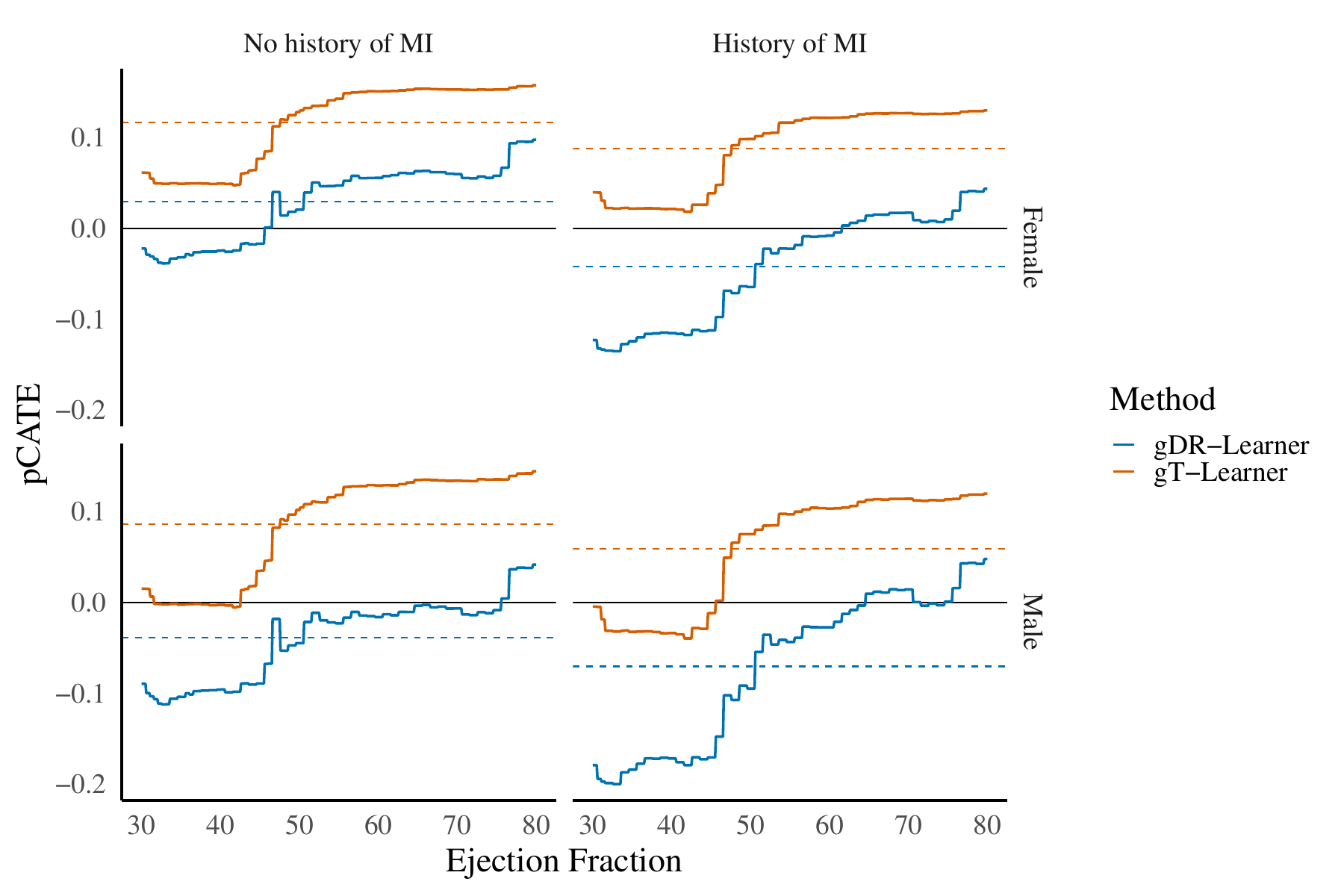}
\caption{CASS: non-parametric gDR- and gT-Learner pCATE estimates against ejection fraction, by MI history and age (left) or sex (right). Pointwise medians over 50 cross-fitting replications. Horizontal lines indicate subgroup means over the grid.}
    \label{fig:np-ejec_frac-subgroup}
\end{figure}

\begin{figure}
    \centering
    \includegraphics[width=0.9\linewidth]{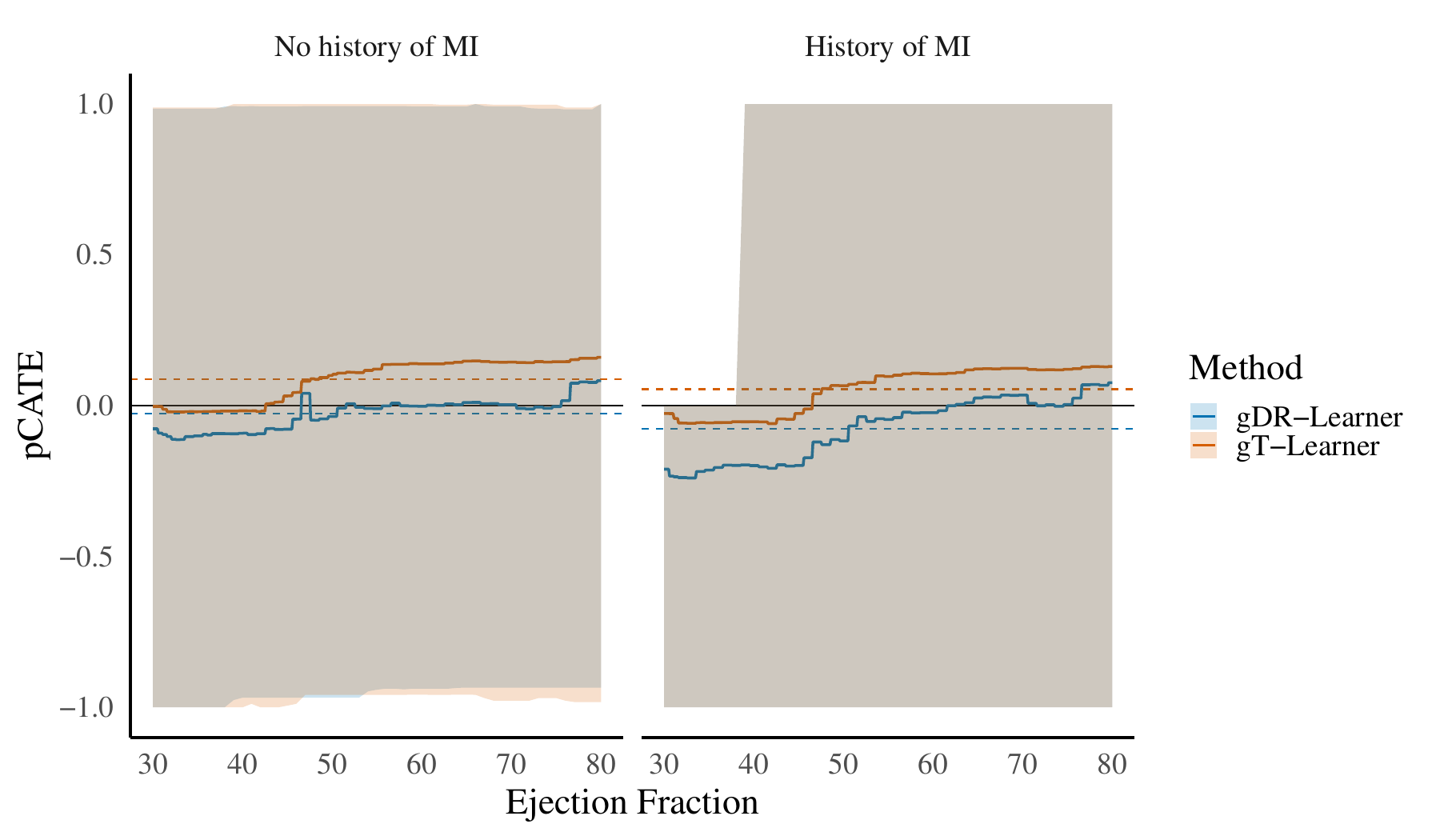}
\caption{CASS: non-parametric gDR- and gT-Learner pCATE estimates against ejection fraction and MI history. As main text Figure~\ref{fig:np-ejec_frac}, with inexact conformal prediction intervals (shaded).}
    \label{fig:np-ejec_frac-conformal}
\end{figure}

\begin{figure}
\centering
    \includegraphics[width=0.7\linewidth]{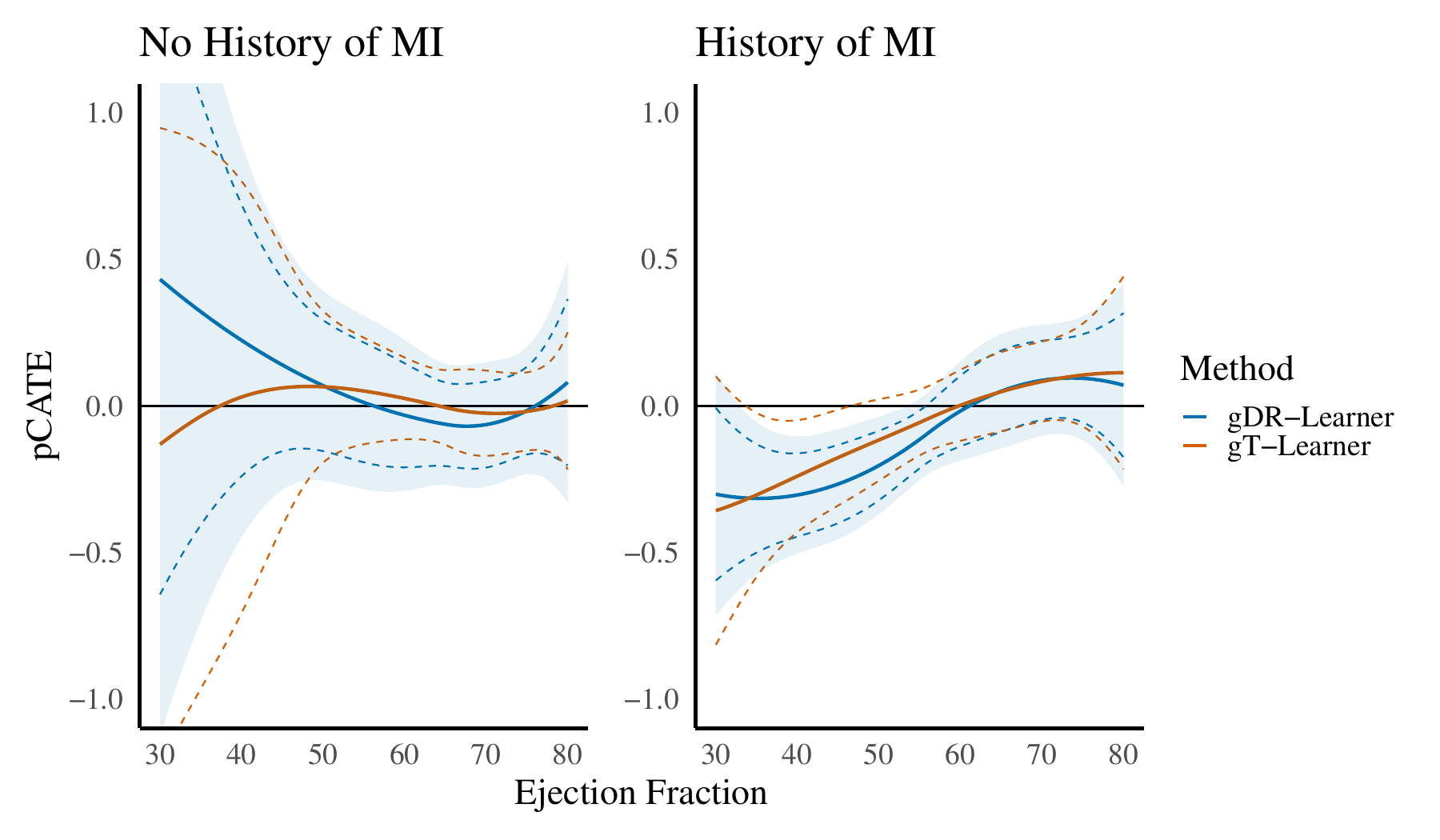}
\caption{CASS: parametric (spline) gDR- and gT-Learner pCATE estimates against ejection fraction, by MI history, with 95\% pointwise confidence intervals (dashed) and 95\% uniform confidence bands for the gDR-Learner (shaded).}
        \label{fig:param-se}
\end{figure}

\begin{figure}
    \centering
    \includegraphics[width=0.75\linewidth]{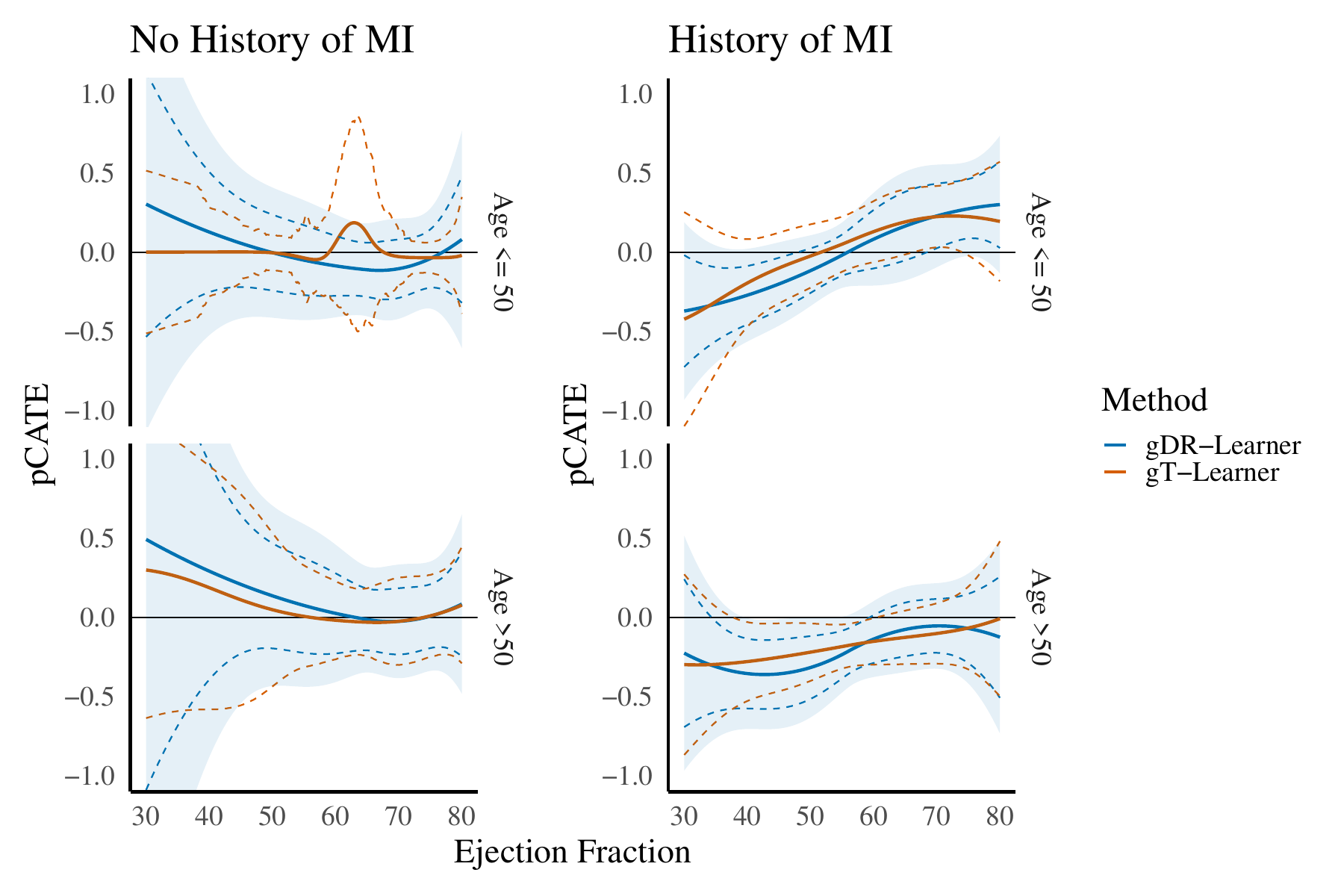}
\caption{CASS: parametric (spline) gDR- and gT-Learner pCATE estimates against ejection fraction, by MI history and age, with 95\% pointwise confidence intervals (dashed) and 95\% uniform confidence bands for the gDR-Learner (shaded).}
    \label{fig:age-param}
\end{figure}

\begin{figure}
    \centering
    \includegraphics[width=0.75\linewidth]{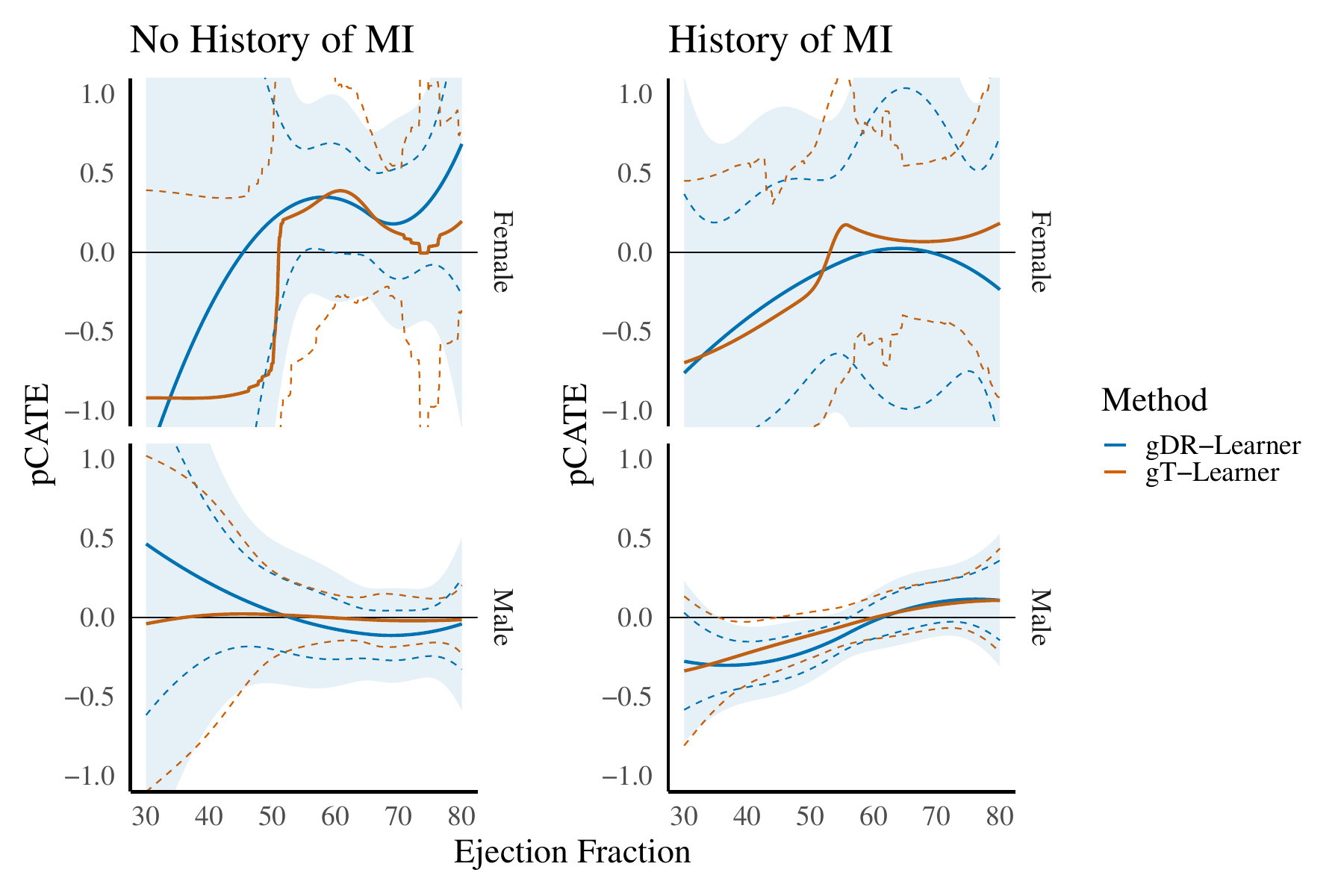}
\caption{CASS: parametric (spline) gDR- and gT-Learner pCATE estimates against ejection fraction, by MI history and sex, with 95\% pointwise confidence intervals (dashed) and 95\% uniform confidence bands for the gDR-Learner (shaded).}
    \label{fig:sex-param}
\end{figure}

This section presents subgroup-stratified \ac{pcate} estimates for the CASS analysis. Web Figure~\ref{fig:np-ejec_frac-subgroup} shows the non-parametric gDR- and gT-Learner estimates across the grid of effect modifiers, stratified by history of myocardial infarction together with age or sex, and Web Figure~\ref{fig:np-ejec_frac-conformal} adds inexact conformal prediction intervals for the estimates over ejection fraction. Web Figures~\ref{fig:param-se}--\ref{fig:sex-param} report the corresponding parametric spline-based estimates: Web Figure~\ref{fig:param-se} gives the pooled \ac{ef} estimate with pointwise confidence intervals and uniform confidence bands, while Web Figures~\ref{fig:age-param} and~\ref{fig:sex-param} stratify by age and sex respectively, with naive conformal prediction sets. 


\end{document}